\documentclass[usenatbib]{mnras}

\usepackage{newtxtext,newtxmath}

\usepackage[T1]{fontenc}

\DeclareRobustCommand{\VAN}[3]{#2}
\let\VANthebibliography\thebibliography
\def\thebibliography{\DeclareRobustCommand{\VAN}[3]{##3}\VANthebibliography}

\usepackage{graphicx}	
\usepackage{amsmath}	
\usepackage{subcaption}
\usepackage{mwe}
\usepackage{siunitx}
\DeclareSIUnit{\angstrom}{\textup{\AA}} 
\usepackage{hyperref}
\usepackage{orcidlink}

\title[NCS - VII. Disk Candidate in Orion]{NGTS clusters survey - VII. An Enigmatic Short-Period Circumsecondary Disk Candidate in Orion}

\author[N Mallaghan et al.]{\parbox{\textwidth}{\Large
Niamh Mallaghan\orcidlink{0009-0004-4749-8173},$^{1}$\thanks{E-mail: nmallaghan01@qub.ac.uk} 
Ernst J. W. de Mooij\orcidlink{0000-0001-6391-9266},$^{1}$ 
Christopher A. Watson\orcidlink{0000-0002-9718-3266},$^{1}$ 
Edward Gillen\orcidlink{0000-0003-2851-3070},$^{2}$ 
Matthew Kenworthy\orcidlink{0000-0002-7064-8270},$^{3}$
Louise D. Nielsen\orcidlink{0000-0002-5254-2499},$^{4}$ 
David R. Anderson\orcidlink{0000-0001-7416-7522},$^{5}$
Matthew P. Battley\orcidlink{0000-0002-1357-9774},$^{2}$ 
Edward M. Bryant\orcidlink{0000-0001-7904-4441},$^{6,7}$ 
Matthew R. Burleigh\orcidlink{0000-0003-0684-7803},$^{8}$ 
Benjamin M. J. Cadell\orcidlink{0009-0006-5883-3138},$^{1}$ 
Sarah L. Casewell\orcidlink{0000-0003-2478-0120},$^{8}$
Alexander Chaushev,$^{9}$ 
V. S. Dhillon\orcidlink{0000-0003-4236-9642},$^{10,11}$
Elsa Ducrot\orcidlink{0000-0002-7008-6888},$^{12,13}$
Jorge Fern\'andez Fern\'andez\orcidlink{0000-0002-1416-2188},$^{6,7}$
Samuel Gill\orcidlink{0000-0002-4259-0155},$^{6,7}$ 
Michaël Gillon\orcidlink{0000-0003-1462-7739},$^{14}$
Michael R. Goad\orcidlink{0000-0002-2908-7360},$^{8}$ 
Faith Hawthorn\orcidlink{0000-0002-8675-182X},$^{6,7,15}$
Katlyn L. Hobbs\orcidlink{0009-0004-4519-5080},$^{1}$ 
James McCormac\orcidlink{0000-0003-1631-4170},$^{6,7}$
Maximiliano Moyano,$^{5}$
Catriona A. Murray\orcidlink{0000-0001-8504-5862},$^{16}$
Toby Rodel\orcidlink{0009-0009-2175-7284},$^{1}$ 
Suman Saha\orcidlink{0000-0001-8018-0264},$^{17,18}$
Ramotholo R. Sefako,$^{19}$ 
John Southworth\orcidlink{0000-0002-3807-3198},$^{20}$
Mathilde Timmermans\orcidlink{0009-0008-2214-5039},$^{21}$
Amaury H. M. J. Triaud\orcidlink{0000-0002-5510-8751},$^{21}$
Jose Vines\orcidlink{0000-0002-2135-9018},$^{5}$
Richard G. West\orcidlink{0000-0001-6604-5533},$^{6,7}$
Peter J. Wheatley\orcidlink{0000-0003-1452-2240},$^{6,7}$
Tafadzwa Zivave\orcidlink{0009-0001-8055-995X}$^{6,7}$
}
\vspace{3mm}
\\
$^{1}$Astrophysics Research Centre, Queen's University Belfast, Northern Ireland, BT7 1NN, UK \\
$^{2}$Astronomy Unit, Queen Mary University of London, Mile End Road, London E1 4NS, UK \\
$^{3}$Leiden Observatory, Leiden University, Einsteinweg 55, NL-2333 CC Leiden, The Netherlands \\
$^{4}$University Observatory, Faculty of Physics, Ludwig-Maximilians-Universit{\"a}t M{\"u}nchen, Scheinerstr. 1, 81679 Munich, Germany\\
$^{5}$Instituto de Astronom\'ia, Universidad Cat\'olica del Norte, Angamos 0610, 1270709, Antofagasta, Chile \\
$^{6}$Department of Physics, University of Warwick, Gibbet Hill Road, Coventry CV4 7AL, UK\\
$^{7}$Centre for Exoplanets and Habitability, University of Warwick, Gibbet Hill Road, Coventry CV4 7AL, UK \\
$^{8}$School of Physics and Astronomy, University of Leicester, University Road, Leicester, LE1 7RH, UK\\
$^{9}$University of California Irvine, Department of Physics and Astronomy, Irvine, California, United States\\
$^{10}$Astrophysics Research Cluster, School of Mathematical and Physical Sciences, University of Sheffield, Sheffield S3 7RH, UK\\
$^{11}$Instituto de Astrof\'isica de Canarias, E-38205 La Laguna, Tenerife, Spain\\
$^{12}$LESIA, Observatoire de Paris, CNRS, Université Paris Diderot, Université Pierre et Marie Curie, Meudon, France\\
$^{13}$Université Paris-Saclay, Université Paris Cité, CEA, CNRS, AIM, Gif-sur-Yvette, France\\
$^{14}$Astrobiology Research Unit, Université de Liège, 19C Allée du 6 Août, 4000 Liège, Belgium\\
$^{15}$Rugby School, Lawrence Sheriff St, Rugby, Warwickshire, CV22 5EH, UK\\
$^{16}$Department of Earth, Atmospheric and Planetary Sciences, Massachusetts Institute of Technology, 77 Massachusetts Avenue, Cambridge, MA 02139, USA\\
$^{17}$Instituto de Estudios Astrofísicos, Facultad de Ingeniería y Ciencias, Universidad Diego Portales, Av. Ejército Libertador 441, Santiago, Chile\\
$^{18}$Centro de Excelencia en Astrofísica y Tecnologías Afines (CATA), Camino El Observatorio 1515, Las Condes, Santiago, Chile\\
$^{19}$South African Astronomical Observatory, P.O. Box 9, Observatory, Cape Town 7935, South Africa\\
$^{20}$Astrophysics Group, Keele University, Staffordshire, ST5 5BG, UK\\
$^{21}$School of Physics \& Astronomy, University of Birmingham, Edgbaston, Birmingham B15 2TT, United Kingdom
}

\date{Accepted XXX. Received YYY; in original form ZZZ}

\pubyear{\the\year{}}

\begin{document}
\label{firstpage}
\pagerange{\pageref{firstpage}--\pageref{lastpage}}
\maketitle

\begin{abstract}

Young stellar systems provide key insights into the processes governing star/planet formation, often exhibiting complex and evolving photometric variability. In this paper, we present an analysis of the object around NOI\,105872, observed by the Next Generation Transit Survey (NGTS) to eclipse a young ($4.1\pm0.4$\,Myr) M-type star in the Orion Nebula Cluster. The system displays deep, asymmetric eclipses with a period of $\sim0.69$\,days, alongside variability in morphology. We combine photometric observations spanning approximately a decade from multiple ground- and space-based facilities to investigate the nature of this system. We explore a range of physical scenarios to reproduce the eclipse profiles and their evolution. After analysis, our preferred explanation is a circumsecondary disk surrounding a companion in orbit around the M-type primary. We adopt a six-parameter disk model and fit the eclipse profiles independently to each dataset. The best-fitting solution yields a disk radius of $\sim 0.005\,\mathrm{au}$ or $0.98\,R_{*}$, with broadly consistent orientations across epochs, but with variations that may be attributed to disk precession or intrinsic stellar variability. Constraints from the Hill radius imply a companion mass consistent with a very low-mass stellar companion. These results establish NOI\,105872 as a candidate for a circumsecondary disk system in a young stellar environment. Continued photometric monitoring and spectroscopic follow-up will be essential to confirm this interpretation. If confirmed, systems such as NOI\,105872 provide valuable benchmarks for understanding disk dynamics, companion formation, and the origin of complex eclipse phenomena in young stars.

\end{abstract}




\section{Introduction}



Young stellar clusters provide some of the most powerful laboratories for understanding the processes that govern star and planet formation. In these environments, populations of coeval stars (stars that formed at the same time and from the same molecular cloud) with similar initial chemical compositions, but a range of masses and evolutionary stages, allow us to isolate the physical mechanisms that shape stellar evolution \citep{Lada2003,Portegies2010}. In particular, open clusters and star-forming regions offer a statistically rich sample for investigating early stellar variability, circumstellar disc evolution, and the onset of planet formation.

Wide-field time-domain surveys of clusters have significantly advanced this field by enabling high-precision, long-baseline photometric monitoring of young stellar populations. In particular, surveys such as the Next Generation Transit Survey (NGTS; \citealt{NGTS}) have demonstrated the value of cluster observations in identifying and characterising variable young stars and substellar companions \citep{Gillen2020, Jackman2020, Smith2021, Moulton2023, Smith2023}. These datasets have revealed complex and often unexpected light curve morphologies that are closely linked to the presence of circumstellar material and dynamical interactions in young systems.

One of the most intriguing outcomes of such surveys is the identification of unusual eclipse and dimming behaviour in young stellar objects. A prominent class of these are ``dipper'' stars, which exhibit quasi-periodic flux dips thought to arise from occultations by circumstellar disc material located at or near the inner disc edge \citep{Cody2018, Andsell2016}. These systems provide direct insight into disc structure, inclination effects, and dust evolution in planet-forming environments.

In addition to dipper stars, a broader range of complex periodic variables have been identified in young clusters, exhibiting non-sinusoidal, evolving, or multi-component variability \citep{McQuillan2014, Rebull2014, Cody2014}. These behaviours are often attributed to a combination of starspot modulation, accretion variability, warped inner discs, and occulting material in asymmetric configurations. Such systems highlight the dynamic and often unstable nature of early stellar environments.


Even more extreme examples of unusual eclipse signatures have been observed in young and pre-main sequence systems where circumstellar or circumbinary dust produces large-amplitude, often highly structured photometric variability. The best-known example is J1407b, which exhibited a deep, long-duration and complex eclipse interpreted as the transit of an extensive ring system or circumplanetary material \citep{mamajek2012}. Similar dust-related eclipses have been reported in V928 Tau, although no clear periodicity has yet been established \citep{vandam2020}. Other pre-main sequence variables, including V718 Per and CHS 7797, display approximately sinusoidal brightness variations attributed to obscuration by circumstellar material, despite having markedly different periods of 4.7 yr and 17.8 d, respectively \citep{grinin2008,rodriguezledesma2013}.

Several of these objects are often compared to KH 15D, the prototype of a class of eclipsing pre-main sequence binaries in which a misaligned circumbinary disc periodically obscures one or both stellar components \citep{kearns1998}. Photometric and spectroscopic studies have demonstrated that the disc in KH 15D is significantly tilted relative to its highly eccentric central binary, producing the system's distinctive long-duration eclipses \citep[e.g.,][]{winn2004,chiang2004}. More recently, searches for analogous systems have identified additional candidates, including ZTF J202055.22+381323.1 and ZTF J071445.39$-$090152.1, whose variability is likewise consistent with obscuration by circumbinary material \citep{zhu2022}. Long-term monitoring has shown that the amplitudes of these systems can evolve substantially with time, reflecting changes in viewing geometry and disc structure. Collectively, these objects demonstrate the wide range of eclipse morphologies produced by circumstellar and circumbinary dust and highlight the importance of time-domain surveys in identifying new examples of disc-related eclipsing systems.


Together, these classes of objects demonstrate that variability in young stellar clusters can uncover a rich diagnostic of circumstellar structure, disk evolution, and early planetary system architecture. In this work, we present an analysis of the enigmatic NOI\,105872, situated within this broader context of young stellar variability and eclipse phenomenology. First identified in the Orion Nebula Cluster (ONC) by NGTS in August 2017, observations revealed a persistent periodic and asymmetric eclipsing signal.

In this work, we combine photometric data from multiple telescopes to model the light curves of NOI\,105872 in a range of different scenarios, including orbital dynamics (highly eccentric orbit), stellar effects (stellar activity, ``dipper'' stars, gravity darkening, complex periodic variables) and exotic companions (evaporating planet, tilted disk system). One of the most plausible hypotheses for explaining its unusual light curves is the presence of a tilted circumsecondary disk around a companion object. We investigate the geometry and properties of the hypothesized disk. By fitting parametric disk models to the observed eclipses, we aim to constrain its size, orientation, and variability, and to explore the implications for disk evolution and planet formation around low-mass stars.

\section{Photometric Observations}
\label{sec:photometry}

NOI\,105872 was initially observed by NGTS in 2017 as part of a survey field and is identified in GAIA by Gaia DR3 3023459583984210304 with a mean magnitude of G=15.5 mag. It is also identified in the 2MASS catalogue by 2MASS J05395645-0420140, with a K-band magnitude of 11.9 mag, and in \textit{TESS} by TIC 11356619. There were subsequently a number of eclipses observed both during other telescope surveys and as targeted follow-up. A summary of these photometric observations is presented in Table \ref{tab:telescopes}. 

\begin{table*}
	\centering
	\caption{Summary of photometric observations of NOI\,105872. Eclipse depth values quoted are approximate values for comparison between epochs.}
	\label{tab:telescopes}
	\begin{tabular}{ccccccc} 
		\hline
		Telescope & Obs Dates & Wavelengths (nm) & Exp Time (s) & $N_\text{points}$ & Eclipse Depth \\
		\hline
		NGTS & 2017 Aug 16 - 2018 Mar 17 & 520-890 & 10 & 162139 & $6-8\%$ \\
        \textit{TESS} Sector 6 & 2018 Dec 15 - 2019 Jan 06 & 600-1000 & 1800 & 987 & $13\%$ \\
        SAAO & 2020 Jan 03 & 700-900 & 60 & 350 & $7\%$ \\
        SPECULOOS & 2020 Jan 28 - 2020 Jan 29 & 750-1000 & 50 & 335 & $7.5\%$ \\
        ULTRASPEC & 2020 Feb 02 & 400-540 & 15 & 1239 & $11.5\%$ \\
        \textit{TESS} Sector 32 & 2020 Nov 20 - 2020 Dec 16 & 600-1000 & 600 & 3459 & $8.5\%$ \\
        NGTS & 2021 Dec 09 - 2022 May 22 & 520-890 & 10 & 34098 & $9-10\%$ \\
        ATLAS & 2015 - 2025 & 560-820 & 30 & 2279 & $5-10\%$\\
		\hline
	\end{tabular}
\end{table*}

\subsection{NGTS}
\label{subsec:NGTS}

\begin{figure}
	\includegraphics[width=\columnwidth]{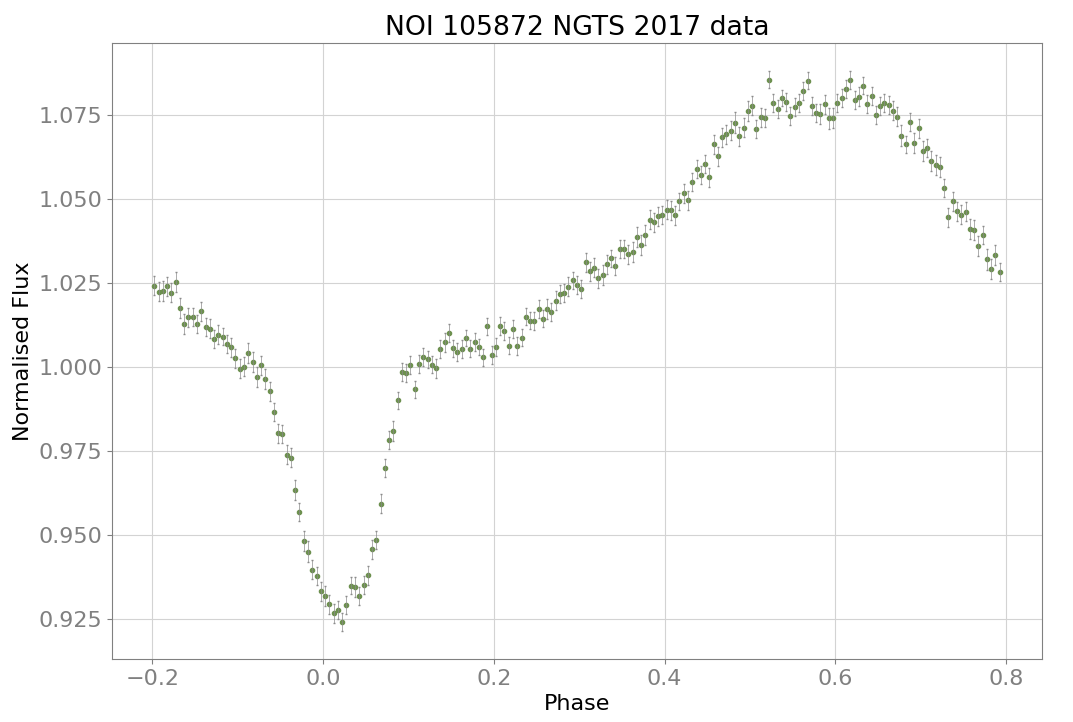}
    \caption{NGTS phase-folded light curve from the original observed data in 2017-18, using a period of 0.691 days. The data are binned into 200 data points.}
    \label{fig:OG NGTS curve}
\end{figure}

The Next Generation Transit Survey (NGTS) is a collection of 12 robotic ground-based telescopes situated at the ESO Paranal Observatory \citep{NGTS}. Each telescope is 20cm in diameter, with a wide field of view ($2.8 \times 2.8$ degrees) and uses a custom filter that covers wavelengths from 520nm to 890nm. Specifically, this bandpass is particularly sensitive to early M type stars, which is beneficial in the case of NOI\,105872. Observations by NGTS were reduced by a custom-built modular pipeline as described in \citet{NGTS}. 

NOI\,105872 was first observed and discovered during a survey of Orion in 2017. We detected multiple eclipses and it was flagged as an NGTS Object of Interest (NOI). Objects in the NGTS data are searched through for transits/eclipses and rotation signals. This object (with both clear eclipses and rotation) was detected. Between 2017 August 16 and 2018 March 17 it was observed for a total of 213 nights, the longest period of continuous observation for this object to date, and Figure \ref{fig:OG NGTS curve} shows the phase-folded light curve. Initially, NOI\,105872 was not flagged as anything peculiar, however, as more data from other telescopes was collected this object became more difficult to constrain. Between 2021 Dec 08 and 2022 Feb 24 the object was observed for a further 78 nights. The eclipse depth changes from $6\%$ to $10\%$, and the eclipse duration in the second dataset appears to be lengthened compared to the first, as seen in Epoch 1 and 5 of Figure \ref{fig:chunks}.

\subsection{TESS}
\label{subsec:TESS}

The Transiting Exoplanet Survey Satellite (\textit{TESS}) is a NASA space telescope launched in 2018 to complete an all-sky survey of nearby bright stars with the aim of finding transiting exoplanets \citep{TESS}. The spacecraft houses four wide-field cameras with a bandpass of 600-1000nm with observations being split into sectors, periods of around 27 days of nearly continuous observation, depending on which part of the sky is being surveyed at that time.

For our analysis of NOI\,105872, we generated custom light curves directly from the \textit{TESS} Science Processing Operations Centre \citep[SPOC;][]{Jenkins2016TESSSPOC} Full-Frame images. This was done as the SPOC and Quick Look Pipeline \citep[QLP;][]{Huang2020QLPdoi} lightcurves for this star were not always available when we needed them for our analysis. It also allowed us to tailor the photometric aperture and background pixel mask to suit NOI\,105872 in each TESS sector. We used the \texttt{lightkurve} package \citep{Lightkurve2018lightkurve} to download the \textit{TESS} target pixel files for NOI\,105872 filtered using the SPOC quality flag $\mathrm{QUALITY} = 0$ \citep{Jenkins2016TESSSPOC, TESSSPOC2020}. Pixel-level aperture was constructed as a two-tier mask: background pixels were defined as the 30\,\% lowest-flux pixels in a reference cadence, while target pixels were selected using a threshold mask applied to the summed flux across cadences. The target flux was computed by summing all pixels in the target aperture and subtracting the median background flux per pixel multiplied by the number of target pixels. The flux uncertainties were estimated using the median absolute deviation of the simple aperture photometry \texttt{SAP} flux or the summed background, providing robust per-cadence error estimates. To remove long-term trends while preserving transit signals, the lightcurve was split into segments when gaps exceeded 2.4\,h, corresponding to spacecraft data downloads and SPOC-flagged regions. Each segment was normalised individually and subsequently smoothed using an iterative Savitzky–Golay filter (five iterations with a 3$\sigma$ clipping threshold) and a 48\,h window size, which efficiently removes stellar activity while setting an upper limit on the duration of detectable transit events \citep{Savitzky_1964, Hattori_2022}.

Currently there are data for NOI\,105872 within Sector 6 and Sector 32 of the \textit{TESS} observations. The Sector 6 data were taken between 2018 Dec 15 and 2019 Jan 06, and the Sector 32 data was taken between 2020 Nov 20 and 2020 Dec 16, with the Sector 32 data having a higher cadence (20 second cadence) than previously observed in Sector 6 (2 minute cadence). Sector 6 has a greater transit depth than Sector 32, $13\%$ compared to $8.5\%$, and we present all of the \textit{TESS} photometry in Epoch 2 and 4 of Figure \ref{fig:chunks}.

\subsection{SAAO}
\label{subsec:SAAO}

The South African Astronomical Observatory (SAAO), has a number of research themes, one of which is the search for and characterisation of transiting exoplanets. NOI\,105872 was observed at the SAAO 1m with the Sutherland High Speed Optical Camera (SHOC, specifically SHOC’n’Awe, \citealt{SHOC}) on 2020 Jan 03 in the I filter with 60s exposures. The data were bias and flat field corrected via the standard procedure, using the \textsc{SAFPhot} Python package\footnote{https://github.com/apchsh/SAFPhot}. \textsc{SAFPhot} was also used to carry out differential photometry, by extracting aperture photometry from the target as well as comparison stars using the 'SEP' package \citep{Barbary2016}. SEP also measured and subtracted the sky background, adopting a box size and filter width which minimised the background residuals measured across the frame after the stars had been masked out. A few comparison stars in the $2.85\arcmin \times 2.85\arcmin$ field of view were used to perform differential photometry on the target.

This was a targeted observation that lasted almost 6 hours in total, but was ended prematurely due to high humidity at the observatory. Therefore, it only spans from phase -0.10 to 0.25, so does not include an out-of-eclipse observation. The maximum depth of the SAAO lightcurve is $7\%$, and it can be seen in Epoch 3 of Figure \ref{fig:chunks}. 

\subsection{SPECULOOS}
\label{subsec:SPECULOOS}

The Search for habitable Planets EClipsing ULtra-cOOl Stars (SPECULOOS) is a network of 1m telescopes across the globe (ESO Paranal Observatory in Chile, Teide Observatory in Tenerife and San Pedro Mártir observatory in Mexico) that aims to find transiting exoplanets around ultracool dwarf stars \citep{SPECULOOS2, SPECULOOS}. 

The SPECULOOS observations were taken using the Callisto telescope in Chile between 2020 Jan 28 and 2020 Jan 29 in the i- and z-bands, and have a similar maximum eclipse depth as the SAAO data, taken a few weeks before, at $\sim 7.5\%$. The data were reduced as per \citet{Murray2020}. In Epoch 3 of Figure \ref{fig:chunks} the SPECULOOS data are split into two sets, Cambridge and Li\`{e}ge. These represent the same data, however they have been reduced by two different institutional pipelines and we have included both in this paper for completeness.

\subsection{ULTRASPEC}
\label{subsec:ULTRASPEC}

ULTRASPEC is a high-speed imaging photometer mounted on the 2.4 m Thai National Telescope (TNT) at the Thai National Observatory \citep{ULTRASPEC}. ULTRASPEC offers a selection of broad and narrow-band filters covering the optical range, with a typical field of view of approximately 7.7 × 7.7 arcminutes when using the full frame. 

The full ULTRASPEC eclipse of NOI\,105872 was observed on 2020 Feb 02 using the g' filter. Something of note is that the ULTRASPEC light curve is a lot deeper ($11.5\%$) than SAAO and SPECULOOS, even though they are taken within a few weeks of each other, as seen in Epoch 3 of Figure \ref{fig:chunks}. These features will be discussed in Section \ref{subsec:inital photometry summary}.




\subsection{ATLAS}
\label{subsec:ATLAS}

The Asteroid Terrestrial-impact Last Alert System (ATLAS) is a ground-based telescope survey consisting of two independent units situated on Hawaiian islands \citep{ATLAS}. The telescopes have three filters, cyan (420-650\,nm), orange (560-820\,nm) and red (560-975\,nm), so overall, the wavelength coverage ranges from around 400\,nm to 1000\,nm. ATLAS was designed as an all-sky survey to discover changes in light curves that signify moving or variable objects, such as asteroids or supernovae, and hence also lends itself to the detection of eclipsing stars or planets. It is for this reason that we looked at the ATLAS database to see whether eclipses of NOI\,105872 had been captured. 

The data points for this object observed using ATLAS are sparse, as it has a typical cadence of two days with four exposures over a one-hour period, but they give us a broad idea of how the eclipse changes over time. We chose to only use the data points from the orange filter due to the higher data quality for this filter. Furthermore, the 10 years worth of data is valuable for tracking longer term changes, and therefore these data are mainly used to back-up the other higher cadence data that we have for these periods of time. Due to the long duration of the total observations, there are a range of eclipse depths quoted in Table \ref{tab:telescopes}. Since the ATLAS data spans around a decade, we sectioned the data into years, and this can be seen in Figure \ref{fig:chunks}.

\subsection{Initial photometry results}
\label{subsec:inital photometry summary}

\begin{figure*}
        \centering
        \begin{subfigure}[b]{0.95\textwidth}
            \centering
            \includegraphics[width=\textwidth]{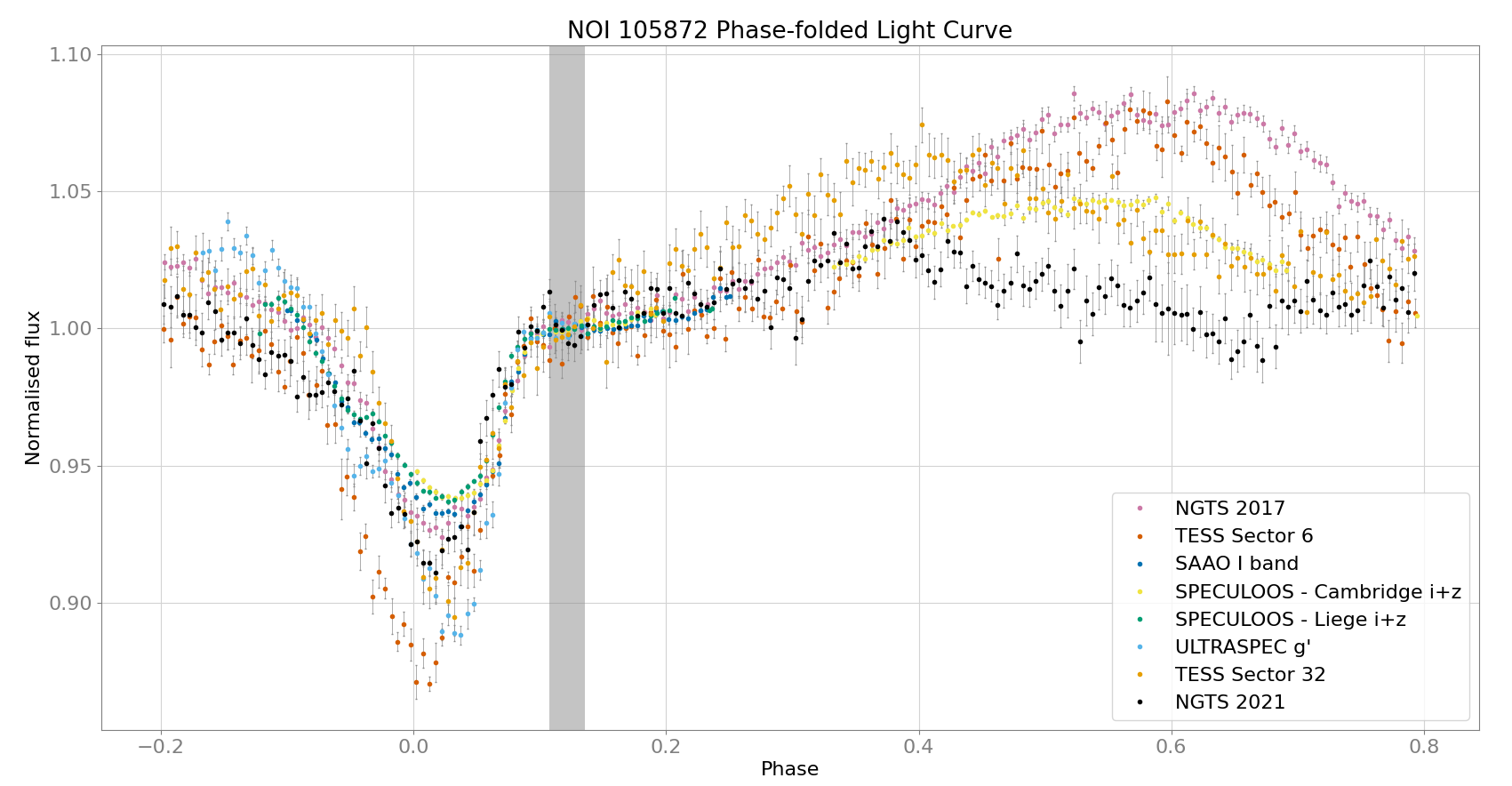}
            \caption[]%
            {{\small Normalised phase plot}}    
            \label{fig:main_phase_plot}
        \end{subfigure}
        \vfill
        \begin{subfigure}[b]{0.95\textwidth}  
            \centering 
            \includegraphics[width=\textwidth]{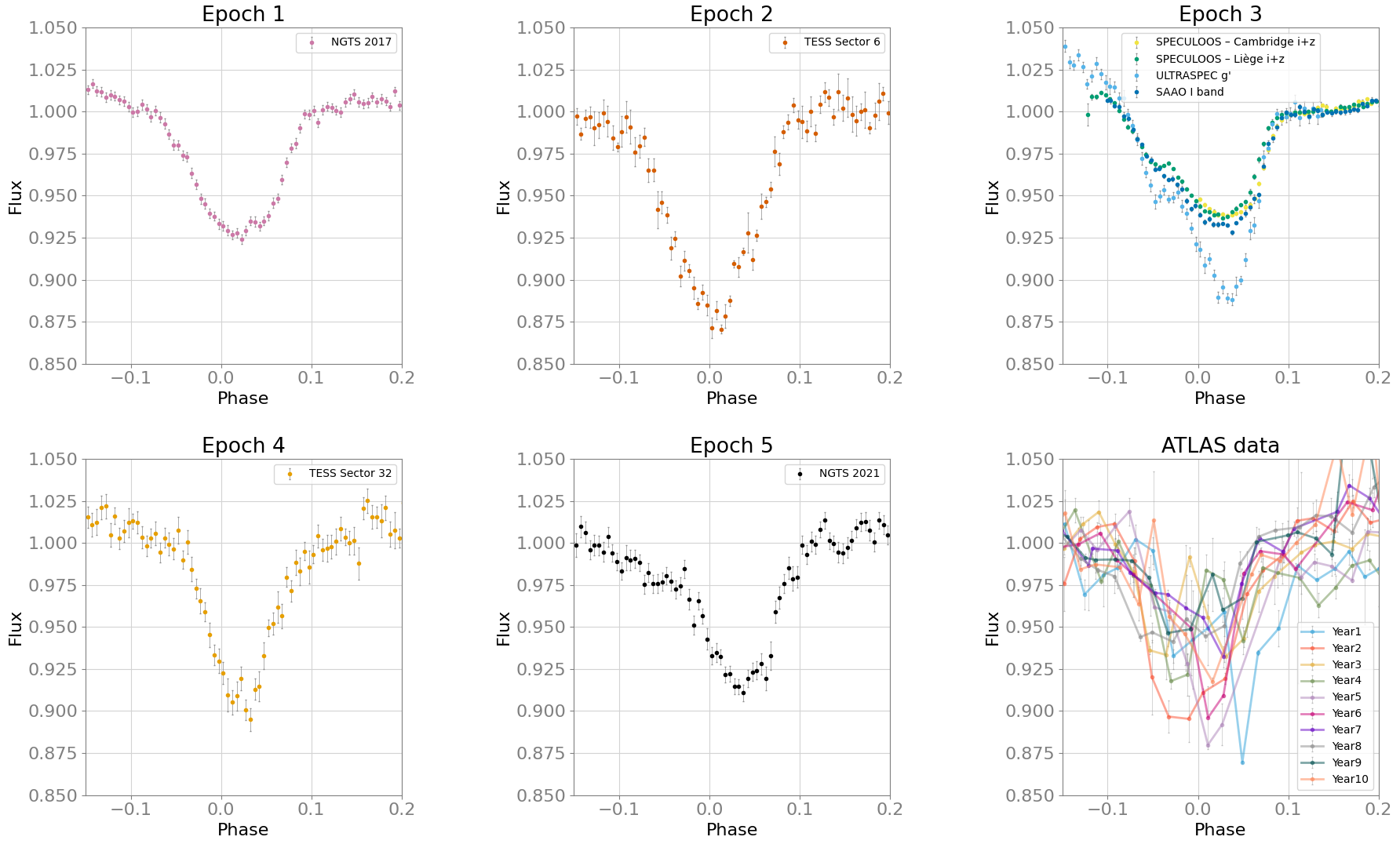}
            \caption[]%
            {{\small Phase-folded light curves for each data set shown as a progression over time}}    
            \label{fig:chunks}
        \end{subfigure}
        \caption[ ]
        {\small Normalised eclipse light curves of NOI\,105872 as a function of orbital phase, observed by a variety of different telescopes (see legend). These datasets span 5 years and display significant variability in both the eclipse depth and phase-curve shape. The shaded grey region shows the data range that was used to normalise the light curves. This region was chosen because the variation in gradients between the different datasets was the smallest. The data are also binned into 200 data points for each telescope dataset. Sub-figure (b) show the progression of the eclipse light curve with each epoch of observation.} 
        \label{fig:main_plots}
    \end{figure*}

Figure \ref{fig:main_phase_plot} shows the light curves for NOI\,105872 taken with all of these facilities over the course of 5 years. The orbital period of the object was determined using a Lomb–Scargle periodogram, with the period corresponding to the strongest peak in the power spectrum. The orbital period of this object was found to be $0.691\pm0.005$ days and is phase-folded with an epoch of 2458591.744 BJD. The eclipses vary and have depths ranging from $7\%$ to $13\%$. 

These photometric data reveal a number of interesting features. The first is that the eclipse is asymmetric, with the ingress generally lasting longer than the egress, as well as being much deeper than a normal transiting exoplanet. 

Secondly, the light curves (Fig. \ref{fig:main_phase_plot}) also display an increase in brightness towards orbital phases of $\sim$0.5 (approximately half an orbit after the eclipse). This feature also replaces any potential secondary eclipse that we might expect to see for an eclipsing object, which raises questions on the nature of the primary eclipse, as well as the nature and composition of the eclipsing object, perhaps causing some sort of reflection effect. Finally, we observe clear variability when we take a closer look at how the light curve changes over time, as seen in Fig. \ref{fig:chunks}. One of the main differences is a feature that we see appear in the SAAO, SPECULOOS and ULTRASPEC data: an increase in flux, or `bump', in the ingress, as seen in Epoch 3, between phase $-0.065$ and $-0.025$. These observations were taken within a few weeks of each other, and whilst the bump is not immediately obvious in the SAAO data, from our modelling (as seen in Section \ref{sec:tilted disk}) we can see the beginnings of this feature. For the bump to appear at the same phase in each light curve, the period and stellar rotation would have to be well-matched. With good spectra we can try to measure this. The eclipses also appear to be changing in terms of depth (see Table \ref{tab:telescopes}) and level of asymmetry. Table \ref{tab:BIS} shows the Bisector Inverse Span (BIS) for each of the eclipses, with the code used for this being adapted from similar methods used to identify asymmetries in spectral lines. Positive values indicate that the ingress is slower than the egress, however, certain datasets (NGTS 2017 and both \textit{TESS} sectors) have a relatively negligible asymmetry, while the other datasets show a clearer asymmetric trend. The fact that these values also change is further testament to the variability of the light curve. Together, all of these features paint a complicated picture that we need to disentangle and solve to be able to characterise this enigmatic object.

It should be noted that when normalising the light curve we had to be careful due to its variable nature. Generally, when normalising an eclipse we would mask out the eclipse and then use the out-of-eclipse part of the phase curve to find a weighted mean or median and divide the full light curve by this amount. In the case of NOI 105872, the out-of-eclipse flux is much more variable than a standard light curve. We therefore identified the phase range over which the gradient of the light-curve varied least across the different epochs. The phase range used for normalisation was 0.1073–0.1355 and is shown by the grey bar in Figure \ref{fig:main_phase_plot}. We used the weighted average flux in this range to normalise each dataset. Although this normalisation is based on the post-egress region only, the pre-ingress region showed significant variation in the local gradient between datasets, which would introduce a dataset-dependent bias into the normalisation, and was therefore not included in the normalisation.

\begin{table}
	\centering
	\caption{Bisector Inverse Span (BIS) values for the main seven datasets included in this paper. These are used as a measure of asymmetry of the eclipses.}
	\label{tab:BIS}
    \setlength{\tabcolsep}{4pt}
	\begin{tabular}{cc} 
		\hline
        Telescope & BIS \\
		\hline
        NGTS (2017) & 0.0039$\pm 0.0025$\\
        \textit{TESS} Sector 6 & 0.0034$\pm 0.0072$\\
        SAAO & 0.01856$\pm 0.0012$\\
        SPECULOOS & 0.0228$\pm 0.0018$\\
        ULTRASPEC & 0.0235$\pm 0.0024$\\
        \textit{TESS} Sector 32 & 0.0033$\pm 0.0074$\\
        NGTS (2021) & 0.0197$\pm 0.0069$\\
		\hline
	\end{tabular}
\end{table}

\section{Spectroscopic observations \& host star characterisation}
\label{sec:spectroscopy}

\subsection{HARPS}

In order to better characterise the system, we obtained three spectroscopic observations using the HARPS (High Accuracy Radial velocity Planet Searcher) spectrograph, on 2019 Dec 24, 2019 Dec 26 and 2020 Feb 10, under ESO programme ID 0104.C-0588 (P.I Bouchy). HARPS is a fibre-fed, high-resolution \'echelle spectrograph mounted on the ESO 3.6 m telescope at La Silla Observatory \citep{HARPS}. Operating in the visible range (380–690 nm) at a resolving power of $R\approx115,000$, HARPS is optimized for precise radial velocity measurements with a typical radial velocity precision of $\sim\SI{0.5}{m.s^{-1}}$ \citep{Barbieri2023} for bright, slow-rotating stars, and a long-term stability better than $\SI{1}{m.s^{-1}}$ \citep{Lovis2006}.

The spectra were obtained in the high-efficiency mode (EGGS), and with an exposure time of 2700\,s for the first two spectra and 3600\,s for the third spectrum from 2020. HARPS is a fibre-fed spectrograph which enables the user to either monitor a wavelength calibration source or the sky brightness, simultaneously to a science observation. For a faint and fast-rotating source like NOI\,105872, it is advantageous to place fibre B on-sky as we are limited by photon noise and not wavelength stability. 

We accessed the raw data through the ESO data Archive and used the publicly available Data Reduction pipeline espdr version 3.3.9, for the extraction of the spectra, sky-subtracting, telluric correction, and wavelength calibration. The resulting spectra all have low SNR of respectively 4, 3 and 7 per pixel at 551 nm. The one observation in 2020 has a higher SNR, partly due to a slightly longer exposure time, but mainly attributed to moon contamination, which is clearly seen in fibre B. Using the binary masks available as part of the official pipeline, we cross-correlated the spectra with G2, K0, K5 and M2 masks on a wide CCF window of $\pm$250\,km\,s$^{-1}$,  which all produce noisy CCFs which are insufficient to properly constrain the stellar velocity and mass of the orbiting object. The observations were taken at orbital phases 0.63, 0.33 and 0.77, respectively, and so none of them fall within the eclipse itself. Even if these spectra were of a reasonable SNR value, there is not a sufficient sampling across the phase curve to do any analysis on the changes in spectral line profiles.

\begin{figure}
	\includegraphics[width=\columnwidth]{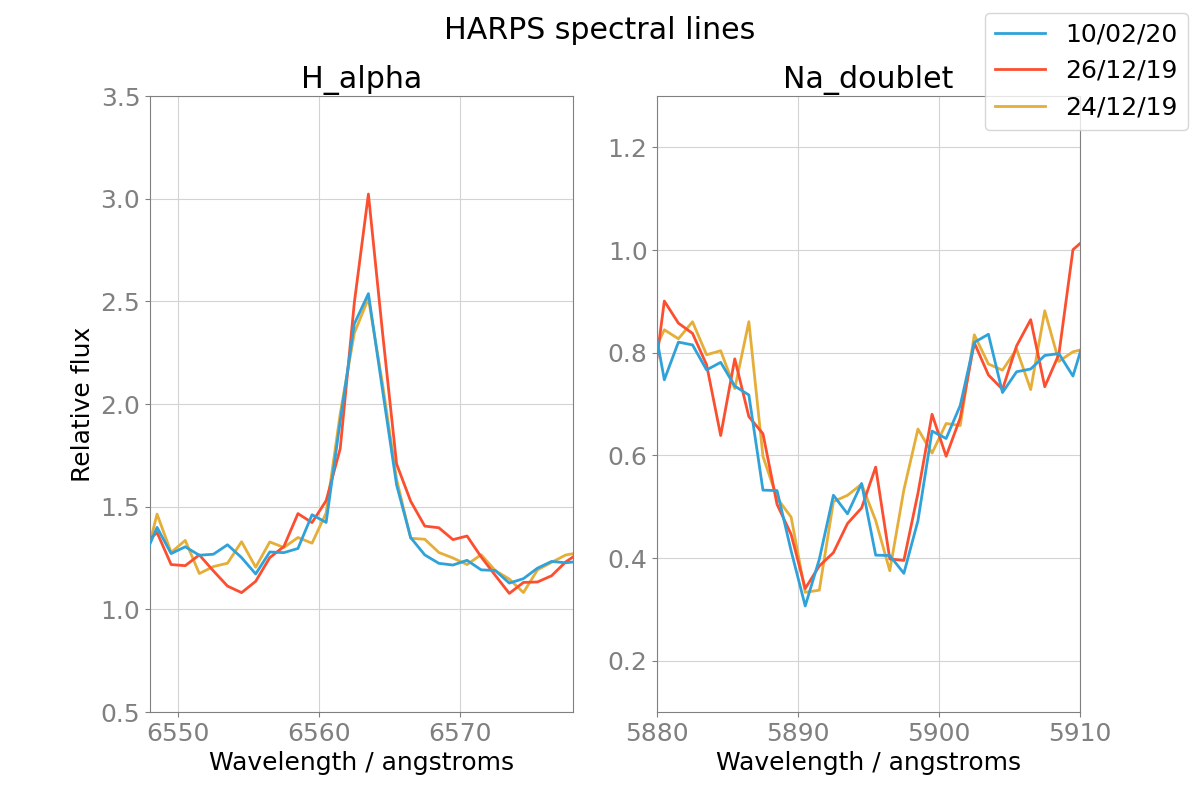}
    \caption{H$\alpha$ emission and Na doublet absorption features in the three HARPS spectra. These observations have been sky-subtracted and binned into $\SI{1}{\angstrom}$ chunks.}
    \label{fig:Halpha emission}
\end{figure}

Due to the very low SNR nature of the spectra, we also performed our own sky subtraction: rather than direct subtraction of the spectra, we needed to perform a scaling because these spectra from the sky and science fibres will not be exactly the same. The strength of the contaminant features needed to be scaled so that they are the same for both fibres and then a subtraction could be performed to completely remove the features not coming from the object we are interested in. To do this, first, we found strong contaminant features within the stellar and sky spectra, in this case, emission features at $\SI{5577}{\angstrom}$ and $\SI{6300}{\angstrom}$. We fit a Gaussian to these lines and then used the amplitudes to create a subtraction coefficient ($S$) between the two spectra by taking the ratio of the amplitude of the line in the science fibre to the amplitude of the line in the sky fibre, and found the average of the two emission lines. This coefficient was then multiplied by the sky flux and subtracted from the science spectrum. The contaminant features are time-variable and so this process is completed separately for each spectrum. The sky-corrected spectra around the H$\alpha$ and Na doublet can be seen in Fig.\,\ref{fig:Halpha emission}.

\subsection{Stellar characterisation}
\label{sec:SED}

\begin{table}
	\centering
	\caption{Magnitudes of the host star.}
	\label{tab:mags}
	\begin{tabular}{cccr} 
		\hline
		  Band & Value & Source\\
		\hline
        u & 19.747$\pm 0.038$ & SDSS\\
        g & 17.380$\pm 0.005$ & SDSS\\
        r & 15.904$\pm 0.004$ & SDSS\\
        i & 14.854$\pm 0.004$ & SDSS\\
        z & 14.224$\pm 0.004$ & SDSS\\
        \hline
        g & 17.071$\pm 0.007$ & PanSTARRS\\
        r & 15.874$\pm 0.003$ & PanSTARRS\\
        i & 14.750$\pm 0.004$ & PanSTARRS\\
        z & 14.317$\pm 0.001$ & PanSTARRS\\
        y & 14.012$\pm 0.005$ & PanSTARRS\\
        \hline
        g & 16.756$\pm 0.012$ & SkyMapper\\
        r & 15.875$\pm 0.027$ & SkyMapper\\
        i & 14.671$\pm 0.013$ & SkyMapper\\
        z & 14.179$\pm 0.006$ & SkyMapper\\
        \hline
        BP & 16.727$\pm 0.014$ & Gaia DR2\\
        G & 15.463$\pm 0.004$ & Gaia DR2\\
        RP & 14.348$\pm 0.008$ & Gaia DR2\\
        \hline
        TESS & 14.356$\pm 0.012$ & TESS \\
		\hline
        J & 12.828$\pm 0.026$ & 2MASS \\
        H & 12.042$\pm 0.021$ & 2MASS \\
        K & 11.864$\pm 0.021$ & 2MASS \\
        \hline
        WISE 3.4 micron & 11.728$\pm 0.024$ & WISE \\
        WISE 4.6 micron & 11.609$\pm 0.021$ & WISE\\
        \hline
	\end{tabular}
\end{table}

\begin{figure}
	\includegraphics[width=\columnwidth]{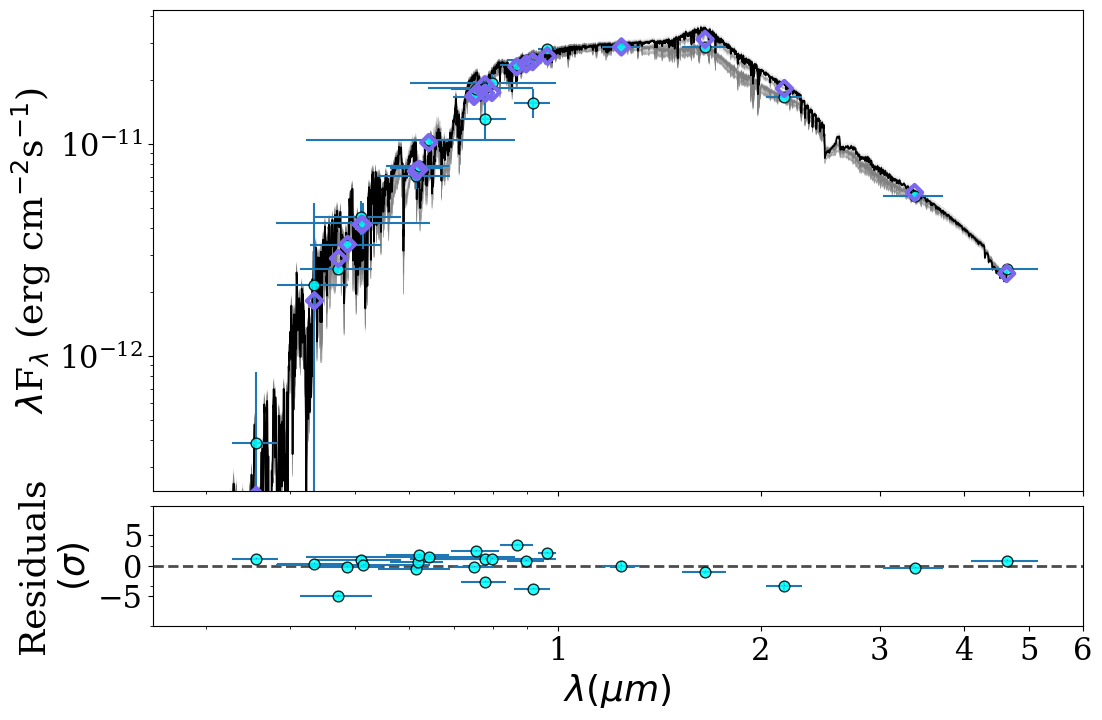}
    \caption{Spectral energy distribution of NOI 105872. Cyan circles show the observed photometry (Table \ref{tab:mags}), purple diamonds show the corresponding synthetic fluxes from the best-fit model. The black line is the best-fit NextGen stellar atmosphere model \citep{Hauschildt1999}, with grey lines showing a sample of models drawn from the posterior distribution to illustrate the fit uncertainty, obtained using the astroARIADNE package \citep{astroARIADNE}. Best-fit parameters are given in Table \ref{tab:SED results}. The lower panel shows the residuals in units of the photometric uncertainty.}
    \label{fig:SED}
\end{figure}

In order for us to fully understand the object, we need to understand the host star first. The target star is a confirmed member of the Orion Nebula Cluster (ONC) \citep{Kounkel2019}. We combine spectral diagnostics and spectral energy distribution (SED) modelling to constrain its fundamental properties. 

We performed the initial SED fitting using the \texttt{astroARIADNE} package \citep{astroARIADNE}, and then refined those values by using them to complete an isochrone fit with \texttt{astroLACHESIS} (Vines et al. submitted to A\&A, 2026), results of which can be seen in Figure \ref{fig:SED} and Table \ref{tab:SED results}. In the \texttt{astroARIADNE} code, there is an initial model filtering process based on initial values for things like the effective temperature. In this case the NextGen stellar atmosphere model \citep{Hauschildt1999} was the only model left after this internal filtering process. The errors found are based on the single-grid fit, and so are less realistic than that of a full multi-grid search. The stellar parameters considered here place the object firmly in the regime of low-mass pre-main-sequence stars. The effective temperature corresponds to an approximate spectral type of M1–M2, while the combination of relatively low mass and inflated radius indicates that the star has not yet reached the main sequence, and is instead still contracting toward the main sequence.

The SED fit yields a distance of 371.7 pc, consistent with the distance inferred from the Gaia parallax ($2.6315$ mas; d $\approx 380$ pc).

When placed on a Hertzsprung–Russell diagram according to the effective temperature and luminosity, this position lies above the main sequence and is consistent with evolution along a Hayashi track, characteristic of fully convective pre-main-sequence stars \citep{Hayahsi1961}. This is further supported by the young age ($4.1$ Myr), which is well within the contraction phase for a $\sim0.4$\(\textup{M}_\odot\) star (e.g., \citealt{Baraffe2015}).

Membership in the ONC reinforces this interpretation. The ONC is a well-studied, very young star-forming region containing a large population of low-mass, pre-main-sequence objects \citep{Hillenbrand1997}. Stars in this environment are typically magnetically active, rapidly rotating to moderate degrees, and often exhibit strong surface inhomogeneities such as starspots and, in some cases, accretion hotspots.

No infrared excess is detected in the SED, indicating the absence of a primordial inner circumstellar disk, our observations do not go to long enough wavelengths to constrain middle/outer disc material. In the SED fit we fix the metallicity value at Fe/H $= 0.0$, as seen in Table \ref{tab:SED results}. The metallicty of the ONC usually sits around $-0.01\pm 0.04$ \citep{DOrazi2009}. It should also be noted that due to the low SNR of the spectra, we cannot get a constraint of the metallicity spectroscopically and so the only way to find it is numerically from the photometry values. This is why we decided to fix it at a solar metallicity value.

The optical spectrum shows signs of Li I $\lambda \SI{6708}{\angstrom}$ (see Figure \ref{limb lines}) in absorption, a well-established indicator of stellar youth in low-mass stars, as lithium is rapidly depleted during pre–main-sequence evolution \citep{Bodenheimer1965, Jeffries2017}. The spectrum also exhibits absorption from the Na I D doublet at $\sim\SI{5890}{\angstrom}$ (Figure \ref{fig:Halpha emission}), together with prominent Ca I absorption features at $\lambda6102$ and $\lambda\SI{6122}{\angstrom}$ (see also Figure \ref{limb lines}). These features are shown as examples of the quality of our spectra and are just other photospheric lines that are visible despite the low SNR. 

Balmer lines are detected in emission in the spectrum, indicative of elevated chromospheric activity commonly observed in young M-type stars. The H$\alpha$ emission line can be seen in Figure \ref{fig:Halpha emission}. Given the absence of an infrared excess, these features are attributed to magnetic activity rather than ongoing disk accretion \citep{White+Basri2003, Fang2009}. The spectrum also exhibits emission from N II and O I. These lines cannot originate in the stellar photosphere at the effective temperature of the source ($T_{\rm eff} \simeq 3574$~K), as the ionisation energies required are far in excess of those produced by a cool M-type star. Instead, these features are most naturally explained as contamination from the surrounding Orion H II region, which is known to produce strong nebular emission lines that are frequently superposed on stellar spectra in the ONC \citep{Osterbrock2006, Baldwin1991, DaRio2009}.

In summary, the combined spectroscopic and photometric evidence identifies the source as a young ($4.1$ Myr), inner diskless pre–main-sequence M1–M2 star in the Orion Nebula Cluster, with chromospheric activity typical of weak-lined T Tauri stars and superposed nebular emission originating from the surrounding H II region.

\begin{table}
	\centering
	\caption{Stellar properties of the host star. The parameters from \texttt{astroLACHESIS} were determined in the modelling for this paper. The results, and therefore errors, produced for this paper are entirely numerical and do not take into account systematic errors. We urge the reader to take this into account when reading these figures.}
	\label{tab:SED results}
	\begin{tabular}{lccr} 
		\hline
		Property & Value & Source\\
		\hline
        NGTS designations & NGTS\,J053956.4-042013 & \\
        & NOI\,105872 & \\
        GAIA designation & Gaia DR3 3023459583984210304 & \\
        2MASS designation & 2MASS J05395645-0420140 & \\
        \textit{TESS} designation & TIC 11356619 & \\
        $\alpha$ [$^{\circ}$] & 84.985216 & GAIA DR3\\
		$\delta$ [$^{\circ}$] & -4.337223 & GAIA DR3\\
		Parallax [mas] & 2.6315$\pm 0.0379$ & GAIA DR3\\
        Distance [pc] & 371.7$^{+4.6}_{-4.9}$ & \texttt{astroLACHESIS}\\ [5pt]
        $T_{\mathrm{eff}}$ [K] & 3574$^{+45}_{-34}$ & \texttt{astroLACHESIS}\\ [5pt]
        $\log{g}$ [cm/s$^2$] & 3.99$\pm0.03$ & \texttt{astroLACHESIS}\\[5pt]
        $A_{V}$ [mag] & 0.197$^{+0.199}_{-0.132}$ & \texttt{astroLACHESIS}\\[5pt]
        $R_{*}$ [\(\textup{R}_\odot\)] & 1.10$\pm0.03$ & \texttt{astroLACHESIS}\\[5pt]
        $M_{*}$ [\(\textup{M}_\odot\)] & 0.42$^{+0.03}_{-0.02}$ & \texttt{astroLACHESIS}\\[5pt]
        $L_{*}$ [{\(\textup{L}_\odot\)}] & 0.17$^{+0.03}_{-0.02}$ & \texttt{astroLACHESIS}\\ [3pt]
        Fe/H [dex] & 0.00 & Fixed \\ [3pt]
        Age [Myrs] & 4.1$\pm0.4$ & \texttt{astroLACHESIS}\\[5pt]
		\hline
	\end{tabular}
\end{table}

\section{Modelling and interpreting the system}
\label{results}

In this paper, our analyses focus on the depth/asymmetry and variability of the eclipse, however we need to take into account the increase in out-of-eclipse flux when considering the full picture of models and hypotheses. A number of hypotheses were then explored for NOI\,105872 to help us explain these features and fully characterise the object. These hypotheses are laid out below and summarised in Table \ref{tab:hypotheses}.

\begin{table*}
	\centering
	\caption{Proposed explanations for the eclipse of NOI\,105872. Inspired by the table in \citealt{Trevor2017}.}
	\label{tab:hypotheses}
	\begin{tabular}{p{2.5cm} p{5.75cm} p{5.75cm} p{2cm}}
		\hline
		Hypothesis & Supporting evidence & Conflicting evidence & Conclusion \\
		\hline
		Stellar activity & Asymmetrical and variable eclipse, with bumps in the profile & Short-lived activity could be a factor influencing the variability, but unable to account for the overall eclipse shape -- should be taken into account in tandem with other hypotheses & Not major contributing factor \\ 
        \hline
		``Dipper'' stars & Deep, asymmetrical and variable eclipse & Less variable than usual dipper behaviour -- no evidence of IR excess from the SED fit to indicate protoplanetary disk & Highly unlikely\\
        \hline
		Gravity darkening & Asymmetrical eclipse light curve & Models unable to produce the same eclipse morphology, don't take into account variability, and convective envelope implies weaker gravity darkening component & Highly unlikely \\
        \hline
		CPVs & Variable eclipse light curve and young M type host star & Morphology and its associated variability much more complex than our object & Unlikely \\
        \hline
        Magnetospheric clouds & Variable eclipse light curve & Short-duration, lack of accretion indicators, and unknown source of replenishment & Somewhat likely \\
        \hline
		Highly eccentric orbit & Asymmetrical eclipse light curve & Does not account for the variability of the eclipse profile & Not a major contributing factor\\
        \hline
		Evaporating planet & Deep, asymmetrical and variable eclipse & Mass-loss rate would have to remain relatively constant and not be affecting the orbital alignment & Highly unlikely \\
        \hline
		Tilted disk system & Deep, asymmetrical and variable eclipse & Models line up well; variability can be explained through precession of the disk and effects of stellar activity on the disk, asymmetry due to the tilted nature of the system & Most likely \\
		\hline
	\end{tabular}
\end{table*}

\subsection{Caused by the star}

\subsubsection{Stellar activity}

Stellar activity, particularly in the form of dark starspots or bright plages, can significantly distort eclipse light curves. When an eclipsing companion crosses over a cooler, darker starspot, the amount of flux blocked by the companion is reduced, because the companion is blocking dimmer regions. This results in a temporary positive ``bump'' in the eclipse profile \citep{Oshagh2013, valio}. Models show that such spot occultations by a companion generate identifiable anomalies whose magnitude increases when the spot is closer to the stellar disk centre and depends on the spot’s size (filling factor) and temperature. These anomalies can bias key eclipse measurements, such as depth, duration, and timing, and in the case of exoplanets leads to underestimates in planet radius by a few percent and causing transit timing variations (TTVs) up to several minutes \citep{Watson2004, Barros2013, Ioannidis2015}. Over multiple eclipses, evolving starspot positions, and their relationship to the orbital and rotational periods, can shift the phase of these bumps, producing asymmetries and time-dependent variability in the light curve, including mid-eclipse asymmetry and apparent shifts in the ingress and egress \citep{sanchis-ojeda2011, Barros2013}. More extreme, long-lived cases, such as TOI-3884, show a massive polar starspot covering about 17\% of the stellar surface, causing distinct double-bump patterns and strongly asymmetric transits due to the companion’s unconventional pole-crossing orbit \citep{almenara2022}.

It is unlikely in the case of NOI\,105872 that stellar activity is the cause of our asymmetric light curve. Short-lived stellar activity, such as starspots, could however cause some features that we see, such as the ingress ``bump'' in Figure \ref{fig:chunks}. Unocculted starspots would also help to explain why the ULTRASPEC eclipse is much deeper than the other two datasets in Figure \ref{fig:chunks}. This is because starspots are cooler material and so emit much less blue light than red, hence appear especially dark at bluer wavelengths. If some of these spots are not crossed by the companion during an eclipse, they reduce the star’s overall brightness, particularly in blue light, while the companion still blocks the same amount of the stellar surface. This makes the drop in light during the transit look larger than it really is, so the transit appears deeper in blue wavelengths compared to red. Another effect could be the sinusoidal nature of the out-of-eclipse part of the phase curve. While this could be associated with starspots and faculae, these events alone could not explain the overall eclipse curve which shows asymmetry and great eclipse depth over the course of 4 years. Long-lived stellar activity, such as the polar spot in \citet{almenara2022}, would be too stable to give the variability in the eclipse curve. The conclusion here is that stellar activity should be taken into account in any further analyses, but does not explain the key features of the light curve.

\subsubsection{``Dipper'' stars}

``Dipper'' stars are a type of young stellar object (YSO) and subclass of T Tauri stars that exhibit dimming events that can be episodic or quasiperiodic. These events can cause the brightness to drop by $\sim10\%$–$50\%$ (e.g., \citealt{Alencar2010, Andsell2016, Rodriguez2017, Cody2018, Roggero2021}). While the light curve of our object shows a deep dimming (up to $\sim13\%$), asymmetric ingress/egress, and a multi‐year stability over $\geq 10$ yr, these features differ in several respects from classical ``dipper'' stars. Dippers in star‐forming regions (e.g.\ Taurus, Upper Sco, etc.) have dips that are usually highly stochastic or quasi‐periodic, with large variation from epoch to epoch in depth, duration, shape, or interval. For example, \citet{Roggero2021} find that among disk‐bearing young stars $\sim30\%$ show dipper behaviour; many are aperiodic, most show changing structure on timescales of one rotation, and dip depths and shapes vary substantially. By contrast, our object maintains a reproducible shape (including asymmetry) over many years; moreover, unless other diagnostics point to a protoplanetary disk (as noted earlier, we see no signs of an IR excess), we have no evidence for the typical protoplanetary disk environment that dipper models require. Therefore, although superficially similar in some respects, the evidence strongly suggests that this object’s occulting structure is likely due to a different geometry.

\subsubsection{Gravity darkening}

Gravity darkening occurs when changes in local effective gravity cause flux variations across the stellar surface and, in the case of rapidly rotating stars, this leads to darker equatorial regions and brighter poles \citep{von_zeipel, gravdarkening}. An object transiting in front of a gravity darkened host star will produce an asymmetric or differently shaped light curve \citep{Barnes2009, gravdarkening_transits}. Structural and environmental properties have important implications for the expected strength of gravity darkening. In classical formulations, gravity darkening follows the von Zeipel relation \citep{von_zeipel}, in which the local effective temperature scales with surface gravity as $T_{\mathrm{eff}} \propto g^{\beta}$, with $\beta \approx 0.25$ for stars with radiative envelopes. However, this prescription does not apply to early M type stars that have a convective envelope. More appropriate treatments (e.g. \citealt{Lucy1967}) show that convective envelopes yield significantly smaller exponents ($\beta \approx 0.08$ or lower), resulting in substantially weaker surface temperature gradients even in the presence of rotationally induced gravity variations. We still modelled the object orbiting a gravity darkened star to see if the shape of the light curve could be explained in this way. We built a custom code to model this for a range of stellar and orbital orientations. 

To test whether the observed eclipse morphology could be reproduced by a planet orbiting a gravity-darkened star, we constructed a simple three-dimensional model of a transit across an oblate stellar photosphere. The model is intended as a qualitative, proof-of-concept framework rather than a fully self-consistent treatment of stellar structure or orbital dynamics.

The stellar surface is represented as an oblate spheroid defined on a spherical polar grid $(\theta,\phi)$. Here, $\theta$ is the colatitude measured from the stellar rotation axis (such that $\theta=0$ at the North pole and $\theta=\pi/2$ at the equator), and $\phi$ is the azimuthal angle in the equatorial plane. In Cartesian coordinates the surface is described by
\begin{equation}
x = R_\star \sin\theta \cos\phi,
\end{equation}
\begin{equation}
y = R_\star \sin\theta \sin\phi,
\end{equation}
\begin{equation}
z = R_\star (1 - f)\cos\theta,
\end{equation}
where $R_\star$ is the equatorial radius and $f$ is an oblateness parameter describing fractional polar compression. The surface is discretised on a uniform angular grid. To allow arbitrary viewing geometries, the stellar surface is rotated via successive Euler rotations, enabling independent specification of the stellar inclination and sky-projected orientation.

Gravity darkening is implemented using a simplified, latitude-dependent parametrisation. In a physically self-consistent treatment, the local effective temperature is related to the effective surface gravity via the von Zeipel relation,
\begin{equation}
T_{\rm eff} \propto g_{\rm eff}^{\beta},
\end{equation}
where $\beta$ is the gravity-darkening exponent and which implies a local flux scaling 
\begin{equation}
    F \propto g_{\rm eff}^{4\beta}
\end{equation}
The effective gravity itself depends on latitude through the combined gravitational and centrifugal potentials, and must be computed from a Roche model for a rotating, oblate star (e.g., \citealt{von_zeipel, gravdarkening}).

In this work, we do not attempt to compute $g_{\rm eff}(\theta)$ self-consistently. Instead, we adopt a simple analytic prescription in which the surface brightness varies smoothly between the poles and equator according to
\begin{equation}
I_{\rm grav} \propto |\cos\theta|^{\beta},
\end{equation}
where $\theta$ is the colatitude measured from the stellar rotation axis. This form reproduces the qualitative behaviour expected from gravity darkening, namely brighter poles and a dimmer equator, while avoiding the need for a full treatment of the stellar potential. The exponent $\beta$ therefore acts as an effective parameter controlling the strength of the latitudinal brightness contrast, rather than a direct physical mapping to the von Zeipel coefficient. Limb darkening is included using a quadratic law, defined by $u_1$ and $u_2$. The final intensity map is constructed as a weighted combination of $I_{\rm grav}$ and $I_{\rm limb}$ and normalised to unity in the absence of occultation.


The companion is modelled as an opaque spherical body of radius $R_{\rm p}$ moving along a linear trajectory across the projected stellar disc. The orbital inclination and semi-major axis determine the impact parameter, while a sky-projected spin-orbit misalignment angle, $\lambda$, is introduced by rotating the transit chord relative to the stellar rotation axis.

At each time step, surface elements whose projected separation from the planet centre satisfies
\begin{equation}
d \leq R_{\rm p}
\end{equation}
are treated as occulted. The observed stellar flux is computed by summing the intensities of all visible, unocculted surface elements and normalising by the unobscured stellar flux.

The model neglects several physical effects, including differential rotation, self-consistent computation of surface gravity, wavelength-dependent limb darkening, finite exposure-time integration, and full Keplerian orbital motion. Nevertheless, it captures the principal geometric and photometric consequences of transiting an oblate, gravity-darkened star.

\begin{figure}
	\includegraphics[width=\columnwidth]{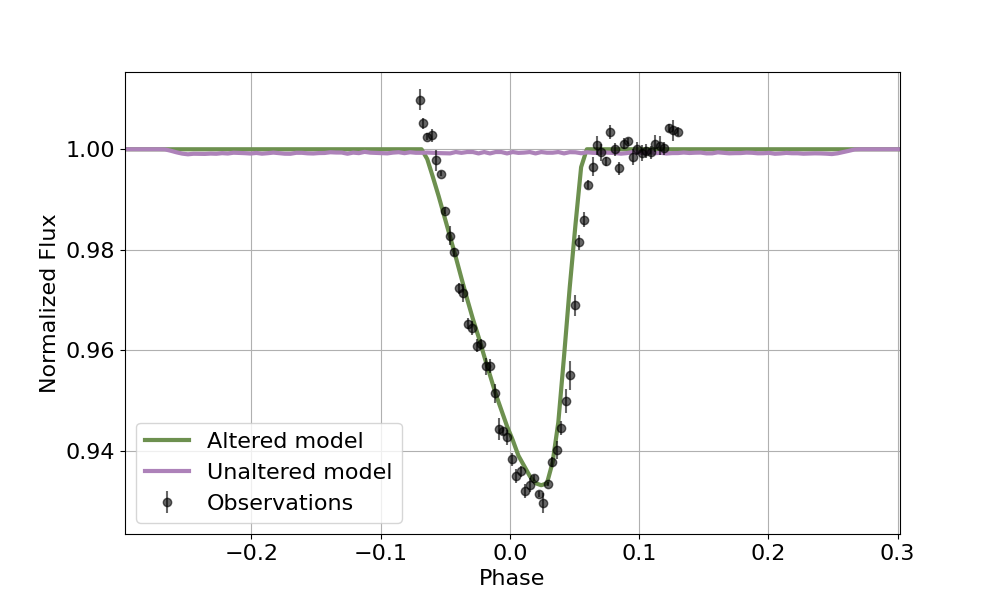}
    \caption{3D gravity darkening model for two scenarios overplotted with the SAAO observations for NOI\,105872. There is an unaltered model which represents the model if we use the calculated orbital separation and correct gravity darkening exponent (0.08) which is shown in purple. The model in green is provided as an example and not a true fit to the data. The parameters adopted to get a reasonable fit are in Table \ref{tab:grav darkening}, and are significantly different to the known, measured or expected parameters of the system. This includes using a value of $\beta = 0.25$ rather than 0.08 as would be expected for this kind of star. To adopt reasonable parameters the model differs significantly from the observed data.}
    \label{fig:grav darkening}
\end{figure}

Figure \ref{fig:grav darkening} shows a sample of the gravity darkening model over the SAAO observational data set. We provide this as an example for how the model can be used, but it should be noted that the already known parameters had to be changed for this model to more closely reflect the eclipse morphology of NOI\,105872. The values input into the model can be seen in Table \ref{tab:grav darkening}. In this table $\theta$ represents the angle that the star is tilted up/down in our line of sight, and $\phi$ is the angle that the star is tilted left/right. Without allowing all of these variables to change, we could not find an appropriate fit through MCMC fitting. Most notably, the semi-major axis is much larger than our already calculated value of $0.011$\,au, and the gravity darkening exponent used is that of a radiative envelope ($\beta = 0.25$) to show a more extreme version of gravity darkening compared to what we would get with $\beta = 0.08$ of a convective star. We did try to run the model fit using $\beta = 0.08$, but the eclipse depth and shape could not be closely recovered. We can see that the asymmetric structures predicted by gravity darkening do not match the form of the measured light curve. From our modelling, along with the literature search, we therefore conclude that the observed eclipse shape is unlikely to arise from a planet orbiting a gravity-darkened star.

\begin{table}
	\centering
	\caption{Model input parameters for the 3D gravity darkening model.}
	\label{tab:grav darkening}
	\begin{tabular}{cc} 
		\hline
        Parameter & Value \\
		\hline
        $R_*$[$R_{\odot}$] & 1.1\\        $R_p$[$R_\text{J}$] & 2.83\\
        $f$ & 0.1\\
        $\beta$ & 0.25\\
        $u_1$ & 0.8\\
        $u_2$ & 0.2\\
        $a$[AU] & 0.069\\
        $i$[$^\circ$] & 84.9\\
        $\lambda$[$^\circ$] & 33.5\\
        $\theta$[$^\circ$] & 65\\
        $\phi$[$^\circ$] & 1.5\\
		\hline
	\end{tabular}
\end{table}

\subsubsection{Complex Periodic Variables (CPVs)}

Complex Periodic Variables (CPVs) are a class of young, rapidly rotating stars, typically late-type M dwarfs, that exhibit complex periodic light curves with periods generally shorter than a few days \citep{Stauffer2017}. Their photometric variability cannot be explained by starspots alone \citep{Stauffer2017, Zhan2019, Koen2021}, indicating the presence of additional structures or processes associated with the stellar environment. The leading explanations invoke spatially concentrated circumstellar material that corotates with the star \citep{Stauffer2017, Gunther2022}. Proposed scenarios include dust clouds orbiting near the Keplerian corotation radius and periodically transiting the stellar disk \citep{Farihi2017, Sanderson2023}, as well as magnetically supported prominences consisting of cool, dense gas embedded within the stellar corona \citep{CollierCameron1989, Jardine2019, Waugh2022}. Both mechanisms can produce the complex and evolving eclipse-like features that characterise CPV light curves.

Recent studies have further established the observational properties of the CPV population \citep{Bouma2024}. CPVs are found predominantly among young M dwarfs with masses of approximately 0.1--0.4\,$M_{\odot}$, while higher-mass hosts are comparatively rare. Their photometric periods are generally stable over timescales of years, although the detailed light-curve morphology can evolve substantially during the same interval. CPVs are also frequently observed to exhibit superposed spot-modulation signals, producing an underlying quasi-sinusoidal component in addition to the more complex eclipse-like variability.

NOI\,105872 shares some characteristics with known CPVs. In particular, its period appears stable over the two-year observing baseline, consistent with the behaviour reported by \citet{Bouma2024}. However, its light-curve morphology is considerably simpler than that typically observed in CPVs, which often display multiple evolving dips and significantly more complex structures. Furthermore, the host star of NOI,105872 has a mass exceeding 0.4\,$M_{\odot}$, placing it just outside the mass range occupied by the majority of known CPVs. While the CPV interpretation therefore cannot be ruled out entirely, the available evidence suggests that NOI\,105872 is unlikely to be a typical member of this class.

\subsubsection{Magnetospheric clouds}

For certain stars, particularly high-mass stars, transiting circumstellar magnetospheric clouds are suggested to be a cause of photometric variability \citep{Groote1982}. The clouds consist of plasma which has been trapped in the magnetosphere of the star and appear most dense at the corotation radius \citep{Townsend2013}. For fully convective, low-mass pre-main sequence stars, such as the host star NOI\,105872, magnetic field strengths are generally in the range 0.1--1\,kG (e.g., \citealt{Johns-Krull2007, Donati2009}), which can cause charged material to become trapped in the magnetosphere once it enters an orbit about the star. Closed field lines around the corotation radius cause the material to settle at this point, forming a co-rotating cloud of plasma (e.g., \citealt{Stauffer2015, CollierCameron1989}). This model has been invoked for a small but well-studied set of young, rapidly rotating stars exhibiting periodic, depth-variable flux dips synchronous with the stellar rotation period, including $\sigma$\,Ori\,E and RIK-210 \citep{Townsend2005, Trevor2017}.

The variable eclipse depth observed in the light curve of NOI\,105872 is qualitatively consistent with this scenario. In magnetospheric cloud models, recurrent processes remove dust and gas from the prominence while chromospheric evaporation replenishes gas from below \citep{DaleyYates2024}, naturally producing long-term variability in the column density of occulting material and therefore in the eclipse depth. The period of 0.69\,days further supports this interpretation, as slingshot prominences are a characteristic feature of rapidly rotating stars, with the canonical example being AB\,Doradus at a comparable rotation period of 0.514\,days \citep{CollierCameron1989}.

The asymmetric eclipse profile offers a further constraint on the cloud morphology. The variable morphology of dimming events in systems such as RIK-210 suggests that the occulting material is not a single, spherical body \citep{Trevor2017}, and asymmetric profiles are expected when the leading and trailing edges of the cloud have different density gradients. 

The magnetospheric cloud model does however face several difficulties in this system. The main issue is the lack of accretion indicators within the spectroscopy, which is not observed here. Furthermore, the long-term increase in eclipse depth between certain epochs implies that material is being periodically replenished on a timescale of years. While the cyclic slingshot and replenishment process provides a qualitative explanation, the decade-long coherence of the signal is more consistent with a stable, long-lived structure than with the typically short-lived prominences seen on other rapidly rotating stars \citep{CollierCameron1989, DaleyYates2024}. Alternatively, the cloud could be precessing, such that the transiting portion of the cloud is different at each epoch. Overall, the magnetospheric cloud model is qualitatively consistent with several features of the light curve of NOI\,105872 but is not modelled quantitatively here, and should therefore be considered with caution as one of several competing hypotheses.

\subsection{Caused by the orbit}

\subsubsection{Eccentricity}

One of the first causes for an asymmetric light curve that we hypothesised and tested was a companion on a highly eccentric orbit. Whenever transiting planets are on an eccentric orbit, the acceleration of the planet as it transits in front of its host star can cause changes in the ingress and egress, which therefore makes the light curve asymmetric \citep{eccentricity}. 

\begin{figure*}
        \centering
        \begin{subfigure}[b]{0.475\textwidth}
            \centering
            \includegraphics[width=\textwidth]{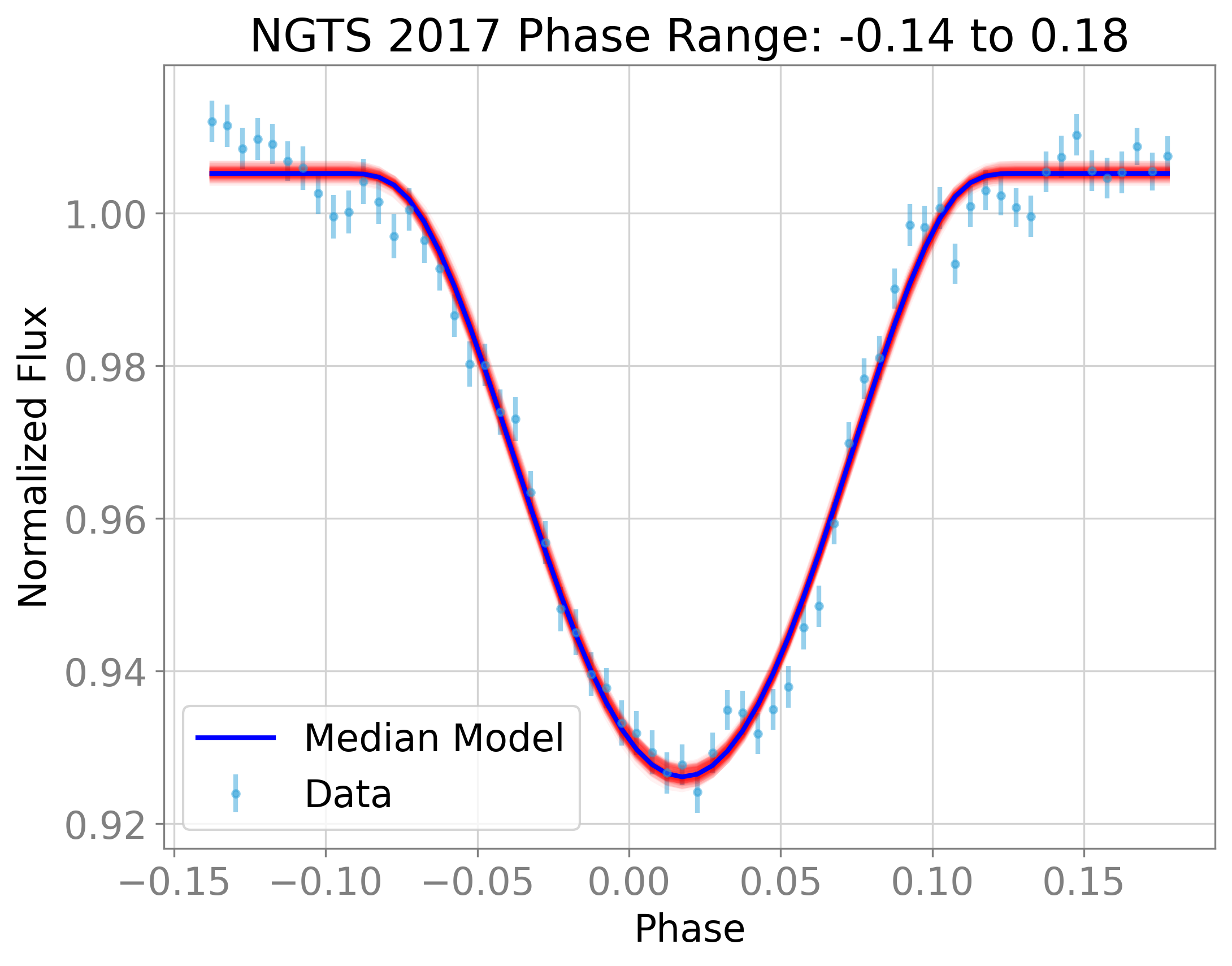}
            \caption[]{}%
            \label{fig:NGTS 2017 batman}
        \end{subfigure}
        \hfill
        \begin{subfigure}[b]{0.475\textwidth}  
            \centering 
            \includegraphics[width=\textwidth]{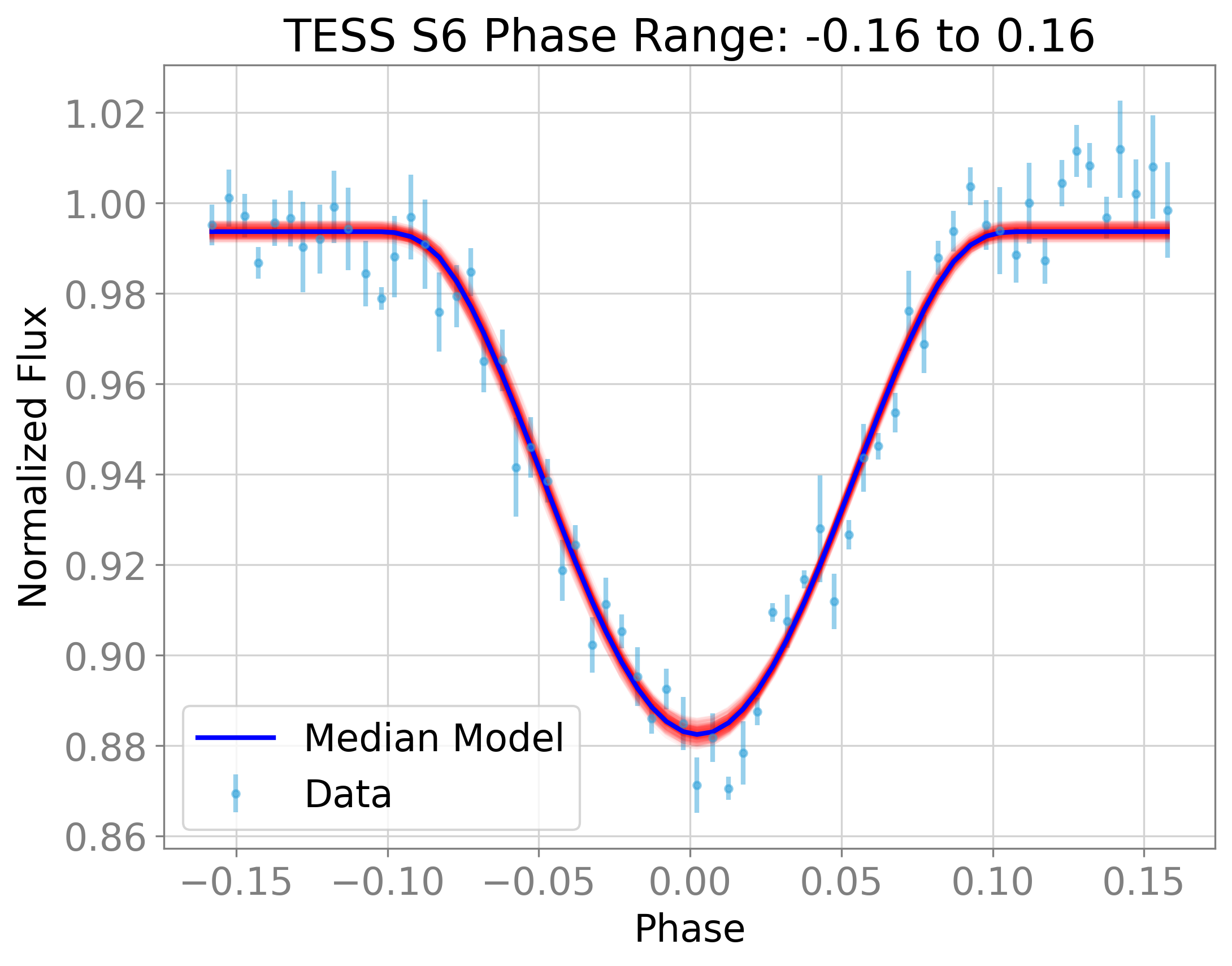}
            \caption[]{}%
            \label{fig:TESS S6 batman}
        \end{subfigure}
        \vskip\baselineskip
        \begin{subfigure}[b]{0.475\textwidth}   
            \centering 
            \includegraphics[width=\textwidth]{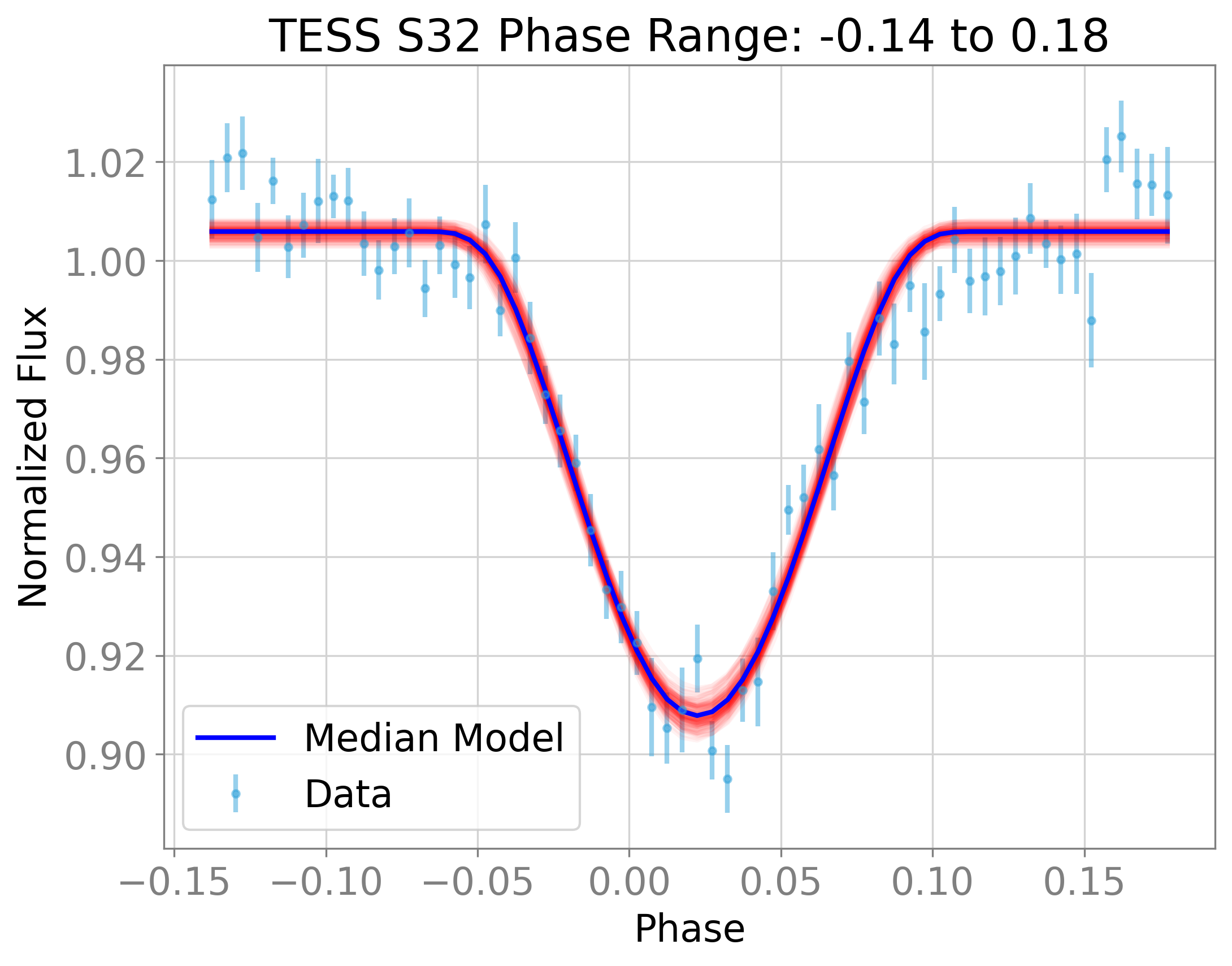}
            \caption[]{}%
            \label{fig:TESS S32 batman}
        \end{subfigure}
        \hfill
        \begin{subfigure}[b]{0.475\textwidth}   
            \centering 
            \includegraphics[width=\textwidth]{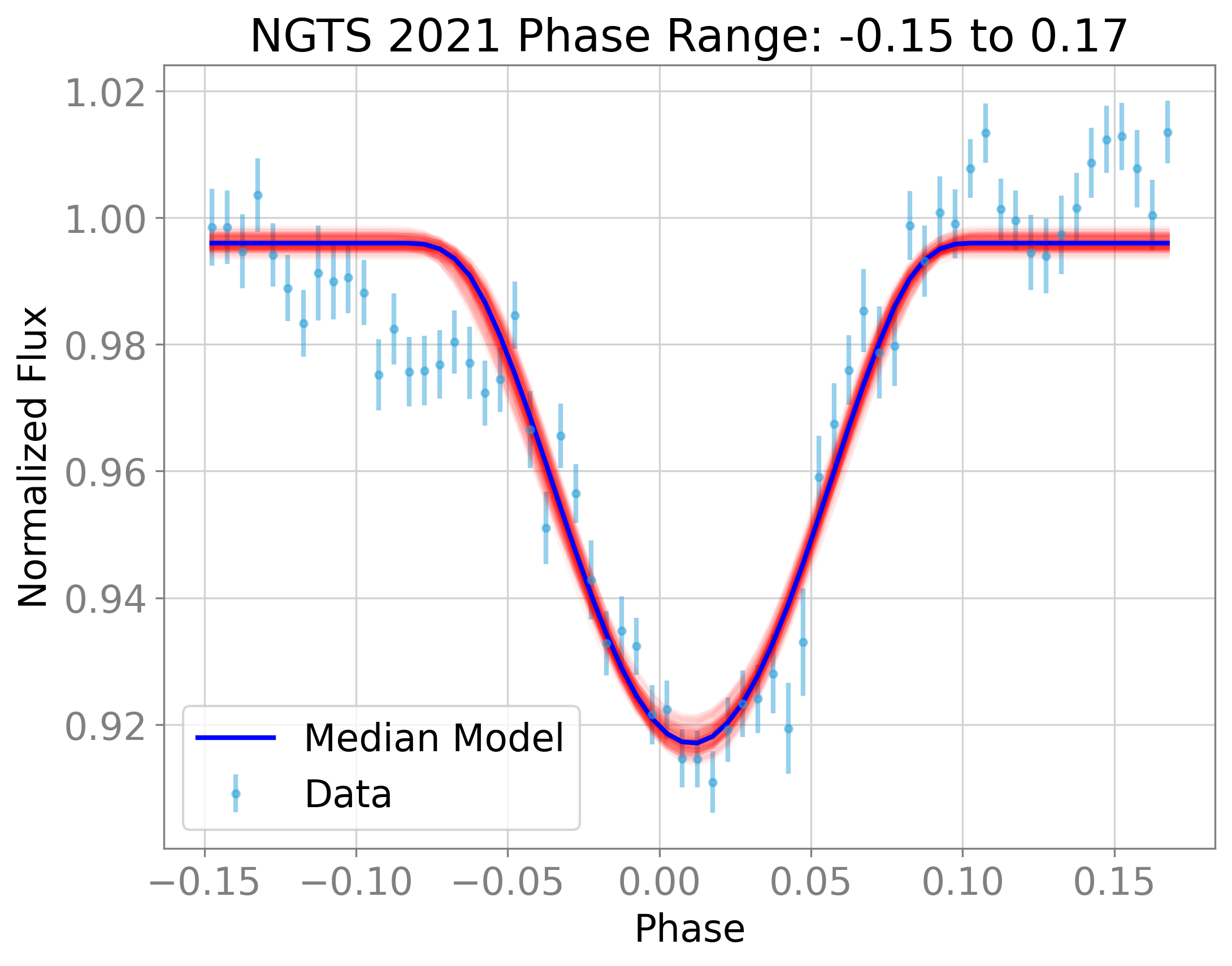}
            \caption[]{}%
            \label{fig:NGTS 2021 batman}
        \end{subfigure}
        \caption[]
        {\small Best-fit MCMC using \texttt{batman} shown with a dark blue line, over-plotted with the NGTS and \textit{TESS} binned data points. The red lines represent the iterations of the MCMC model before the best fit is found.} 
        \label{fig:batman MCMC}
    \end{figure*}

We used the \texttt{batman} python package \citep{batman} to model the eclipse light curve and ran an MCMC fit using \texttt{emcee} \citep{emcee}, allowing all of the following variables to change: radius of the object ($R_p$), semimajor axis ($a$), inclination ($i$), eccentricity ($e$), argument of periastron ($\omega$), time of inferior conjunction ($t_0$) and a model offset that allows the level of the continuum to shift from 1. From this we found the best fit for the \texttt{batman} model for both \textit{TESS} and both NGTS lightcurves, which can be seen in Figure~\ref{fig:batman MCMC}. The best fit parameters found from these fits are given in Table~\ref{tab:batman results}. We chose to fit the lightcurves at a range of $\pm0.16$ to the $t_0$ estimate for consistency, this range is large enough that it encapsulates all of the complete eclipses and includes some of the out of eclipse light as well. At first glance, the eccentricity model appears to fit the data quite well, however there are some caveats that we have to take into account. Firstly, the eccentricity model does not explain the increase in flux in the out-of-eclipse part of the lightcurve, and secondly, we still need to compare these fits to other models, such as the tilted disk model in Section \ref{sec:tilted disk}. We can also see that certain values, such as the semimajor axis, vary between models and, in some cases, are not within errors of one another. It is unlikely that a parameter like this would vary significantly between epochs.

\begin{table}
	\centering
	\caption{Results of MCMC fitting the batman model to the photometric phase curves for NGTS and \textit{TESS}, along with the chi-squared value for each model fit.}
	\label{tab:batman results}
    \setlength{\tabcolsep}{4pt}
	\begin{tabular}{|c|c|c|c|c|} 
		\hline
        Value & NGTS (2017) & \textit{TESS} S6 & \textit{TESS} S32 & NGTS (2021) \\
		\hline
        
        $e$\rule{0pt}{3ex} & 0.484$^{+0.011}_{-0.021}$ & 0.469$^{+0.023}_{-0.042}$ & 0.446$^{+0.038}_{-0.066}$ & 0.467$^{+0.025}_{-0.060}$\\ [5pt]
        $\omega$[$^\circ$] & 280.4$^{+6.5}_{-9.6}$ & 270.4$^{+12.3}_{-13.2}$ & 264.1$^{+21.6}_{-18.1}$ & 278.5$^{+13.0}_{-19.9}$\\ [5pt]
        $R_p$[$R_\text{J}$] & 3.30$^{+0.14}_{-0.06}$ & 4.30$^{+0.48}_{-0.38}$ & 4.17$^{+0.85}_{-0.54}$ & 3.63$^{+0.44}_{-0.30}$\\ [5pt]
        $a$[$R_*$] & 2.18$^{+0.05}_{-0.04}$ & 2.39$^{+0.09}_{-0.11}$ & 2.68$^{+0.18}_{-0.19}$ & 2.45$^{+0.13}_{-0.08}$ \\ [5pt]
        $i$[$^\circ$] & 75.39$^{+0.57}_{-0.29}$ & 75.82$^{+0.97}_{-0.60}$ & 76.15$^{+1.48}_{-0.84}$ & 75.78$^{+1.37}_{-0.58}$ \\ [5pt]
        $t_0$ & 0.021$^{+0.002}_{-0.003}$ & 0.003$\pm{0.003}$ & 0.022$\pm{0.004}$ & 0.013$\pm{0.003}$\\ [5pt]
        $\chi^2$ & 131.2 & 295.2 & 161.9 & 215.5 \\ [3pt]
        BIC & 160.3 & 324.3 & 191.0 & 224.6 \\
		\hline
	\end{tabular}
\end{table}

And so, high orbital eccentricity can be ruled out as the only contributing factor to the asymmetric light curve. This is not to say that an eccentricity in the orbit is not contributing to the asymmetry, but it will not cause all of the features that we see in the photometry.

\subsection{Caused by the companion}

\subsubsection{Evaporating planet}

When exocomets transit in front of a star, they produce a light curve just as a transiting exoplanet would. These light curves tend to be asymmetrical \citep{exocomets}. This is due to comets having a ribbon tail of ice and dust coming off of them, as well as a coma around the nucleus of the comet. This dust will block a portion of the light coming from the star as it passes after the comet, making the egress of the transit curve far more gradual than the ingress. This has also been seen to occur for particular exoplanetary lightcurves, such as KIC 12557548b \citep{budaj2013}, where a comet-like tail is observed to be following along after the planet. This could be attributed to the object having an accompanying variable dust cloud that could have originated from the disintegration of the object \citep{brogi2012}.

However, in the case of NOI\,105872, we can see in Figure \ref{fig:chunks} that, in the epochs that are asymmetric, it is the ingress that is less steep than the egress, which means that if there were a ribbon tail of dust, it would have to be leading the object rather than trailing.

One of the causes of a leading cloud of dust around the object would be if the object's atmosphere is evaporating, or if the object itself is disintegrating as was the case with KIC 12557548b. This could occur due to the high irradiance of the atmosphere from the the activity of the host star. As radiation bombards the atmosphere, it expands past the object's Roche lobe and therefore is no longer bound to the object \citep{bourrier2018, foster2022}. The light curve produced by NOI\,105872 could also be considered reminiscent of disintegrating planets such as K2-22b \citep{Sanchis-Ojeda2015}. In this case, the mass-loss rate would have to be sustainable over the course of our observations, and consistently replenishing. Because dust lifetimes are short, stellar winds could not vary drastically or else the asymmetry may change significantly. The orbital alignment would also have to remain stable even as the object loses mass. It should also be noted that not all of the light curve epochs show the asymmetry that would be expected of an evaporating planet. Therefore, we note that the shape of the eclipse and long-term stability in this case are significantly different from systems like K2-22b, and so is unlikely to be an evaporating planet. 

\subsubsection{Tilted disk system}
\label{sec:tilted disk}

\begin{figure*}
        \centering
        \begin{subfigure}[b]{0.475\textwidth}
            \centering
            \includegraphics[width=\textwidth]{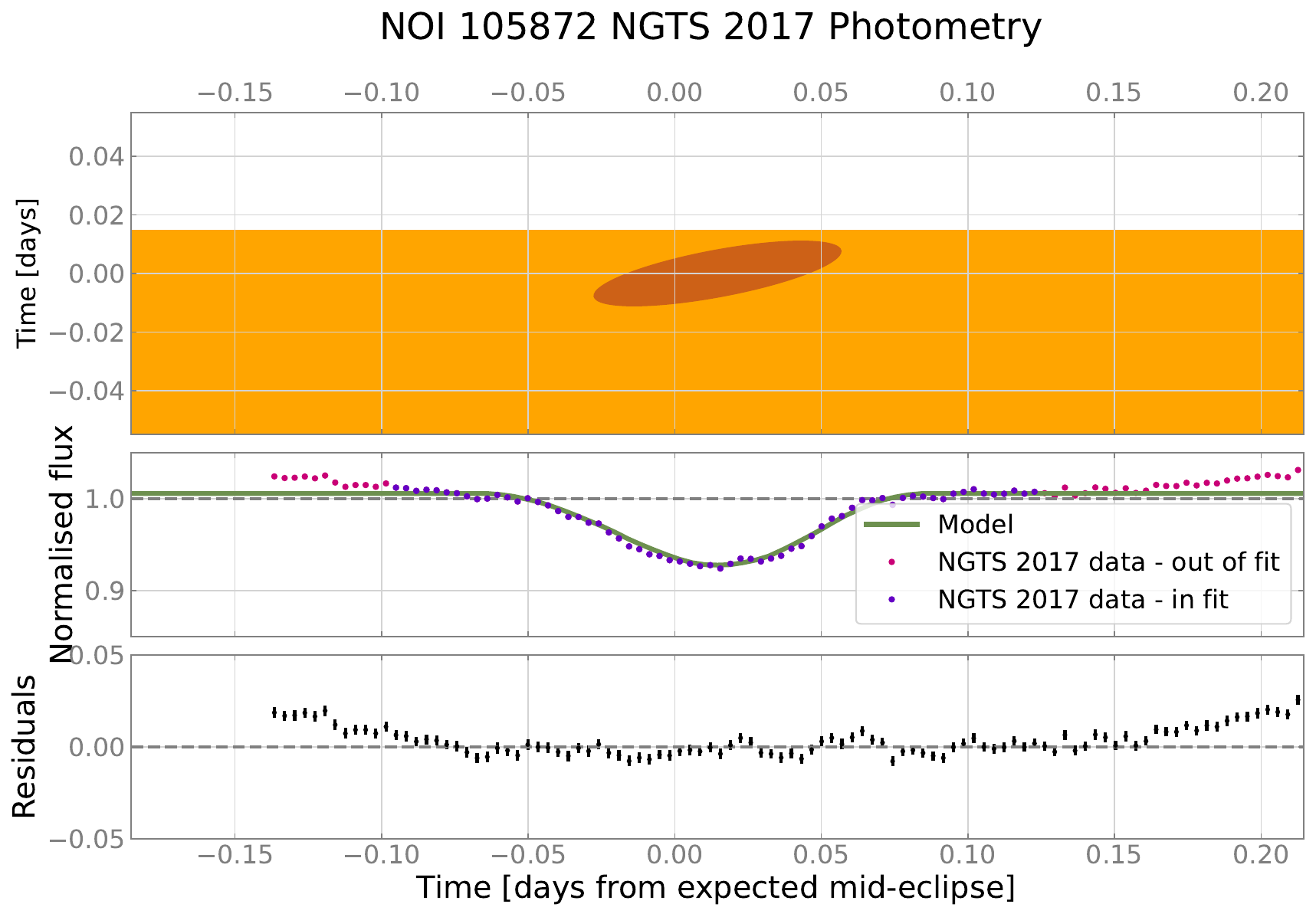}
            \caption[]%
            {}    
            \label{fig:NGTS 2017}
        \end{subfigure}
        \hfill
        \begin{subfigure}[b]{0.475\textwidth}  
            \centering 
            \includegraphics[width=\textwidth]{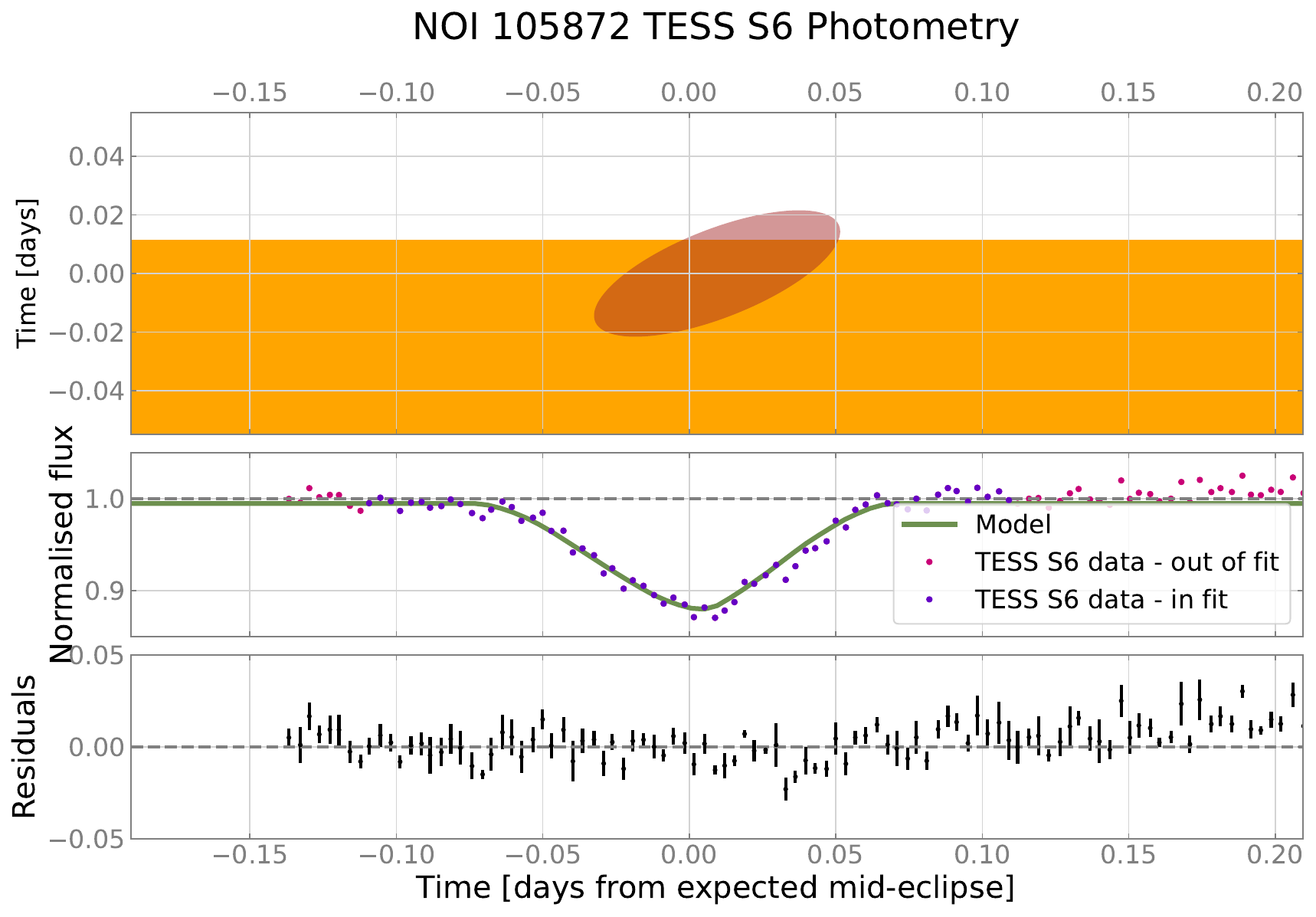}
            \caption[]%
            {}    
            \label{fig:TESS S6}
        \end{subfigure}
        \vskip\baselineskip
        \begin{subfigure}[b]{0.475\textwidth}   
            \centering 
            \includegraphics[width=\textwidth]{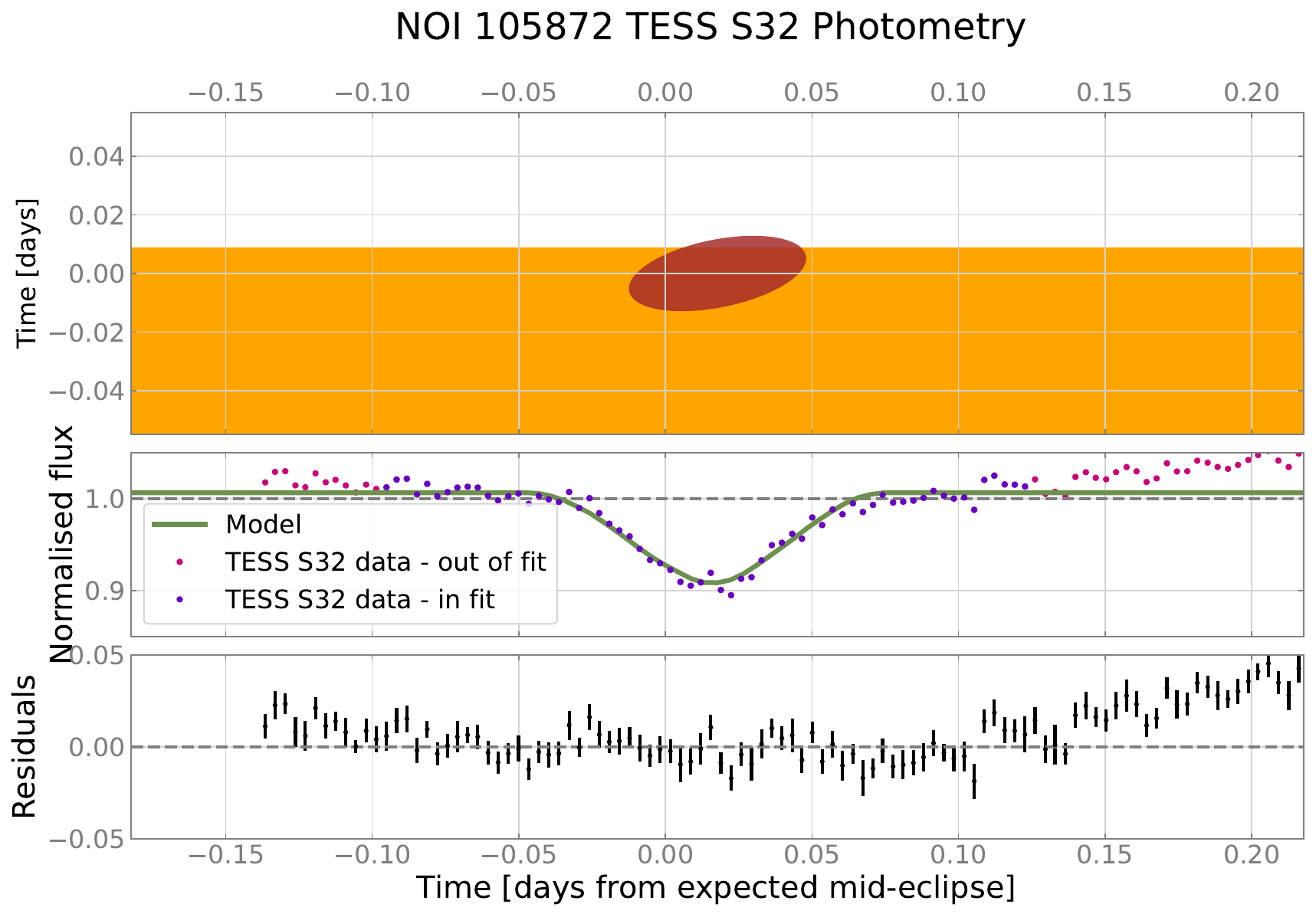}
            \caption[]%
            {}    
            \label{fig:TESS S32}
        \end{subfigure}
        \hfill
        \begin{subfigure}[b]{0.475\textwidth}   
            \centering 
            \includegraphics[width=\textwidth]{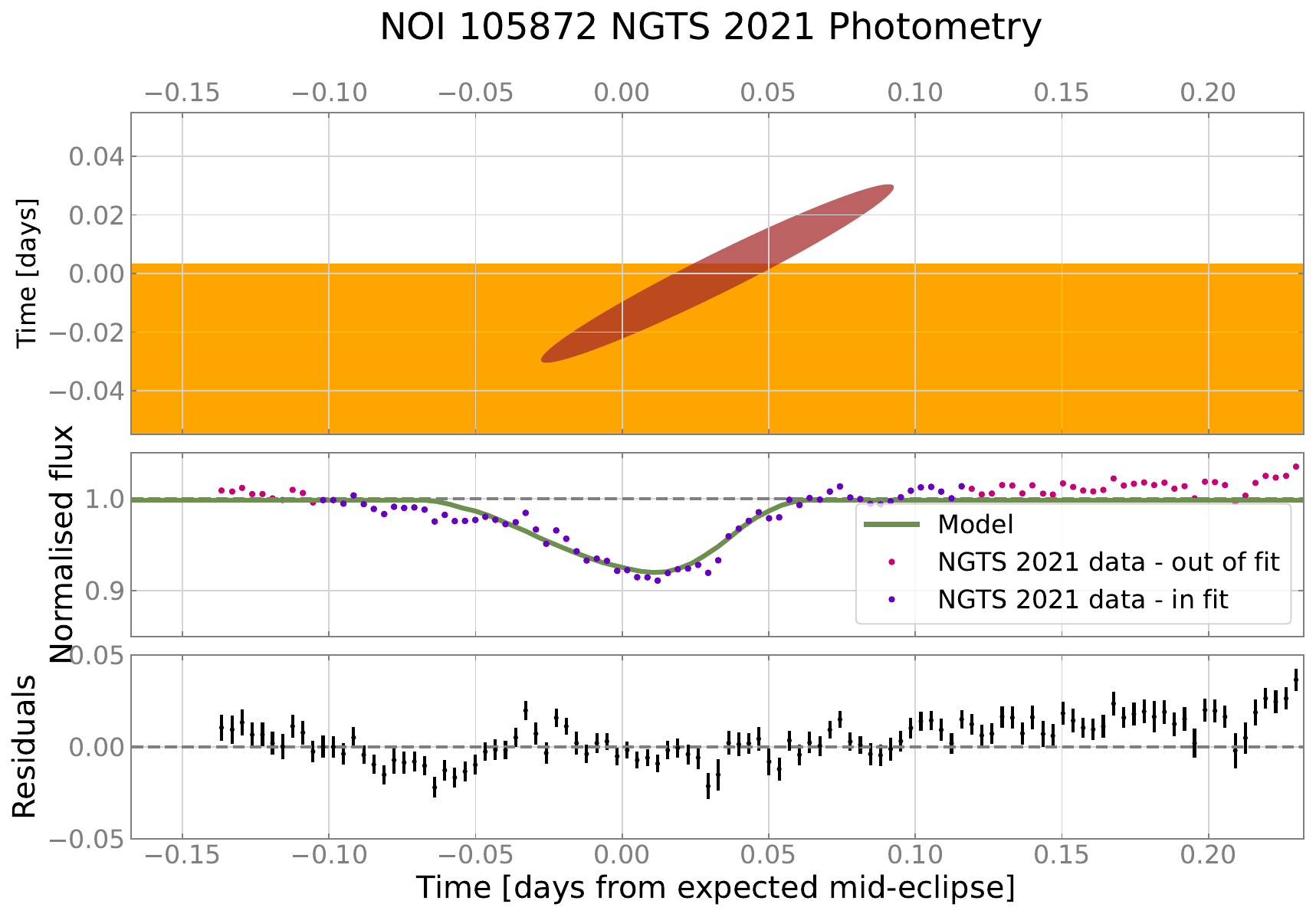}
            \caption[]%
            {}    
            \label{fig:NGTS 2021}
        \end{subfigure}
        \caption[]
        {\small Tilted disk models for four of the datasets of NOI\,105872 after MCMC fitting. The green line represents the model with parameters found from the MCMC fits, and the circles are the actual binned data points from the phase-folded light curves of the respective telescopes, multiplied by the period to convert it into time. The purple circles represent the datapoints that were used to fit the tilted disk model and the pink circles are for the data that were not used in the fit. The yellow band represents the top centre of the star across time, with the width of the band being an arbitrary amount of time so that the band fills the length of the plot area, and the height being the diameter of the star in days. The starting y co-ordinate for the rectangle is $(y-dstar/2)$ so the offset of the top of the star to 0.0 on the y-axis is the modelled $y$ value in days. We are mostly concerned with seeing how the disk covers the star at time of mid-transit. The opacity of the fitted disk is represented by how dark the brown of the disk is in the plot. The NGTS 2021 disk model looks particularly different from the other three as its radius is greatly increased. This is likely due to the effects of stellar activity that were only observed during the model fit. These effects have likely lengthened the eclipse duration and so increased the radius of the disk model fit to the data.} 
        \label{fig:tilted disk modelling}
    \end{figure*}

\begin{table*}
	\centering
	\caption{Results of MCMC fitting the tilted disk model to the phase curves for NGTS and \textit{TESS}, along with the derived values from each of the model fits.}
	\label{tab:tilted disk results}
    \setlength{\tabcolsep}{4pt}
	\begin{tabular}{|c|c|c|c|c|} 
		\hline
        Value & NGTS (2017) & \textit{TESS} S6 & \textit{TESS} S32 & NGTS (2021) \\ 
		\hline 
        $y$ [days]\rule{0pt}{3ex} & $-$0.029$^{+0.004}_{-0.001}$ & $-$0.032$^{+0.006}_{-0.007}$ & $-$0.035$^{+0.005}_{-0.016}$ & $-$0.040$^{+0.006}_{-0.006}$\\ [5pt]
        $i$ [$^\circ$] & 79.1$^{+1.8}_{-1.3}$ & 70.3$^{+3.6}_{-5.1}$ & 67.9$\pm5.1$ & 84.3$^{+1.1}_{-1.8}$\\ [5pt]
        $\phi$ [$^\circ$] & 10.6$^{+7.9}_{-2.4}$ & 21.4$^{+5.9}_{-8.6}$ & 11.4$^{+19.0}_{-9.6}$ & 26.4$^{+4.0}_{-2.4}$\\ [5pt]
        $r$ [days] & 0.043$\pm0.002$ & 0.045$^{+0.009}_{-0.006}$ & 0.031$^{+0.023}_{-0.004}$ & 0.067$\pm0.013$\\ [5pt]
        $t_x$ & 0.450$^{+0.054}_{-0.055}$ & 0.515$^{+0.053}_{-0.095}$ & 0.165$^{+0.172}_{-0.112}$ & 0.265$^{+0.144}_{-0.137}$\\ [5pt]
        $hjd_{\rm central}$ [days] & 0.015$^{+0.002}_{-0.001}$ & 0.009$^{+0.008}_{-0.006}$ & 0.018$^{+0.019}_{-0.002}$ & 0.033$\pm0.011$\\ [5pt]
        \hline
        $r$ [au]\rule{0pt}{3ex} & $\left(5.18\pm0.24\right)\times10^{-3}$ & $\left(5.42^{+1.08}_{-0.72}\right)\times10^{-3}$ & $\left(3.73^{+2.77}_{-0.48}\right)\times10^{-3}$ & $\left(8.07\pm1.56\right)\times10^{-3}$ \\ [5pt]
        $M_{\rm min}$ [$M_\text{J}$] & $120\pm19$ & 142$^{+88}_{-60}$ & 40$^{+89}_{-16}$ & $1154\pm694$ \\ [5pt]
        $\chi^2$ & 129.1 & 222.3 & 95.1 & 169.7 \\ [3pt]
        BIC & 158.2 & 251.4 & 124.2 & 198.8 \\
		\hline
	\end{tabular}
\end{table*}

So far, we have discussed how the geometry of this system has to be relatively atypical for us to see the shape of the eclipse that has been observed, with the variability of the light curve being an added barrier for our models and hypotheses. It is entirely possible that this object is something far more complicated than what can be easily explained by a single hypothesis. Other objects that have very complex and unusual light curves include J1407b \citep{Kenworthy2015} and ASASSN-21js \citep{pramono2024}, both of which are unusually deep and asymmetric, and have since been modelled as giant exoring (rings around exoplanets, analagous to Saturn's rings) systems. Our object is too close to its host star to be a similar system, however perhaps these objects can give us an insight into the geometry and origin of our light curve.

This leads us to our most promising hypothesis so far; we considered a tilted disk around an eclipsing companion. The asymmetry of our light curve could be due to the tilted geometry of the disk system, with the depth of the eclipse being directly related to the radius of the disk around the object. With this hypothesis we can also explain the variability that is seen year upon year. There are two processes that could be causing this; precession and stellar activity. A disk precesses due to radiation, magnetic or gravitational influences on the system over time, and this seems especially likely when we consider how small the distance is between the object and its host star. Even if these changes are gradual, they will eventually cause changes within the eclipse profile. We have already discussed the young and active nature of the host M-star in this system, and how certain stellar activity events, such as star spots, can cause features like the ingress bump previously discussed. Other stellar activity events, such as flares and coronal mass ejections would cause the disk to be bombarded by radiation. Upon absorbing this radiation the disk could expand, or `puff up', changing the size and opacity that is then noticeable within the light curve. A tilted disk composed of dust could also cause reflected light, which would help to explain the increase of flux in the out-of-eclipse part of the lightcurve.

We used a custom code to model a tilted disk orbiting in front of the star \citep{Kenworthy_software, Kenworthy2015}. Once we created a preliminary fit by eye to the data, we developed MCMC fits for each of the light curves, using the package \texttt{emcee} \citep{emcee}, to ensure that the disk parameters were best-fit to the data. The free parameters within the MCMC fit were as follows: offset of disk from centre of star ($y$), inclination of disk in degrees ($i$) where $i=90^\circ$ would be edge on, rotation of the semi-major axis from the x-axis ($\phi$) where the x-axis is the horizontal line left to right along the star, radius of disk ($r$), transmission of disk ($t_x$), the point where the middle of the disk covers the star ($hjd_{\rm central}$) and a model offset that allows the continuum to shift from 1.

When performing the MCMC fits we had to choose carefully which of the datasets to use, especially because of the bump in the ingress that we have already discussed (see Figure \ref{fig:chunks}); this excludes SAAO, SPECULOOS and ULTRASPEC. Without taking stellar activity into account, the fits could be skewed by their effects on the light curve. There are two options for dealing with the ingress bump: 1) mask out the bump and fit the rest of the eclipse, or 2) simultaneously fit a stellar activity model that takes into account the bump while the rest of the eclipse is fit by the disk model. Masking out this bump would be difficult as the limits will be set by eye and this would not be completely accurate. If the fits for these data were needed in the future then an accompanying stellar activity model would need to be fit simultaneously. Instead, we choose to only fit the data to NGTS (2017), \textit{TESS} Sector 6, \textit{TESS} Sector 32 and NGTS (2021) due to their lack of any noticeable bump in the eclipse. 

Upon initial fitting of the data it became clear that the NGTS (2021) light curve is likely also affected by stellar activity, in a similar way to the SAAO data where we did not pick up on the effects by eye, but noticed the change in ingress when we started fitting models to the data. Once the model was fit to the data, a small bump in the ingress became clear, as well as a lengthening of the eclipse duration, starting earlier than the rest of the light curves. This data was taken over the course of several months, so if we separate the data into small chunks, representing around 3-4 weeks at a time, and phase fold it, we can see that some of the chunks do not show as much stellar activity interference. We keep the tilted disk fit in as an example, to show how the effects of stellar activity change the model fit, and hence we still do not choose to fit the SAAO, SPECULOOS and ULTRASPEC datasets, the ones that exhibit the bump as shown in Figure \ref{fig:chunks}.


Figure \ref{fig:tilted disk modelling} and Table \ref{tab:tilted disk results} show the results of the MCMC fitting for the four remaining light curves. Each light curve is modelled over a phase range of $\pm0.16$ from the estimated $t_0$ value, the same range as the \texttt{batman} models for direct comparison. Since we have already discussed the effect of stellar activity on the NGTS (2021) light curve, it is unsurprising to see that this fit is different from the others. Particularly, the increased eclipse duration will affect the radius of the disk needed to produce this shape, which is why the disk is much wider than in previous fits. Overall, though, the orientation and transmission of the disk is relatively similar to the other fits.

At first glance it might appear that the \textit{TESS} Sector 32 fit is also different from the initial two, however when we look at Table \ref{tab:tilted disk results} we can see that most of the \textit{TESS} Sector 32 model parameters are within the errors of the fits for NGTS 2017 and \textit{TESS} Sector 6. This is with the exception of the transmission value, however we can also see that the value for $\phi$ is not constrained very well and so the fitted value of this could be affecting the transmission value. These values are linked because changing both the orientation and opacity of the disk will change the amount of light that is blocked coming from the host star so if they are unconstrained then it is difficult to determine which factor is affecting the eclipse curve more.

\begin{figure}
	\includegraphics[width=\columnwidth]{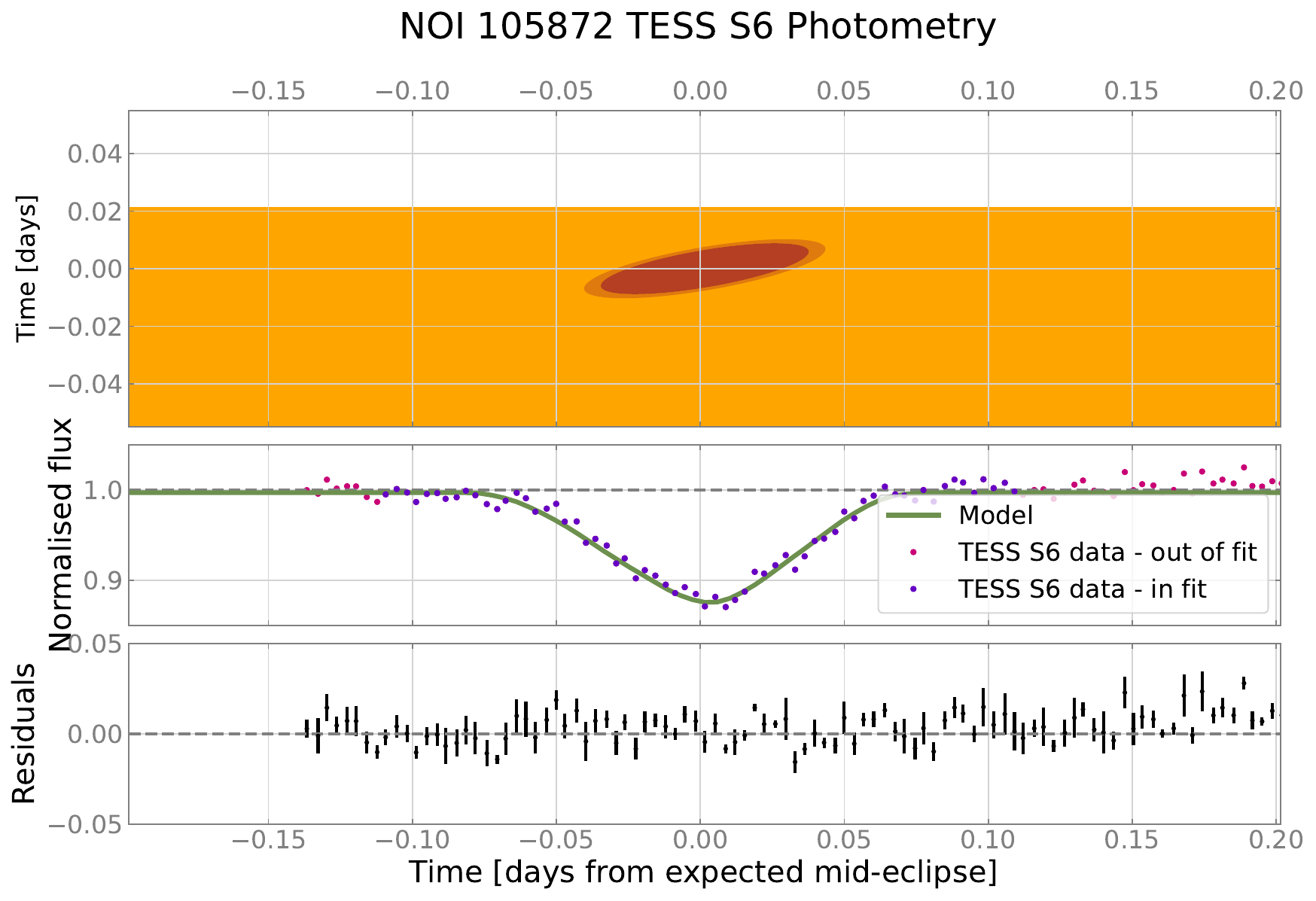}
    \caption{Tilted disk models for \textit{TESS} Sector 6 modelled with original disk, but effects due to stellar activity (expansion of the disk) also modelled.}
    \label{fig:TESS S6 2 disk}
\end{figure}


If we look at the NGTS 2017 and \textit{TESS} Sector 6 data, the disks are incredibly similar, and most parameters are within errors of each other. This is very promising, especially since the two datasets are taken about a year apart and we would not expect to see many differences on those timescales. The slight change in orientation could be representative of the disk precession that was previously discussed. Figure \ref{fig:TESS S6 2 disk} shows the same \textit{TESS} Sector 6 data, this time we fit an inner disk similar to the initial fits, but with an accompanying ``outer disk''. This represents the expansion of the disk as it is bombarded with radiation from stellar activity events, with the outer disk being more transparent than the original inner disk. This hypothesis also fits the data very well and could help to explain why the \textit{TESS} eclipses are deeper than the NGTS eclipses, along with the possible disk precession.

If we take the NGTS 2017 fit then the radius of the disk is $5.18\times10^{-3}$~au, and for this disk to be contained within the Hill sphere of the object then we calculated the object's minimum mass to be $120$\,M$_\text{J}$. If this is the case then the object is likely a small star and not a planet, but the only way to confirm this would be to get a constraint on the object's mass through spectroscopy. We also calculated the temperature of the disk using Equation \ref{eqn:temp}, which is generally used for calculating the equilibrium temperature of exoplanets.

\begin{equation}
    \label{eqn:temp}
    T_{eq} = T_{*} \cdot \sqrt{\frac{R_{*}}{f \cdot a}} \cdot (1-A)^{\frac{1}{4}}
\end{equation}

\noindent where $T_{*}$ is the effective temperature of the host star, $R_{*}$ is the radius of the host star, $f$ is a multiplication factor based on the redistribution of heat across the surface, from day- to night-side, of the object, $a$ is the semimajor axis and $A$ is the albedo. The factor, $f$, we used is 2 because we assume uniform redistribution.

The temperature is found to be $\approx 1660$K. Using this we were able to calculate the likely depth of a secondary eclipse if the object is on a circular orbit, and this value came to $7.32\pm0.13\%$. 


We wanted to check that the lack of a secondary eclipse is not due to the eccentricity and inclination of the orbit itself. By using the equations set out in \citet{Santerne2013} which satisfy the concepts from \citet{Winn2010}, we calculate the probability of seeing a secondary eclipse for this object. The lower limit probability of there being a primary eclipse but no secondary eclipse was calculated to be $\sim23\%$, however it should be noted that the object radius used in this equation assumes that the object is spherical in shape, and the shape of a disk instead could change the value of this probability, although we do not think that this would be a major contributing factor. This calculation shows that it is possible that we see no secondary eclipse due to the orientation of the orbit, however there could still be sinusoidal stellar activity effects as previously mentioned.

There are still a number of unanswered questions with this model. The model itself is relatively flexible, hence possibly why it fits our data so well, but we hope that with further observations, we could try and constrain our understanding of each model better and their plausibility. This is why we are focused on this hypothesis as our main one for now, but it cannot be confirmed or dirproven until we understand the system in its totality.

If we compare the chi-squared values for each dataset across the eccentricity and tilted disk models, all of the datasets have a better fit with the tilted disk model. We made sure to keep the phase range and number of data points the same across each model fit for direct comparison. Because the models have a slightly different number of parameters we also compared the Bayesian Information Criterion (BIC) values for each model since this takes into account that a model with more parameters is more likely to be a better fit. The BIC values are also lower for the tilted disk model compared to the eccentricity model. This further compounds the tilted disk model as our most promising hypothesis so far.



\section{Conclusions}

We have presented a comprehensive photometric and spectroscopic investigation of the object NOI\,105872 , a young, short-period eclipsing system located in the Orion Nebula Cluster. Using multi-epoch photometry spanning nearly a decade from both ground-based and space-based facilities, we explored the origin of its unusually deep, asymmetric, and time-variable eclipse light curves.

The host star is confirmed to be a young ($4.1\pm0.4$ Myr), diskless pre-main-sequence M type star, based on spectroscopic indicators and spectral energy distribution modelling. The absence of infrared excess rules out a primordial circumstellar disk, placing constraints on scenarios involving large-scale dust structures associated with the star itself.

A range of hypotheses were examined to explain the observed eclipse morphology, including stellar activity, dipper-like behaviour, gravity darkening, orbital eccentricity, and atmospheric mass loss from a disintegrating or evaporating planet. While some of these mechanisms may contribute second-order effects (particularly stellar activity), none are able to reproduce the depth, asymmetry, and long-term shape of the eclipse signal in a self-consistent manner.

Our modelling strongly favours a scenario in which the eclipsing structure is an extended, tilted circumsecondary disk surrounding a close-in companion. Parametric disk models fitted independently to multiple datasets consistently reproduce the observed eclipse profiles and yield a characteristic disk radius of $\sim0.005$~au. While the inferred disk orientations vary modestly between epochs, these differences are consistent with slow disk precession and/or changes in disk structure driven by irradiation from the young, active host star. The presence of ingress features in some datasets is likely due to a combination of stellar activity and transient changes in disk opacity or vertical extent.

If the disk is gravitationally bound and confined within the Hill sphere of the companion, the inferred disk size implies a companion mass in the low-mass stellar regime. However, current spectroscopic data are insufficient to place meaningful constraints on the companion mass or radial velocity semi-amplitude. The absence of a detectable secondary eclipse can be explained by the system geometry and is not inconsistent with the proposed disk configuration. The lack of infrared excess can also be explained if we assume that the disk is too small to be seen in the IR data.

This enigmatic object therefore emerges as a candidate for a short-period circumsecondary disk system, but with the caveat that there must be follow-up observations to confirm its nature. Its proximity to the host star, youth, and apparent disk variability make it an important system for studying disk survival, evolution, and dynamics in extreme environments. Continued high-cadence, multi-wavelength photometric monitoring, along with higher signal-to-noise spectroscopic observations, will be crucial for confirming the nature of the companion, probing disk composition, and testing models of disk precession and irradiation-driven evolution.

There are only a handful of circumsecondary disk candidates, of which the shortest period one is orbiting every 5.4 days \citep{David2016,Ohno2022}. If confirmed, this system would provide a valuable benchmark for understanding circumsecondary disks and their role in shaping planetary and satellite systems around low-mass stars.


\section*{Acknowledgements}


This work is based on data collected under the NGTS project at the ESO La Silla Paranal Observatory. The NGTS facility is operated by a consortium institutes with support from the UK Science and Technology Facilities Council (STFC) under projects ST/M001962/1, ST/S002642/1 and ST/W003163/1.
C.A.W. and E.dM would like to acknowledge support from the UK Science and Technology Facilities Council (STFC,
grant number ST/X00094X/1). EG and MPB gratefully acknowledge support from UK Research and Innovation (UKRI) under the UK government’s Horizon Europe funding guarantee for an ERC starting grant [grant number EP/Z000890/1]. NM acknowledges support from the Department for the Economy (DfE) Northern Ireland postgraduate studentship scheme. This paper uses observations made at the South African Astronomical Observatory (SAAO). VSD and ULTRASPEC are supported by STFC grant ST/Z000033/1, a Leverhulme Research Fellowship (RF-2025-297/9) and the Spanish Ministry of Science, Innovation and Universities (PID2023-151588NB-I00).

The ULiege's contribution to SPECULOOS has received funding from the European Research Council under the European Union's Seventh Framework Programme (FP/2007-2013) (grant Agreement n$^\circ$ 336480/SPECULOOS), from the Balzan Prize and Francqui Foundations, from the Belgian Scientific Research Foundation (F.R.S.-FNRS; grant n$^\circ$ T.0109.20), from the University of Liege, and from the ARC grant for Concerted Research Actions financed by the Wallonia-Brussels Federation. MG is F.R.S.-FNRS Research Director.

The Cambridge contribution is supported by a grant from the Simons Foundation (PI Queloz, grant number 327127).

J.d.W. and MIT gratefully acknowledge financial support from the Heising-Simons Foundation, Dr. and Mrs. Colin Masson and Dr. Peter A. Gilman for Artemis, the first telescope of the SPECULOOS network situated in Tenerife, Spain. 

The Bern contribution to SPECULOOS is supported by the Swiss National Science Foundation (PP00P2-163967, PP00P2-190080 and the National Centre for Competence in Research PlanetS). 

The Birmingham contribution to SPECULOOS has received fund from the European Research Council (ERC) under the European Union's Horizon 2020 research and innovation programme (grant agreement n$^\circ$ 803193/BEBOP), from the MERAC foundation, and from the Science and Technology Facilities Council (STFC; grant n$^\circ$ ST/S00193X/1, ST/W002582/1, and ST/Y001710/1) and from the ERC/UKRI Frontier Research Guarantee programme (EP/Z000327/1/CandY).

\section*{Data Availability}


The TESS data is accessible via the MAST (Mikulski Archive for Space Telescopes) portal at \url{https://mast.stsci.edu/portal/
Mashup/Clients/Mast/Portal.html}. Data from NASA/ATLAS are publicly available at \url{https://fallingstar-data.com/forcedphot/}.
SPECULOOS-South data are available via the ESO archive portal at \url{https://archive.eso.org/eso/eso_archive_main.html}.
Any code used for analysis or in producing the plots in this paper can be made available upon reasonable request to the author(s).



\bibliographystyle{mnras}
\bibliography{bibliography} 

@ARTICLE{astroARIADNE,
       author = {{Vines}, Jose I. and {Jenkins}, James S.},
        title = "{ARIADNE: Measuring accurate and precise stellar parameters through SED fitting}",
      journal = {\mnras},
         year = 2022,
        month = apr,
          doi = {10.1093/mnras/stac956},
archivePrefix = {arXiv},
       eprint = {2204.03769},
 primaryClass = {astro-ph.SR},
       adsurl = {https://ui.adsabs.harvard.edu/abs/2022MNRAS.tmp..920V}
}

@ARTICLE{eccentricity,
       author = {{Barnes}, Jason W.},
        title = "{Effects of Orbital Eccentricity on Extrasolar Planet Transit Detectability and Light Curves}",
      journal = {\pasp},
         year = 2007,
        month = sep,
       volume = {119},
       number = {859},
        pages = {986-993},
          doi = {10.1086/522039},
archivePrefix = {arXiv},
       eprint = {0708.0243},
 primaryClass = {astro-ph},
       adsurl = {https://ui.adsabs.harvard.edu/abs/2007PASP..119..986B}
}

@ARTICLE{batman,
       author = {{Kreidberg}, Laura},
        title = "{batman: BAsic Transit Model cAlculatioN in Python}",
      journal = {\pasp},
         year = 2015,
        month = nov,
       volume = {127},
       number = {957},
        pages = {1161},
          doi = {10.1086/683602},
archivePrefix = {arXiv},
       eprint = {1507.08285},
 primaryClass = {astro-ph.EP},
       adsurl = {https://ui.adsabs.harvard.edu/abs/2015PASP..127.1161K}
}

@ARTICLE{gravdarkening,
       author = {{Espinosa Lara}, F. and {Rieutord}, M.},
        title = "{Gravity darkening in binary stars}",
      journal = {\aap},
         year = 2012,
        month = nov,
       volume = {547},
          eid = {A32},
        pages = {A32},
          doi = {10.1051/0004-6361/201219942},
archivePrefix = {arXiv},
       eprint = {1210.4004},
 primaryClass = {astro-ph.SR},
       adsurl = {https://ui.adsabs.harvard.edu/abs/2012A&A...547A..32E}
}

@ARTICLE{gravdarkening_transits,
       author = {{Dholakia}, Shashank and {Luger}, Rodrigo and {Dholakia}, Shishir},
        title = "{Efficient and Precise Transit Light Curves for Rapidly Rotating, Oblate Stars}",
      journal = {\apj},
         year = 2022,
        month = feb,
       volume = {925},
       number = {2},
          eid = {185},
        pages = {185},
          doi = {10.3847/1538-4357/ac33aa},
archivePrefix = {arXiv},
       eprint = {2109.03250},
 primaryClass = {astro-ph.EP},
       adsurl = {https://ui.adsabs.harvard.edu/abs/2022ApJ...925..185D}
}

@ARTICLE{exocomets,
       author = {{Luk'yanyk}, I. and {Kulyk}, I. and {Shubina}, O. and {Pavlenko}, Ya. and {Vasylenko}, M. and {Dobrycheva}, D. and {Korsun}, P.},
        title = "{Numerical simulations of exocomet transits: Insights from {\ensuremath{\beta}} Pic and KIC 3542116}",
      journal = {\aap},
         year = 2024,
        month = aug,
       volume = {688},
          eid = {A65},
        pages = {A65},
          doi = {10.1051/0004-6361/202348498},
archivePrefix = {arXiv},
       eprint = {2407.17961},
 primaryClass = {astro-ph.EP},
       adsurl = {https://ui.adsabs.harvard.edu/abs/2024A&A...688A..65L}
}

@ARTICLE{NGTS,
       author = {{Wheatley}, Peter J. and {West}, Richard G. and {Goad}, Michael R. and {Jenkins}, James S. and {Pollacco}, Don L. and {Queloz}, Didier and {Rauer}, Heike and {Udry}, St{\'e}phane and {Watson}, Christopher A. and {Chazelas}, Bruno and {Eigm{\"u}ller}, Philipp and {Lambert}, Gregory and {Genolet}, Ludovic and {McCormac}, James and {Walker}, Simon and {Armstrong}, David J. and {Bayliss}, Daniel and {Bento}, Joao and {Bouchy}, Fran{\c{c}}ois and {Burleigh}, Matthew R. and {Cabrera}, Juan and {Casewell}, Sarah L. and {Chaushev}, Alexander and {Chote}, Paul and {Csizmadia}, Szil{\'a}rd and {Erikson}, Anders and {Faedi}, Francesca and {Foxell}, Emma and {G{\"a}nsicke}, Boris T. and {Gillen}, Edward and {Grange}, Andrew and {G{\"u}nther}, Maximilian N. and {Hodgkin}, Simon T. and {Jackman}, James and {Jord{\'a}n}, Andr{\'e}s and {Louden}, Tom and {Metrailler}, Lionel and {Moyano}, Maximiliano and {Nielsen}, Louise D. and {Osborn}, Hugh P. and {Poppenhaeger}, Katja and {Raddi}, Roberto and {Raynard}, Liam and {Smith}, Alexis M.~S. and {Soto}, Maritza and {Titz-Weider}, Ruth},
        title = "{The Next Generation Transit Survey (NGTS)}",
      journal = {\mnras},
         year = 2018,
        month = apr,
       volume = {475},
       number = {4},
        pages = {4476-4493},
          doi = {10.1093/mnras/stx2836},
archivePrefix = {arXiv},
       eprint = {1710.11100},
 primaryClass = {astro-ph.EP},
       adsurl = {https://ui.adsabs.harvard.edu/abs/2018MNRAS.475.4476W}
}

@ARTICLE{TESS,
       author = {{Ricker}, George R. and {Winn}, Joshua N. and {Vanderspek}, Roland and {Latham}, David W. and {Bakos}, G{\'a}sp{\'a}r {\'A}. and {Bean}, Jacob L. and {Berta-Thompson}, Zachory K. and {Brown}, Timothy M. and {Buchhave}, Lars and {Butler}, Nathaniel R. and {Butler}, R. Paul and {Chaplin}, William J. and {Charbonneau}, David and {Christensen-Dalsgaard}, J{\o}rgen and {Clampin}, Mark and {Deming}, Drake and {Doty}, John and {De Lee}, Nathan and {Dressing}, Courtney and {Dunham}, Edward W. and {Endl}, Michael and {Fressin}, Francois and {Ge}, Jian and {Henning}, Thomas and {Holman}, Matthew J. and {Howard}, Andrew W. and {Ida}, Shigeru and {Jenkins}, Jon M. and {Jernigan}, Garrett and {Johnson}, John Asher and {Kaltenegger}, Lisa and {Kawai}, Nobuyuki and {Kjeldsen}, Hans and {Laughlin}, Gregory and {Levine}, Alan M. and {Lin}, Douglas and {Lissauer}, Jack J. and {MacQueen}, Phillip and {Marcy}, Geoffrey and {McCullough}, Peter R. and {Morton}, Timothy D. and {Narita}, Norio and {Paegert}, Martin and {Palle}, Enric and {Pepe}, Francesco and {Pepper}, Joshua and {Quirrenbach}, Andreas and {Rinehart}, Stephen A. and {Sasselov}, Dimitar and {Sato}, Bun'ei and {Seager}, Sara and {Sozzetti}, Alessandro and {Stassun}, Keivan G. and {Sullivan}, Peter and {Szentgyorgyi}, Andrew and {Torres}, Guillermo and {Udry}, Stephane and {Villasenor}, Joel},
        title = "{Transiting Exoplanet Survey Satellite (TESS)}",
      journal = {Journal of Astronomical Telescopes, Instruments, and Systems},
         year = 2015,
        month = jan,
       volume = {1},
          eid = {014003},
        pages = {014003},
          doi = {10.1117/1.JATIS.1.1.014003},
       adsurl = {https://ui.adsabs.harvard.edu/abs/2015JATIS...1a4003R}
}

@INPROCEEDINGS{SPECULOOS,
       author = {{Delrez}, Laetitia and {Gillon}, Micha{\"e}l. and {Queloz}, Didier and {Demory}, Brice-Olivier and {Almleaky}, Yaseen and {de Wit}, Julien and {Jehin}, Emmanu{\"e}l. and {Triaud}, Amaury H.~M.~J. and {Barkaoui}, Khalid and {Burdanov}, Artem and {Burgasser}, Adam J. and {Ducrot}, Elsa and {McCormac}, James and {Murray}, Catriona and {Silva Fernandes}, Catarina and {Sohy}, Sandrine and {Thompson}, Samantha J. and {Van Grootel}, Val{\'e}rie and {Alonso}, Roi and {Benkhaldoun}, Zouhair and {Rebolo}, Rafael},
        title = "{SPECULOOS: a network of robotic telescopes to hunt for terrestrial planets around the nearest ultracool dwarfs}",
    booktitle = {Ground-based and Airborne Telescopes VII},
         year = 2018,
       editor = {{Marshall}, Heather K. and {Spyromilio}, Jason},
       series = {Society of Photo-Optical Instrumentation Engineers (SPIE) Conference Series},
       volume = {10700},
        month = jul,
          eid = {107001I},
        pages = {107001I},
          doi = {10.1117/12.2312475},
archivePrefix = {arXiv},
       eprint = {1806.11205},
 primaryClass = {astro-ph.IM},
       adsurl = {https://ui.adsabs.harvard.edu/abs/2018SPIE10700E..1ID}
}

@ARTICLE{ATLAS,
       author = {{Tonry}, J.~L. and {Denneau}, L. and {Heinze}, A.~N. and {Stalder}, B. and {Smith}, K.~W. and {Smartt}, S.~J. and {Stubbs}, C.~W. and {Weiland}, H.~J. and {Rest}, A.},
        title = "{ATLAS: A High-cadence All-sky Survey System}",
      journal = {\pasp},
         year = 2018,
        month = jun,
       volume = {130},
       number = {988},
        pages = {064505},
          doi = {10.1088/1538-3873/aabadf},
archivePrefix = {arXiv},
       eprint = {1802.00879},
 primaryClass = {astro-ph.IM},
       adsurl = {https://ui.adsabs.harvard.edu/abs/2018PASP..130f4505T}
}

@ARTICLE{ULTRASPEC,
       author = {{Dhillon}, V.~S. and {Marsh}, T.~R. and {Atkinson}, D.~C. and {Bezawada}, N. and {Bours}, M.~C.~P. and {Copperwheat}, C.~M. and {Gamble}, T. and {Hardy}, L.~K. and {Hickman}, R.~D.~H. and {Irawati}, P. and {Ives}, D.~J. and {Kerry}, P. and {Leckngam}, A. and {Littlefair}, S.~P. and {McLay}, S.~A. and {O'Brien}, K. and {Peacocke}, P.~T. and {Poshyachinda}, S. and {Richichi}, A. and {Soonthornthum}, B. and {Vick}, A.},
        title = "{ULTRASPEC: a high-speed imaging photometer on the 2.4-m Thai National Telescope}",
      journal = {\mnras},
         year = 2014,
        month = nov,
       volume = {444},
       number = {4},
        pages = {4009-4021},
          doi = {10.1093/mnras/stu1660},
archivePrefix = {arXiv},
       eprint = {1408.2733},
 primaryClass = {astro-ph.IM},
       adsurl = {https://ui.adsabs.harvard.edu/abs/2014MNRAS.444.4009D}
}

@ARTICLE{von_zeipel,
       author = {{von Zeipel}, H.},
        title = "{The radiative equilibrium of a rotating system of gaseous masses}",
      journal = {\mnras},
         year = 1924,
        month = jun,
       volume = {84},
        pages = {665-683},
          doi = {10.1093/mnras/84.9.665},
       adsurl = {https://ui.adsabs.harvard.edu/abs/1924MNRAS..84..665V}
}

@ARTICLE{mamajek2012,
       author = {{Mamajek}, Eric E. and {Quillen}, Alice C. and {Pecaut}, Mark J. and {Moolekamp}, Fred and {Scott}, Erin L. and {Kenworthy}, Matthew A. and {Collier Cameron}, Andrew and {Parley}, Neil R.},
        title = "{Planetary Construction Zones in Occultation: Discovery of an Extrasolar Ring System Transiting a Young Sun-like Star and Future Prospects for Detecting Eclipses by Circumsecondary and Circumplanetary Disks}",
      journal = {\aj},
         year = 2012,
        month = mar,
       volume = {143},
       number = {3},
          eid = {72},
        pages = {72},
          doi = {10.1088/0004-6256/143/3/72},
archivePrefix = {arXiv},
       eprint = {1108.4070},
 primaryClass = {astro-ph.SR},
       adsurl = {https://ui.adsabs.harvard.edu/abs/2012AJ....143...72M}
}

@ARTICLE{SPECULOOS2,
       author = {{Gillon}, Micha{\"e}l},
        title = "{Searching for red worlds}",
      journal = {Nature Astronomy},
         year = 2018,
        month = apr,
       volume = {2},
        pages = {344-344},
          doi = {10.1038/s41550-018-0443-y},
       adsurl = {https://ui.adsabs.harvard.edu/abs/2018NatAs...2..344G}
}

@ARTICLE{HARPS,
       author = {{Mayor}, M. and {Pepe}, F. and {Queloz}, D. and {Bouchy}, F. and {Rupprecht}, G. and {Lo Curto}, G. and {Avila}, G. and {Benz}, W. and {Bertaux}, J. -L. and {Bonfils}, X. and {Dall}, Th. and {Dekker}, H. and {Delabre}, B. and {Eckert}, W. and {Fleury}, M. and {Gilliotte}, A. and {Gojak}, D. and {Guzman}, J.~C. and {Kohler}, D. and {Lizon}, J. -L. and {Longinotti}, A. and {Lovis}, C. and {Megevand}, D. and {Pasquini}, L. and {Reyes}, J. and {Sivan}, J. -P. and {Sosnowska}, D. and {Soto}, R. and {Udry}, S. and {van Kesteren}, A. and {Weber}, L. and {Weilenmann}, U.},
        title = "{Setting New Standards with HARPS}",
      journal = {The Messenger},
         year = 2003,
        month = dec,
       volume = {114},
        pages = {20-24},
       adsurl = {https://ui.adsabs.harvard.edu/abs/2003Msngr.114...20M}
}

@ARTICLE{SHOC,
       author = {{Coppejans}, R. and {Gulbis}, A.~A.~S. and {Kotze}, M.~M. and {Coppejans}, D.~L. and {Worters}, H.~L. and {Woudt}, P.~A. and {Whittal}, H. and {Cloete}, J. and {Fourie}, P.},
        title = "{Characterizing and Commissioning the Sutherland High-Speed Optical Cameras (SHOC)}",
      journal = {\pasp},
         year = 2013,
        month = aug,
       volume = {125},
       number = {930},
        pages = {976},
          doi = {10.1086/672156},
       adsurl = {https://ui.adsabs.harvard.edu/abs/2013PASP..125..976C}
}

@ARTICLE{emcee,
       author = {{Foreman-Mackey}, Daniel and {Hogg}, David W. and {Lang}, Dustin and {Goodman}, Jonathan},
        title = "{emcee: The MCMC Hammer}",
      journal = {\pasp},
         year = 2013,
        month = mar,
       volume = {125},
       number = {925},
        pages = {306},
          doi = {10.1086/670067},
archivePrefix = {arXiv},
       eprint = {1202.3665},
 primaryClass = {astro-ph.IM},
       adsurl = {https://ui.adsabs.harvard.edu/abs/2013PASP..125..306F}
}

@ARTICLE{Oshagh2013,
       author = {{Oshagh}, M. and {Santos}, N.~C. and {Boisse}, I. and {Bou{\'e}}, G. and {Montalto}, M. and {Dumusque}, X. and {Haghighipour}, N.},
        title = "{Effect of stellar spots on high-precision transit light-curve}",
      journal = {\aap},
         year = 2013,
        month = aug,
       volume = {556},
          eid = {A19},
        pages = {A19},
          doi = {10.1051/0004-6361/201321309},
archivePrefix = {arXiv},
       eprint = {1306.0739},
 primaryClass = {astro-ph.EP},
       adsurl = {https://ui.adsabs.harvard.edu/abs/2013A&A...556A..19O}
}

@INPROCEEDINGS{valio,
       author = {{Valio}, Adriana},
        title = "{Starspots properties and stellar activity from planetary transits}",
    booktitle = {Living Around Active Stars},
         year = 2017,
       editor = {{Nandy}, D. and {Valio}, A. and {Petit}, P.},
       series = {IAU Symposium},
       volume = {328},
        month = oct,
        pages = {69-76},
          doi = {10.1017/S1743921317004094},
       adsurl = {https://ui.adsabs.harvard.edu/abs/2017IAUS..328...69V}
}

@article{Ioannidis2015,
  title = {How do starspots influence the transit timing variations of exoplanets? Simulations of individual and consecutive transits},
  volume = {585},
  ISSN = {1432-0746},
  url = {http://dx.doi.org/10.1051/0004-6361/201527184},
  DOI = {10.1051/0004-6361/201527184},
  journal = {Astronomy &amp; Astrophysics},
  publisher = {EDP Sciences},
  author = {Ioannidis,  P. and Huber,  K. F. and Schmitt,  J. H. M. M.},
  year = {2015},
  month = dec,
  pages = {A72}
}

@ARTICLE{Barros2013,
       author = {{Barros}, S.~C.~C. and {Bou{\'e}}, G. and {Gibson}, N.~P. and {Pollacco}, D.~L. and {Santerne}, A. and {Keenan}, F.~P. and {Skillen}, I. and {Street}, R.~A.},
        title = "{Transit timing variations in WASP-10b induced by stellar activity}",
      journal = {\mnras},
         year = 2013,
        month = apr,
       volume = {430},
       number = {4},
        pages = {3032-3047},
          doi = {10.1093/mnras/stt111},
archivePrefix = {arXiv},
       eprint = {1301.3760},
 primaryClass = {astro-ph.EP},
       adsurl = {https://ui.adsabs.harvard.edu/abs/2013MNRAS.430.3032B}
}

@ARTICLE{sanchis-ojeda2011,
       author = {{Sanchis-Ojeda}, Roberto and {Winn}, Joshua N. and {Holman}, Matthew J. and {Carter}, Joshua A. and {Osip}, David J. and {Fuentes}, Cesar I.},
        title = "{Starspots and Spin-orbit Alignment in the WASP-4 Exoplanetary System}",
      journal = {\apj},
         year = 2011,
        month = jun,
       volume = {733},
       number = {2},
          eid = {127},
        pages = {127},
          doi = {10.1088/0004-637X/733/2/127},
archivePrefix = {arXiv},
       eprint = {1103.4859},
 primaryClass = {astro-ph.EP},
       adsurl = {https://ui.adsabs.harvard.edu/abs/2011ApJ...733..127S}
}

@ARTICLE{almenara2022,
       author = {{Almenara}, J.~M. and {Bonfils}, X. and {Forveille}, T. and {Astudillo-Defru}, N. and {Ciardi}, D.~R. and {Schwarz}, R.~P. and {Collins}, K.~A. and {Cointepas}, M. and {Lund}, M.~B. and {Bouchy}, F. and {Charbonneau}, D. and {D{\'\i}az}, R.~F. and {Delfosse}, X. and {Kidwell}, R.~C. and {Kunimoto}, M. and {Latham}, D.~W. and {Lissauer}, J.~J. and {Murgas}, F. and {Ricker}, G. and {Seager}, S. and {Vezie}, M. and {Watanabe}, D.},
        title = "{TOI-3884 b: A rare 6-R$_{E}$ planet that transits a low-mass star with a giant and likely polar spot}",
      journal = {\aap},
         year = 2022,
        month = nov,
       volume = {667},
          eid = {L11},
        pages = {L11},
          doi = {10.1051/0004-6361/202244791},
archivePrefix = {arXiv},
       eprint = {2210.10909},
 primaryClass = {astro-ph.EP},
       adsurl = {https://ui.adsabs.harvard.edu/abs/2022A&A...667L..11A}
}

@ARTICLE{budaj2013,
       author = {{Budaj}, J.},
        title = "{Light-curve analysis of KIC 12557548b: an extrasolar planet with a comet-like tail}",
      journal = {\aap},
         year = 2013,
        month = sep,
       volume = {557},
          eid = {A72},
        pages = {A72},
          doi = {10.1051/0004-6361/201220260},
archivePrefix = {arXiv},
       eprint = {1208.3693},
 primaryClass = {astro-ph.EP},
       adsurl = {https://ui.adsabs.harvard.edu/abs/2013A&A...557A..72B}
}

@ARTICLE{brogi2012,
       author = {{Brogi}, M. and {Keller}, C.~U. and {de Juan Ovelar}, M. and {Kenworthy}, M.~A. and {de Kok}, R.~J. and {Min}, M. and {Snellen}, I.~A.~G.},
        title = "{Evidence for the disintegration of KIC 12557548 b}",
      journal = {\aap},
         year = 2012,
        month = sep,
       volume = {545},
          eid = {L5},
        pages = {L5},
          doi = {10.1051/0004-6361/201219762},
archivePrefix = {arXiv},
       eprint = {1208.2988},
 primaryClass = {astro-ph.EP},
       adsurl = {https://ui.adsabs.harvard.edu/abs/2012A&A...545L...5B}
}

@INCOLLECTION{bourrier2018,
       author    = {Bourrier, Vincent and Lecavelier des Etangs, Alain},
       title     = {Characterizing Evaporating Atmospheres of Exoplanets},
       booktitle = {Handbook of Exoplanets},
       editor    = {Deeg, Hans J. and Belmonte, Juan Antonio},
       publisher = {Springer International Publishing},
       year      = {2018},
       eid       = {148},
       pages     = {148},
       doi       = {10.1007/978-3-319-55333-7_148},
       adsurl    = {https://ui.adsabs.harvard.edu/abs/2018haex.bookE.148B}
}

@ARTICLE{foster2022,
       author = {{Foster}, G. and {Poppenhaeger}, K. and {Ilic}, N. and {Schwope}, A.},
        title = "{Exoplanet X-ray irradiation and evaporation rates with eROSITA}",
      journal = {\aap},
         year = 2022,
        month = may,
       volume = {661},
          eid = {A23},
        pages = {A23},
          doi = {10.1051/0004-6361/202141097},
archivePrefix = {arXiv},
       eprint = {2106.14550},
 primaryClass = {astro-ph.EP},
       adsurl = {https://ui.adsabs.harvard.edu/abs/2022A&A...661A..23F}
}

@ARTICLE{Kenworthy2015,
       author = {{Kenworthy}, M.~A. and {Mamajek}, E.~E.},
        title = "{Modeling Giant Extrasolar Ring Systems in Eclipse and the Case of J1407b: Sculpting by Exomoons?}",
      journal = {\apj},
         year = 2015,
        month = feb,
       volume = {800},
       number = {2},
          eid = {126},
        pages = {126},
          doi = {10.1088/0004-637X/800/2/126},
archivePrefix = {arXiv},
       eprint = {1501.05652},
 primaryClass = {astro-ph.SR},
       adsurl = {https://ui.adsabs.harvard.edu/abs/2015ApJ...800..126K}
}

@ARTICLE{pramono2024,
       author = {{Pramono}, T.~H. and {Kenworthy}, M.~A. and {van Boekel}, R.},
        title = "{ASASSN-21js: A multi-year transit of a ringed disc}",
      journal = {\aap},
         year = 2024,
        month = aug,
       volume = {688},
          eid = {L11},
        pages = {L11},
          doi = {10.1051/0004-6361/202450288},
archivePrefix = {arXiv},
       eprint = {2408.06744},
 primaryClass = {astro-ph.SR},
       adsurl = {https://ui.adsabs.harvard.edu/abs/2024A&A...688L..11P}
}

@ARTICLE{Murray2020,
       author = {{Murray}, C.~A. and {Delrez}, L. and {Pedersen}, P.~P. and {Queloz}, D. and {Gillon}, M. and {Burdanov}, A. and {Ducrot}, E. and {Garcia}, L.~J. and {Lienhard}, F. and {Demory}, B.~O. and {Jehin}, E. and {McCormac}, J. and {Sebastian}, D. and {Sohy}, S. and {Thompson}, S.~J. and {Triaud}, A.~H.~M.~J. and {Van Grootel}, V. and {G{\"u}nther}, M.~N. and {Huang}, C.~X.},
        title = "{Photometry and performance of SPECULOOS-South}",
      journal = {\mnras},
         year = 2020,
        month = jun,
       volume = {495},
       number = {2},
        pages = {2446-2457},
          doi = {10.1093/mnras/staa1283},
archivePrefix = {arXiv},
       eprint = {2005.02423},
 primaryClass = {astro-ph.EP},
       adsurl = {https://ui.adsabs.harvard.edu/abs/2020MNRAS.495.2446M}
}

@ARTICLE{Santerne2013,
       author = {{Santerne}, A. and {Fressin}, F. and {D{\'\i}az}, R.~F. and {Figueira}, P. and {Almenara}, J. -M. and {Santos}, N.~C.},
        title = "{The contribution of secondary eclipses as astrophysical false positives to exoplanet transit surveys}",
      journal = {\aap},
         year = 2013,
        month = sep,
       volume = {557},
          eid = {A139},
        pages = {A139},
          doi = {10.1051/0004-6361/201321475},
archivePrefix = {arXiv},
       eprint = {1307.2003},
 primaryClass = {astro-ph.EP},
       adsurl = {https://ui.adsabs.harvard.edu/abs/2013A&A...557A.139S}
}

@INCOLLECTION{Winn2010,
       author    = {Winn, J.~N.},
       title     = {Exoplanet Transits and Occultations},
       booktitle = {Exoplanets},
       editor    = {Seager, S.},
       publisher = {University of Arizona Press},
       year      = {2010},
       pages     = {55--77},
       doi       = {10.48550/arXiv.1001.2010},
       adsurl    = {https://ui.adsabs.harvard.edu/abs/2010exop.book...55W}
}

@misc{Kenworthy_software,
       author = {{Kenworthy}, Matthew A. and {Mamajek}, Eric E.},
        title = "{Exorings: Exoring modelling software}",
 howpublished = {Astrophysics Source Code Library, record ascl:1501.012},
         year = 2015,
        month = jan,
          eid = {ascl:1501.012},
       adsurl = {https://ui.adsabs.harvard.edu/abs/2015ascl.soft01012K}
}

@ARTICLE{Roggero2021,
       author = {{Roggero}, Noemi and {Bouvier}, J{\'e}r{\^o}me and {Rebull}, Luisa M. and {Cody}, Ann Marie},
        title = "{The dipper population of Taurus seen with K2}",
      journal = {\aap},
         year = 2021,
        month = jul,
       volume = {651},
          eid = {A44},
        pages = {A44},
          doi = {10.1051/0004-6361/202140646},
archivePrefix = {arXiv},
       eprint = {2106.02064},
 primaryClass = {astro-ph.SR},
       adsurl = {https://ui.adsabs.harvard.edu/abs/2021A&A...651A..44R}
}

@ARTICLE{Alencar2010,
       author = {{Alencar}, S.~H.~P. and {Teixeira}, P.~S. and {Guimar{\~a}es}, M.~M. and {McGinnis}, P.~T. and {Gameiro}, J.~F. and {Bouvier}, J. and {Aigrain}, S. and {Flaccomio}, E. and {Favata}, F.},
        title = "{Accretion dynamics and disk evolution in NGC 2264: a study based on CoRoT photometric observations}",
      journal = {\aap},
         year = 2010,
        month = sep,
       volume = {519},
          eid = {A88},
        pages = {A88},
          doi = {10.1051/0004-6361/201014184},
archivePrefix = {arXiv},
       eprint = {1005.4384},
 primaryClass = {astro-ph.SR},
       adsurl = {https://ui.adsabs.harvard.edu/abs/2010A&A...519A..88A}
}

@ARTICLE{Cody2014,
       author = {{Cody}, Ann Marie and {Stauffer}, John and {Baglin}, Annie and {Micela}, Giuseppina and {Rebull}, Luisa M. and {Flaccomio}, Ettore and {Morales-Calder{\'o}n}, Mar{\'\i}a and {Aigrain}, Suzanne and {Bouvier}, J{\`e}r{\^o}me and {Hillenbrand}, Lynne A. and {Gutermuth}, Robert and {Song}, Inseok and {Turner}, Neal and {Alencar}, Silvia H.~P. and {Zwintz}, Konstanze and {Plavchan}, Peter and {Carpenter}, John and {Findeisen}, Krzysztof and {Carey}, Sean and {Terebey}, Susan and {Hartmann}, Lee and {Calvet}, Nuria and {Teixeira}, Paula and {Vrba}, Frederick J. and {Wolk}, Scott and {Covey}, Kevin and {Poppenhaeger}, Katja and {G{\"u}nther}, Hans Moritz and {Forbrich}, Jan and {Whitney}, Barbara and {Affer}, Laura and {Herbst}, William and {Hora}, Joseph and {Barrado}, David and {Holtzman}, Jon and {Marchis}, Franck and {Wood}, Kenneth and {Medeiros Guimar{\~a}es}, Marcelo and {Lillo Box}, Jorge and {Gillen}, Ed and {McQuillan}, Amy and {Espaillat}, Catherine and {Allen}, Lori and {D'Alessio}, Paola and {Favata}, Fabio},
        title = "{CSI 2264: Simultaneous Optical and Infrared Light Curves of Young Disk-bearing Stars in NGC 2264 with CoRoT and Spitzer{\textemdash}Evidence for Multiple Origins of Variability}",
      journal = {\aj},
         year = 2014,
        month = apr,
       volume = {147},
       number = {4},
          eid = {82},
        pages = {82},
          doi = {10.1088/0004-6256/147/4/82},
archivePrefix = {arXiv},
       eprint = {1401.6582},
 primaryClass = {astro-ph.SR},
       adsurl = {https://ui.adsabs.harvard.edu/abs/2014AJ....147...82C}
}

@ARTICLE{Stauffer2015,
       author = {{Stauffer}, John and {Cody}, Ann Marie and {McGinnis}, Pauline and {Rebull}, Luisa and {Hillenbrand}, Lynne A. and {Turner}, Neal J. and {Carpenter}, John and {Plavchan}, Peter and {Carey}, Sean and {Terebey}, Susan and {Morales-Calder{\'o}n}, Mar{\'\i}a and {Alencar}, Silvia H.~P. and {Bouvier}, Jerome and {Venuti}, Laura and {Hartmann}, Lee and {Calvet}, Nuria and {Micela}, Giusi and {Flaccomio}, Ettore and {Song}, Inseok and {Gutermuth}, Rob and {Barrado}, David and {Vrba}, Frederick J. and {Covey}, Kevin and {Padgett}, Debbie and {Herbst}, William and {Gillen}, Edward and {Lyra}, Wladimir and {Medeiros Guimaraes}, Marcelo and {Bouy}, Herve and {Favata}, Fabio},
        title = "{CSI 2264: Characterizing Young Stars in NGC 2264 With Short-Duration Periodic Flux Dips in Their Light Curves}",
      journal = {\aj},
         year = 2015,
        month = apr,
       volume = {149},
       number = {4},
          eid = {130},
        pages = {130},
          doi = {10.1088/0004-6256/149/4/130},
archivePrefix = {arXiv},
       eprint = {1501.06609},
 primaryClass = {astro-ph.SR},
       adsurl = {https://ui.adsabs.harvard.edu/abs/2015AJ....149..130S}
}

@ARTICLE{Andsell2016,
       author = {{Ansdell}, M. and {Gaidos}, E. and {Rappaport}, S.~A. and {Jacobs}, T.~L. and {LaCourse}, D.~M. and {Jek}, K.~J. and {Mann}, A.~W. and {Wyatt}, M.~C. and {Kennedy}, G. and {Williams}, J.~P. and {Boyajian}, T.~S.},
        title = "{Young ``Dipper'' Stars in Upper Sco and Oph Observed by K2}",
      journal = {\apj},
         year = 2016,
        month = jan,
       volume = {816},
       number = {2},
          eid = {69},
        pages = {69},
          doi = {10.3847/0004-637X/816/2/69},
archivePrefix = {arXiv},
       eprint = {1510.08853},
 primaryClass = {astro-ph.EP},
       adsurl = {https://ui.adsabs.harvard.edu/abs/2016ApJ...816...69A}
}

@ARTICLE{Rodriguez2017,
       author = {{Rodriguez}, Joseph E. and {Ansdell}, Megan and {Oelkers}, Ryan J. and {Cargile}, Phillip A. and {Gaidos}, Eric and {Cody}, Ann Marie and {Stevens}, Daniel J. and {Somers}, Garrett and {James}, David and {Beatty}, Thomas G. and {Siverd}, Robert J. and {Lund}, Michael B. and {Kuhn}, Rudolf B. and {Gaudi}, B. Scott and {Pepper}, Joshua and {Stassun}, Keivan G.},
        title = "{Identification of Young Stellar Variables with KELT for K2. I. Taurus Dippers and Rotators}",
      journal = {\apj},
         year = 2017,
        month = oct,
       volume = {848},
       number = {2},
          eid = {97},
        pages = {97},
          doi = {10.3847/1538-4357/aa8c78},
archivePrefix = {arXiv},
       eprint = {1703.02522},
 primaryClass = {astro-ph.SR},
       adsurl = {https://ui.adsabs.harvard.edu/abs/2017ApJ...848...97R}
}

@ARTICLE{Sanchis-Ojeda2015,
       author = {{Sanchis-Ojeda}, R. and {Rappaport}, S. and {Pall{\`e}}, E. and {Delrez}, L. and {DeVore}, J. and {Gandolfi}, D. and {Fukui}, A. and {Ribas}, I. and {Stassun}, K.~G. and {Albrecht}, S. and {Dai}, F. and {Gaidos}, E. and {Gillon}, M. and {Hirano}, T. and {Holman}, M. and {Howard}, A.~W. and {Isaacson}, H. and {Jehin}, E. and {Kuzuhara}, M. and {Mann}, A.~W. and {Marcy}, G.~W. and {Miles-P{\'a}ez}, P.~A. and {Monta{\~n}{\'e}s-Rodr{\'\i}guez}, P. and {Murgas}, F. and {Narita}, N. and {Nowak}, G. and {Onitsuka}, M. and {Paegert}, M. and {Van Eylen}, V. and {Winn}, J.~N. and {Yu}, L.},
        title = "{The K2-ESPRINT Project I: Discovery of the Disintegrating Rocky Planet K2-22b with a Cometary Head and Leading Tail}",
      journal = {\apj},
         year = 2015,
        month = oct,
       volume = {812},
       number = {2},
          eid = {112},
        pages = {112},
          doi = {10.1088/0004-637X/812/2/112},
archivePrefix = {arXiv},
       eprint = {1504.04379},
 primaryClass = {astro-ph.EP},
       adsurl = {https://ui.adsabs.harvard.edu/abs/2015ApJ...812..112S}
}

@ARTICLE{Bodenheimer1965,
       author = {{Bodenheimer}, Peter},
        title = "{Studies in Stellar Evolution. II. Lithium Depletion during the Pre-Main Contraction.}",
      journal = {\apj},
         year = 1965,
        month = aug,
       volume = {142},
        pages = {451},
          doi = {10.1086/148310},
       adsurl = {https://ui.adsabs.harvard.edu/abs/1965ApJ...142..451B}
}

@ARTICLE{Jeffries2017,
       author = {{Jeffries}, R.~D. and {Jackson}, R.~J. and {Franciosini}, E. and {Randich}, S. and {Barrado}, D. and {Frasca}, A. and {Klutsch}, A. and {Lanzafame}, A.~C. and {Prisinzano}, L. and {Sacco}, G.~G. and {Gilmore}, G. and {Vallenari}, A. and {Alfaro}, E.~J. and {Koposov}, S.~E. and {Pancino}, E. and {Bayo}, A. and {Casey}, A.~R. and {Costado}, M.~T. and {Damiani}, F. and {Hourihane}, A. and {Lewis}, J. and {Jofre}, P. and {Magrini}, L. and {Monaco}, L. and {Morbidelli}, L. and {Worley}, C.~C. and {Zaggia}, S. and {Zwitter}, T.},
        title = "{The Gaia-ESO Survey: lithium depletion in the Gamma Velorum cluster and inflated radii in low-mass pre-main-sequence stars}",
      journal = {\mnras},
         year = 2017,
        month = jan,
       volume = {464},
       number = {2},
        pages = {1456-1465},
          doi = {10.1093/mnras/stw2458},
archivePrefix = {arXiv},
       eprint = {1609.07150},
 primaryClass = {astro-ph.SR},
       adsurl = {https://ui.adsabs.harvard.edu/abs/2017MNRAS.464.1456J}
}

@ARTICLE{White+Basri2003,
       author = {{White}, Russel J. and {Basri}, Gibor},
        title = "{Very Low Mass Stars and Brown Dwarfs in Taurus-Auriga}",
      journal = {\apj},
         year = 2003,
        month = jan,
       volume = {582},
       number = {2},
        pages = {1109-1122},
          doi = {10.1086/344673},
archivePrefix = {arXiv},
       eprint = {astro-ph/0209164},
 primaryClass = {astro-ph},
       adsurl = {https://ui.adsabs.harvard.edu/abs/2003ApJ...582.1109W}
}

@ARTICLE{Fang2009,
       author = {{Fang}, M. and {van Boekel}, R. and {Wang}, W. and {Carmona}, A. and {Sicilia-Aguilar}, A. and {Henning}, Th.},
        title = "{Star and protoplanetary disk properties in Orion's suburbs}",
      journal = {\aap},
         year = 2009,
        month = sep,
       volume = {504},
       number = {2},
        pages = {461-489},
          doi = {10.1051/0004-6361/200912468},
archivePrefix = {arXiv},
       eprint = {0907.2380},
 primaryClass = {astro-ph.SR},
       adsurl = {https://ui.adsabs.harvard.edu/abs/2009A&A...504..461F}
}

@BOOK{Osterbrock2006,
       author    = {Osterbrock, Donald E. and Ferland, Gary J.},
       title     = {Astrophysics of Gaseous Nebulae and Active Galactic Nuclei},
       publisher = {University Science Books},
       year      = {2006},
       adsurl    = {https://ui.adsabs.harvard.edu/abs/2006agna.book.....O}
}

@ARTICLE{Baldwin1991,
       author = {{Baldwin}, Jack A. and {Ferland}, Gary J. and {Martin}, P.~G. and {Corbin}, Michael R. and {Cota}, Stephen A. and {Peterson}, Bradley M. and {Slettebak}, Arne},
        title = "{Physical Conditions in the Orion Nebula and an Assessment of Its Helium Abundance}",
      journal = {\apj},
         year = 1991,
        month = jun,
       volume = {374},
        pages = {580},
          doi = {10.1086/170146},
       adsurl = {https://ui.adsabs.harvard.edu/abs/1991ApJ...374..580B}
}

@ARTICLE{DaRio2009,
       author = {{Da Rio}, N. and {Robberto}, M. and {Soderblom}, D.~R. and {Panagia}, N. and {Hillenbrand}, L.~A. and {Palla}, F. and {Stassun}, K.},
        title = "{A Multi-color Optical Survey of the Orion Nebula Cluster. I. The Catalog}",
      journal = {\apjs},
         year = 2009,
        month = aug,
       volume = {183},
       number = {2},
        pages = {261-277},
          doi = {10.1088/0067-0049/183/2/261},
archivePrefix = {arXiv},
       eprint = {0906.4336},
 primaryClass = {astro-ph.GA},
       adsurl = {https://ui.adsabs.harvard.edu/abs/2009ApJS..183..261D}
}

@ARTICLE{Baraffe2015,
       author = {{Baraffe}, Isabelle and {Homeier}, Derek and {Allard}, France and {Chabrier}, Gilles},
        title = "{New evolutionary models for pre-main sequence and main sequence low-mass stars down to the hydrogen-burning limit}",
      journal = {\aap},
         year = 2015,
        month = may,
       volume = {577},
          eid = {A42},
        pages = {A42},
          doi = {10.1051/0004-6361/201425481},
archivePrefix = {arXiv},
       eprint = {1503.04107},
 primaryClass = {astro-ph.SR},
       adsurl = {https://ui.adsabs.harvard.edu/abs/2015A&A...577A..42B}
}

@ARTICLE{Hillenbrand1997,
       author = {{Hillenbrand}, Lynne A.},
        title = "{On the Stellar Population and Star-Forming History of the Orion Nebula Cluster}",
      journal = {\aj},
         year = 1997,
        month = may,
       volume = {113},
        pages = {1733-1768},
          doi = {10.1086/118389},
       adsurl = {https://ui.adsabs.harvard.edu/abs/1997AJ....113.1733H}
}

@ARTICLE{Trevor2017,
       author = {{David}, Trevor J. and {Petigura}, Erik A. and {Hillenbrand}, Lynne A. and {Cody}, Ann Marie and {Collier Cameron}, Andrew and {Stauffer}, John R. and {Fulton}, B.~J. and {Isaacson}, Howard T. and {Howard}, Andrew W. and {Howell}, Steve B. and {Everett}, Mark E. and {Wang}, Ji and {Benneke}, Bj{\"o}rn and {Hellier}, Coel and {West}, Richard G. and {Pollacco}, Don and {Anderson}, David R.},
        title = "{A Transient Transit Signature Associated with the Young Star RIK-210}",
      journal = {\apj},
         year = 2017,
        month = feb,
       volume = {835},
       number = {2},
          eid = {168},
        pages = {168},
          doi = {10.3847/1538-4357/835/2/168},
archivePrefix = {arXiv},
       eprint = {1612.03907},
 primaryClass = {astro-ph.SR},
       adsurl = {https://ui.adsabs.harvard.edu/abs/2017ApJ...835..168D}
}

@ARTICLE{Barnes2009,
       author = {{Barnes}, Jason W.},
        title = "{Transit Lightcurves of Extrasolar Planets Orbiting Rapidly Rotating Stars}",
      journal = {\apj},
         year = 2009,
        month = nov,
       volume = {705},
       number = {1},
        pages = {683-692},
          doi = {10.1088/0004-637X/705/1/683},
archivePrefix = {arXiv},
       eprint = {0909.1752},
 primaryClass = {astro-ph.EP},
       adsurl = {https://ui.adsabs.harvard.edu/abs/2009ApJ...705..683B}
}

@ARTICLE{Hauschildt1999,
       author = {{Hauschildt}, Peter H. and {Allard}, France and {Baron}, E.},
        title = "{The NextGen Model Atmosphere Grid for 3000<=T$_{eff}$<=10,000 K}",
      journal = {\apj},
         year = 1999,
        month = feb,
       volume = {512},
       number = {1},
        pages = {377-385},
          doi = {10.1086/306745},
archivePrefix = {arXiv},
       eprint = {astro-ph/9807286},
 primaryClass = {astro-ph},
       adsurl = {https://ui.adsabs.harvard.edu/abs/1999ApJ...512..377H}
}

@ARTICLE{Stauffer2017,
       author = {{Stauffer}, John and {Collier Cameron}, Andrew and {Jardine}, Moira and {David}, Trevor J. and {Rebull}, Luisa and {Cody}, Ann Marie and {Hillenbrand}, Lynne A. and {Barrado}, David and {Wolk}, Scott and {Davenport}, James and {Pinsonneault}, Marc},
        title = "{Orbiting Clouds of Material at the Keplerian Co-rotation Radius of Rapidly Rotating Low-mass WTTs in Upper Sco}",
      journal = {\aj},
         year = 2017,
        month = apr,
       volume = {153},
       number = {4},
          eid = {152},
        pages = {152},
          doi = {10.3847/1538-3881/aa5eb9},
archivePrefix = {arXiv},
       eprint = {1702.01797},
 primaryClass = {astro-ph.SR},
       adsurl = {https://ui.adsabs.harvard.edu/abs/2017AJ....153..152S}
}

@ARTICLE{Zhan2019,
       author = {{Zhan}, Z. and {G{\"u}nther}, M.~N. and {Rappaport}, S. and {Ol{\'a}h}, K. and {Mann}, A. and {Levine}, A.~M. and {Winn}, J. and {Dai}, F. and {Zhou}, G. and {Huang}, Chelsea X. and {Bouma}, L.~G. and {Ireland}, M.~J. and {Ricker}, G. and {Vanderspek}, R. and {Latham}, D. and {Seager}, S. and {Jenkins}, J. and {Caldwell}, D.~A. and {Doty}, J.~P. and {Essack}, Z. and {Furesz}, G. and {Leidos}, M.~E.~R. and {Rowden}, P. and {Smith}, J.~C. and {Stassun}, K.~G. and {Vezie}, M.},
        title = "{Complex Rotational Modulation of Rapidly Rotating M Stars Observed with TESS}",
      journal = {\apj},
         year = 2019,
        month = may,
       volume = {876},
       number = {2},
          eid = {127},
        pages = {127},
          doi = {10.3847/1538-4357/ab158c},
archivePrefix = {arXiv},
       eprint = {1903.02061},
 primaryClass = {astro-ph.SR},
       adsurl = {https://ui.adsabs.harvard.edu/abs/2019ApJ...876..127Z}
}

@ARTICLE{Koen2021,
       author = {{Koen}, C.},
        title = "{Can complex T Tauri star light curves be modelled with star-spots?}",
      journal = {\mnras},
         year = 2021,
        month = jan,
       volume = {500},
       number = {1},
        pages = {1366-1379},
          doi = {10.1093/mnras/staa3347},
       adsurl = {https://ui.adsabs.harvard.edu/abs/2021MNRAS.500.1366K}
}

@ARTICLE{Gunther2022,
       author = {{G{\"u}nther}, Maximilian N. and {Berardo}, David A. and {Ducrot}, Elsa and {Murray}, Catriona A. and {Stassun}, Keivan G. and {Olah}, Katalin and {Bouma}, L.~G. and {Rappaport}, Saul and {Winn}, Joshua N. and {Feinstein}, Adina D. and {Matthews}, Elisabeth C. and {Sebastian}, Daniel and {Rackham}, Benjamin V. and {Seli}, B{\'a}lint and {Triaud}, Amaury H.~M.~J. and {Gillen}, Edward and {Levine}, Alan M. and {Demory}, Brice-Olivier and {Gillon}, Micha{\"e}l and {Queloz}, Didier and {Ricker}, George R. and {Vanderspek}, Roland K. and {Seager}, Sara and {Latham}, David W. and {Jenkins}, Jon M. and {Brasseur}, C.~E. and {Col{\'o}n}, Knicole D. and {Daylan}, Tansu and {Delrez}, Laetitia and {Fausnaugh}, Michael and {Garcia}, Lionel J. and {Jayaraman}, Rahul and {Jehin}, Emmanuel and {Kristiansen}, Martti H. and {Kruijssen}, J.~M. Diederik and {Pedersen}, Peter Pihlmann and {Pozuelos}, Francisco J. and {Rodriguez}, Joseph E. and {Wohler}, Bill and {Zhan}, Zhuchang},
        title = "{Complex Modulation of Rapidly Rotating Young M Dwarfs: Adding Pieces to the Puzzle}",
      journal = {\aj},
         year = 2022,
        month = apr,
       volume = {163},
       number = {4},
          eid = {144},
        pages = {144},
          doi = {10.3847/1538-3881/ac503c},
archivePrefix = {arXiv},
       eprint = {2008.11681},
 primaryClass = {astro-ph.SR},
       adsurl = {https://ui.adsabs.harvard.edu/abs/2022AJ....163..144G}
}

@ARTICLE{Bouma2024,
       author = {{Bouma}, Luke G. and {Jayaraman}, Rahul and {Rappaport}, Saul and {Rebull}, Luisa M. and {Hillenbrand}, Lynne A. and {Winn}, Joshua N. and {David-Uraz}, Alexandre and {Bakos}, G{\'a}sp{\'a}r {\'A}.},
        title = "{Transient Corotating Clumps around Adolescent Low-mass Stars from Four Years of TESS}",
      journal = {\aj},
         year = 2024,
        month = jan,
       volume = {167},
       number = {1},
          eid = {38},
        pages = {38},
          doi = {10.3847/1538-3881/ad0c4c},
archivePrefix = {arXiv},
       eprint = {2309.06471},
 primaryClass = {astro-ph.SR},
       adsurl = {https://ui.adsabs.harvard.edu/abs/2024AJ....167...38B}
}

@ARTICLE{Farihi2017,
       author = {{Farihi}, J. and {von Hippel}, T. and {Pringle}, J.~E.},
        title = "{Magnetospherically-trapped dust and a possible model for the unusual transits at WD 1145+017}",
      journal = {\mnras},
         year = 2017,
        month = oct,
       volume = {471},
       number = {1},
        pages = {L145-L149},
          doi = {10.1093/mnrasl/slx122},
archivePrefix = {arXiv},
       eprint = {1707.09474},
 primaryClass = {astro-ph.SR},
       adsurl = {https://ui.adsabs.harvard.edu/abs/2017MNRAS.471L.145F}
}

@ARTICLE{Sanderson2023,
       author = {{Sanderson}, H. and {Jardine}, M. and {Collier Cameron}, A. and {Morin}, J. and {Donati}, J.-F.},
        title = "{Can scallop-shell stars trap dust in their magnetic fields?}",
      journal = {\mnras},
         year = 2023,
        month = jan,
       volume = {518},
       number = {3},
        pages = {4734-4745},
          doi = {10.1093/mnras/stac3302},
archivePrefix = {arXiv},
       eprint = {2211.04765},
 primaryClass = {astro-ph.SR},
       adsurl = {https://ui.adsabs.harvard.edu/abs/2023MNRAS.518.4734S}
}

@ARTICLE{CollierCameron1989,
       author = {{Collier Cameron}, A. and {Robinson}, R.~D.},
        title = "{Fast H{\ensuremath{\alpha}} variations on a rapidly rotating, cool main-sequence star-II. Cloud formation and ejection.}",
      journal = {\mnras},
         year = 1989,
        month = may,
       volume = {238},
        pages = {657-674},
          doi = {10.1093/mnras/238.2.657},
       adsurl = {https://ui.adsabs.harvard.edu/abs/1989MNRAS.238..657C}
}

@ARTICLE{Jardine2019,
       author = {{Jardine}, Moira and {Collier Cameron}, Andrew},
        title = "{Slingshot prominences: nature's wind gauges}",
      journal = {\mnras},
         year = 2019,
        month = jan,
       volume = {482},
       number = {3},
        pages = {2853-2860},
          doi = {10.1093/mnras/sty2872},
archivePrefix = {arXiv},
       eprint = {1810.09319},
 primaryClass = {astro-ph.SR},
       adsurl = {https://ui.adsabs.harvard.edu/abs/2019MNRAS.482.2853J}
}

@ARTICLE{Waugh2022,
       author = {{Waugh}, Rose F.~P. and {Jardine}, Moira M.},
        title = "{Magnetic confinement of dense plasma inside (and outside) stellar coronae}",
      journal = {\mnras},
         year = 2022,
        month = aug,
       volume = {514},
       number = {4},
        pages = {5465-5477},
          doi = {10.1093/mnras/stac1698},
       adsurl = {https://ui.adsabs.harvard.edu/abs/2022MNRAS.514.5465W}
}

@ARTICLE{Hayahsi1961,
       author = {{Hayashi}, Chushiro},
        title = "{Stellar Evolution in Early Phases of Gravitational Contraction}",
      journal = {\pasj},
         year = 1961,
        month = dec,
       volume = {13},
       number = {4},
        pages = {450-452},
          doi = {10.1093/pasj/13.4.450},
       adsurl = {https://ui.adsabs.harvard.edu/abs/1961PASJ...13..450H}
}

@ARTICLE{Kounkel2019,
       author = {{Kounkel}, Marina and {Covey}, Kevin},
        title = "{Untangling the Galaxy. I. Local Structure and Star Formation History of the Milky Way}",
      journal = {\aj},
         year = 2019,
        month = sep,
       volume = {158},
       number = {3},
          eid = {122},
        pages = {122},
          doi = {10.3847/1538-3881/ab339a},
archivePrefix = {arXiv},
       eprint = {1907.07709},
 primaryClass = {astro-ph.GA},
       adsurl = {https://ui.adsabs.harvard.edu/abs/2019AJ....158..122K}
}

@ARTICLE{DOrazi2009,
       author = {{D'Orazi}, V. and {Randich}, S. and {Flaccomio}, E. and {Palla}, F. and {Sacco}, G.~G. and {Pallavicini}, R.},
        title = "{Metallicity of low-mass stars in Orion}",
      journal = {\aap},
         year = 2009,
        month = jul,
       volume = {501},
       number = {3},
        pages = {973-983},
          doi = {10.1051/0004-6361/200811241},
archivePrefix = {arXiv},
       eprint = {0905.1840},
 primaryClass = {astro-ph.SR},
       adsurl = {https://ui.adsabs.harvard.edu/abs/2009A&A...501..973D}
}

@ARTICLE{Lucy1967,
       author = {{Lucy}, L.~B.},
        title = "{Gravity-Darkening for Stars with Convective Envelopes}",
      journal = {\zap},
         year = 1967,
        month = jan,
       volume = {65},
        pages = {89},
       adsurl = {https://ui.adsabs.harvard.edu/abs/1967ZA.....65...89L}
}

@ARTICLE{Portegies2010,
       author = {{Portegies Zwart}, Simon F. and {McMillan}, Stephen L.~W. and {Gieles}, Mark},
        title = "{Young Massive Star Clusters}",
      journal = {\araa},
         year = 2010,
        month = sep,
       volume = {48},
        pages = {431-493},
          doi = {10.1146/annurev-astro-081309-130834},
archivePrefix = {arXiv},
       eprint = {1002.1961},
 primaryClass = {astro-ph.GA},
       adsurl = {https://ui.adsabs.harvard.edu/abs/2010ARA&A..48..431P}
}

@ARTICLE{Lada2003,
       author = {{Lada}, Charles J. and {Lada}, Elizabeth A.},
        title = "{Embedded Clusters in Molecular Clouds}",
      journal = {\araa},
         year = 2003,
        month = jan,
       volume = {41},
        pages = {57-115},
          doi = {10.1146/annurev.astro.41.011802.094844},
archivePrefix = {arXiv},
       eprint = {astro-ph/0301540},
 primaryClass = {astro-ph},
       adsurl = {https://ui.adsabs.harvard.edu/abs/2003ARA&A..41...57L}
}

@ARTICLE{Gillen2020,
       author = {{Gillen}, Edward and {Briegal}, Joshua T. and {Hodgkin}, Simon T. and {Foreman-Mackey}, Daniel and {Van Leeuwen}, Floor and {Jackman}, James A.~G. and {McCormac}, James and {West}, Richard G. and {Queloz}, Didier and {Bayliss}, Daniel and {Goad}, Michael R. and {Watson}, Christopher A. and {Wheatley}, Peter J. and {Belardi}, Claudia and {Burleigh}, Matthew R. and {Casewell}, Sarah L. and {Jenkins}, James S. and {Raynard}, Liam and {Smith}, Alexis M.~S. and {Tilbrook}, Rosanna H. and {Vines}, Jose I.},
        title = "{NGTS clusters survey - I. Rotation in the young benchmark open cluster Blanco 1}",
      journal = {\mnras},
         year = 2020,
        month = feb,
       volume = {492},
       number = {1},
        pages = {1008-1024},
          doi = {10.1093/mnras/stz3251},
archivePrefix = {arXiv},
       eprint = {1911.09705},
 primaryClass = {astro-ph.SR},
       adsurl = {https://ui.adsabs.harvard.edu/abs/2020MNRAS.492.1008G}
}

@ARTICLE{Smith2023,
       author = {{Smith}, Gareth D. and {Gillen}, Edward and {Hodgkin}, Simon T. and {Alves}, Douglas R. and {Anderson}, David R. and {Battley}, Matthew P. and {Burleigh}, Matthew R. and {Casewell}, Sarah L. and {Gill}, Samuel and {Goad}, Michael R. and {Henderson}, Beth A. and {Jenkins}, James S. and {Kendall}, Alicia and {Moyano}, Maximiliano and {Ramsay}, Gavin and {Tilbrook}, Rosanna H. and {Vines}, Jose I. and {West}, Richard G. and {Wheatley}, Peter J.},
        title = "{NGTS clusters survey - V. Rotation in the Orion star-forming complex}",
      journal = {\mnras},
         year = 2023,
        month = jul,
       volume = {523},
       number = {1},
        pages = {169-188},
          doi = {10.1093/mnras/stad1435},
archivePrefix = {arXiv},
       eprint = {2305.04621},
 primaryClass = {astro-ph.SR},
       adsurl = {https://ui.adsabs.harvard.edu/abs/2023MNRAS.523..169S}
}

@ARTICLE{Cody2018,
       author = {{Cody}, Ann Marie and {Hillenbrand}, Lynne A.},
        title = "{The Many-faceted Light Curves of Young Disk-bearing Stars in Upper Sco -- Oph Observed by K2 Campaign 2}",
      journal = {\aj},
         year = 2018,
        month = aug,
       volume = {156},
       number = {2},
          eid = {71},
        pages = {71},
          doi = {10.3847/1538-3881/aacead},
archivePrefix = {arXiv},
       eprint = {1802.06409},
 primaryClass = {astro-ph.SR},
       adsurl = {https://ui.adsabs.harvard.edu/abs/2018AJ....156...71C}
}

@ARTICLE{McQuillan2014,
       author = {{McQuillan}, A. and {Mazeh}, T. and {Aigrain}, S.},
        title = "{Rotation Periods of 34,030 Kepler Main-sequence Stars: The Full Autocorrelation Sample}",
      journal = {\apjs},
         year = 2014,
        month = apr,
       volume = {211},
       number = {2},
          eid = {24},
        pages = {24},
          doi = {10.1088/0067-0049/211/2/24},
archivePrefix = {arXiv},
       eprint = {1402.5694},
 primaryClass = {astro-ph.SR},
       adsurl = {https://ui.adsabs.harvard.edu/abs/2014ApJS..211...24M}
}

@ARTICLE{Rebull2014,
       author = {{Rebull}, L.~M. and {Cody}, A.~M. and {Covey}, K.~R. and {G{\"u}nther}, H.~M. and {Hillenbrand}, L.~A. and {Plavchan}, P. and {Poppenhaeger}, K. and {Stauffer}, J.~R. and {Wolk}, S.~J. and {Gutermuth}, R. and {Morales-Calder{\'o}n}, M. and {Song}, I. and {Barrado}, D. and {Bayo}, A. and {James}, D. and {Hora}, J.~L. and {Vrba}, F.~J. and {Alves de Oliveira}, C. and {Bouvier}, J. and {Carey}, S.~J. and {Carpenter}, J.~M. and {Favata}, F. and {Flaherty}, K. and {Forbrich}, J. and {Hernandez}, J. and {McCaughrean}, M.~J. and {Megeath}, S.~T. and {Micela}, G. and {Smith}, H.~A. and {Terebey}, S. and {Turner}, N. and {Allen}, L. and {Ardila}, D. and {Bouy}, H. and {Guieu}, S.},
        title = "{Young Stellar Object VARiability (YSOVAR): Long Timescale Variations in the Mid-infrared}",
      journal = {\aj},
         year = 2014,
        month = nov,
       volume = {148},
       number = {5},
          eid = {92},
        pages = {92},
          doi = {10.1088/0004-6256/148/5/92},
archivePrefix = {arXiv},
       eprint = {1408.6756},
 primaryClass = {astro-ph.SR},
       adsurl = {https://ui.adsabs.harvard.edu/abs/2014AJ....148...92R}
}

@INPROCEEDINGS{Jenkins2016TESSSPOC,
       author = {{Jenkins}, Jon M. and {Twicken}, Joseph D. and {McCauliff}, Sean and {Campbell}, Jennifer and {Sanderfer}, Dwight and {Lung}, David and {Mansouri-Samani}, Masoud and {Girouard}, Forrest and {Tenenbaum}, Peter and {Klaus}, Todd and {Smith}, Jeffrey C. and {Caldwell}, Douglas A. and {Chacon}, A.~D. and {Henze}, Christopher and {Heiges}, Cory and {Latham}, David W. and {Morgan}, Edward and {Swade}, Daryl and {Rinehart}, Stephen and {Vanderspek}, Roland},
        title = "{The TESS science processing operations center}",
    booktitle = {Software and Cyberinfrastructure for Astronomy IV},
         year = 2016,
       editor = {{Chiozzi}, Gianluca and {Guzman}, Juan C.},
       series = {Society of Photo-Optical Instrumentation Engineers (SPIE) Conference Series},
       volume = {9913},
        month = aug,
          eid = {99133E},
        pages = {99133E},
          doi = {10.1117/12.2233418},
       adsurl = {https://ui.adsabs.harvard.edu/abs/2016SPIE.9913E..3EJ}
}

@ARTICLE{Huang2020QLPdoi,
       author = {{Huang}, Chelsea X. and {Vanderburg}, Andrew and {P{\'a}l}, Andras and {Sha}, Lizhou and {Yu}, Liang and {Fong}, Willie and {Fausnaugh}, Michael and {Shporer}, Avi and {Guerrero}, Natalia and {Vanderspek}, Roland and {Ricker}, George},
        title = "{Photometry of 10 Million Stars from the First Two Years of TESS Full Frame Images: Part II}",
      journal = {Research Notes of the American Astronomical Society},
         year = 2020,
        month = nov,
       volume = {4},
       number = {11},
          eid = {206},
        pages = {206},
          doi = {10.3847/2515-5172/abca2d},
       adsurl = {https://ui.adsabs.harvard.edu/abs/2020RNAAS...4..206H}
}

@misc{Lightkurve2018lightkurve,
       author = {{Lightkurve Collaboration} and {Cardoso}, Jos{\'e} Vin{\'\i}cius de Miranda and {Hedges}, Christina and {Gully-Santiago}, Michael and {Saunders}, Nicholas and {Cody}, Ann Marie and {Barclay}, Thomas and {Hall}, Oliver and {Sagear}, Sheila and {Turtelboom}, Emma and {Zhang}, Johnny and {Tzanidakis}, Andy and {Mighell}, Ken and {Coughlin}, Jeff and {Bell}, Keaton and {Berta-Thompson}, Zach and {Williams}, Peter and {Dotson}, Jessie and {Barentsen}, Geert},
        title = "{Lightkurve: Kepler and TESS time series analysis in Python}",
 howpublished = {Astrophysics Source Code Library, record ascl:1812.013},
         year = 2018,
        month = dec,
          eid = {ascl:1812.013},
archivePrefix = {ascl},
       eprint = {1812.013},
       adsurl = {https://ui.adsabs.harvard.edu/abs/2018ascl.soft12013L}
}

@INPROCEEDINGS{TESSSPOC2020,
       author = {{Smith}, J. and {Caldwell}, D. and {Jenkins}, J. and {Morris}, R. and {Rose}, M. and {Tenenbaum}, P. and {Ting}, E. and {Twicken}, J.},
        title = "{Finding Every Planet We Can: Optimizing TESS SPOC Pipeline Transit Detection}",
    booktitle = {American Astronomical Society Meeting Abstracts \#235},
         year = 2020,
       series = {American Astronomical Society Meeting Abstracts},
       volume = {235},
        month = jan,
          eid = {174.02},
        pages = {174.02},
       adsurl = {https://ui.adsabs.harvard.edu/abs/2020AAS...23517402S}
}

@ARTICLE{Savitzky_1964,
       author = {{Savitzky}, A. and {Golay}, M.~J.~E.},
        title = "{Smoothing and differentiation of data by simplified least squares procedures}",
      journal = {Analytical Chemistry},
         year = 1964,
        month = jan,
       volume = {36},
        pages = {1627-1639},
          doi = {10.1021/ac60214a047},
       adsurl = {https://ui.adsabs.harvard.edu/abs/1964AnaCh..36.1627S}
}

@ARTICLE{Hattori_2022,
       author = {{Hattori}, Soichiro and {Foreman-Mackey}, Daniel and {Hogg}, David W. and {Montet}, Benjamin T. and {Angus}, Ruth and {Pritchard}, T.~A. and {Curtis}, Jason L. and {Sch{\"o}lkopf}, Bernhard},
        title = "{The unpopular Package: A Data-driven Approach to Detrending TESS Full-frame Image Light Curves}",
      journal = {\aj},
         year = 2022,
        month = jun,
       volume = {163},
       number = {6},
          eid = {284},
        pages = {284},
          doi = {10.3847/1538-3881/ac625a},
archivePrefix = {arXiv},
       eprint = {2106.15063},
 primaryClass = {astro-ph.IM},
       adsurl = {https://ui.adsabs.harvard.edu/abs/2022AJ....163..284H}
}

@ARTICLE{Watson2004,
       author = {{Watson}, C.~A. and {Dhillon}, V.~S.},
        title = "{The effect of star-spots on eclipse timings of binary stars}",
      journal = {\mnras},
         year = 2004,
        month = jun,
       volume = {351},
       number = {1},
        pages = {110-116},
          doi = {10.1111/j.1365-2966.2004.07763.x},
archivePrefix = {arXiv},
       eprint = {astro-ph/0402670},
 primaryClass = {astro-ph},
       adsurl = {https://ui.adsabs.harvard.edu/abs/2004MNRAS.351..110W}
}

@ARTICLE{David2016,
       author = {{David}, Trevor J. and {Hillenbrand}, Lynne A. and {Petigura}, Erik A. and {Carpenter}, John M. and {Crossfield}, Ian J.~M. and {Hinkley}, Sasha and {Ciardi}, David R. and {Howard}, Andrew W. and {Isaacson}, Howard T. and {Cody}, Ann Marie and {Schlieder}, Joshua E. and {Beichman}, Charles A. and {Barenfeld}, Scott A.},
        title = "{A Neptune-sized transiting planet closely orbiting a 5-10-million-year-old star}",
      journal = {\nat},
         year = 2016,
        month = jun,
       volume = {534},
       number = {7609},
        pages = {658-661},
          doi = {10.1038/nature18293},
archivePrefix = {arXiv},
       eprint = {1606.06729},
 primaryClass = {astro-ph.EP},
       adsurl = {https://ui.adsabs.harvard.edu/abs/2016Natur.534..658D}
}

@ARTICLE{Ohno2022,
       author = {{Ohno}, Kazumasa and {Thao}, Pa Chia and {Mann}, Andrew W. and {Fortney}, Jonathan J.},
        title = "{A Circumplanetary Dust Ring May Explain the Extreme Spectral Slope of the 10 Myr Young Exoplanet K2-33b}",
      journal = {\apjl},
         year = 2022,
        month = dec,
       volume = {940},
       number = {2},
          eid = {L30},
        pages = {L30},
          doi = {10.3847/2041-8213/ac9f3f},
archivePrefix = {arXiv},
       eprint = {2211.07706},
 primaryClass = {astro-ph.EP},
       adsurl = {https://ui.adsabs.harvard.edu/abs/2022ApJ...940L..30O}
}

@ARTICLE{Barbieri2023,
       author = {{Barbieri}, Mauro},
        title = "{ESO/HARPS Radial Velocities Catalog}",
      journal = {arXiv e-prints},
         year = 2023,
        month = dec,
          eid = {arXiv:2312.06586},
        pages = {arXiv:2312.06586},
          doi = {10.48550/arXiv.2312.06586},
archivePrefix = {arXiv},
       eprint = {2312.06586},
 primaryClass = {astro-ph.SR},
       adsurl = {https://ui.adsabs.harvard.edu/abs/2023arXiv231206586B}
}

@INPROCEEDINGS{Lovis2006,
       author = {{Lovis}, Christophe and {Pepe}, Francesco and {Bouchy}, Fran{\c{c}}ois and {Lo Curto}, Gaspare and {Mayor}, Michel and {Pasquini}, Luca and {Queloz}, Didier and {Rupprecht}, Gero and {Udry}, St{\'e}phane and {Zucker}, Shay},
        title = "{The exoplanet hunter HARPS: unequalled accuracy and perspectives toward 1 cm s $^{-1}$ precision}",
    booktitle = {Ground-based and Airborne Instrumentation for Astronomy},
         year = 2006,
       editor = {{McLean}, Ian S. and {Iye}, Masanori},
       series = {Society of Photo-Optical Instrumentation Engineers (SPIE) Conference Series},
       volume = {6269},
        month = jun,
          eid = {62690P},
        pages = {62690P},
          doi = {10.1117/12.669991},
       adsurl = {https://ui.adsabs.harvard.edu/abs/2006SPIE.6269E..0PL}
}

@ARTICLE{Jackman2020,
       author = {{Jackman}, James A.~G. and {Wheatley}, Peter J. and {Acton}, Jack S. and {Anderson}, David R. and {Belardi}, Claudia and {Burleigh}, Matthew R. and {Casewell}, Sarah L. and {Eigm{\"u}ller}, Philipp and {Gill}, Samuel and {Gillen}, Edward and {Goad}, Michael R. and {Grange}, Andrew and {Hodgkin}, Simon T. and {Jenkins}, James S. and {McCormac}, James and {Moyano}, Maximiliano and {Queloz}, Didier and {Raynard}, Liam and {Tilbrook}, Rosanna H. and {Watson}, Christopher A. and {West}, Richard G.},
        title = "{NGTS clusters survey - II. White-light flares from the youngest stars in Orion}",
      journal = {\mnras},
         year = 2020,
        month = sep,
       volume = {497},
       number = {1},
        pages = {809-817},
          doi = {10.1093/mnras/staa1971},
archivePrefix = {arXiv},
       eprint = {2007.01553},
 primaryClass = {astro-ph.SR},
       adsurl = {https://ui.adsabs.harvard.edu/abs/2020MNRAS.497..809J}
}

@ARTICLE{Smith2021,
       author = {{Smith}, Gareth D. and {Gillen}, Edward and {Queloz}, Didier and {Hillenbrand}, Lynne A. and {Acton}, Jack S. and {Alves}, Douglas R. and {Anderson}, David R. and {Bayliss}, Daniel and {Briegal}, Joshua T. and {Burleigh}, Matthew R. and {Casewell}, Sarah L. and {Delrez}, Laetitia and {Dransfield}, Georgina and {Ducrot}, Elsa and {Gill}, Samuel and {Gillon}, Micha{\"e}l and {Goad}, Michael R. and {G{\"u}nther}, Maximilian N. and {Henderson}, Beth A. and {Jenkins}, James S. and {Jehin}, Emmanu{\"e}l and {Moyano}, Maximiliano and {Murray}, Catriona A. and {Pedersen}, Peter P. and {Sebastian}, Daniel and {Thompson}, Samantha and {Tilbrook}, Rosanna H. and {Triaud}, Amaury H.~M.~J. and {Vines}, Jose I. and {Wheatley}, Peter J.},
        title = "{NGTS clusters survey - III. A low-mass eclipsing binary in the Blanco 1 open cluster spanning the fully convective boundary}",
      journal = {\mnras},
         year = 2021,
        month = nov,
       volume = {507},
       number = {4},
        pages = {5991-6011},
          doi = {10.1093/mnras/stab2374},
archivePrefix = {arXiv},
       eprint = {2109.00836},
 primaryClass = {astro-ph.SR},
       adsurl = {https://ui.adsabs.harvard.edu/abs/2021MNRAS.507.5991S}
}

@ARTICLE{Moulton2023,
       author = {{Moulton}, Tyler and {Hodgkin}, Simon T. and {Smith}, Gareth D. and {Briegal}, Joshua T. and {Gillen}, Edward and {Acton}, Jack S. and {Battley}, Matthew P. and {Burleigh}, Matthew R. and {Casewell}, Sarah L. and {Gill}, Samuel and {Goad}, Michael R. and {Henderson}, Beth A. and {Kendall}, Alicia and {Ramsay}, Gavin and {Tilbrook}, Rosanna H. and {Wheatley}, Peter J.},
        title = "{NGTS clusters survey - IV. Search for Dipper stars in the Orion Nebular Cluster}",
      journal = {\mnras},
         year = 2023,
        month = may,
       volume = {521},
       number = {2},
        pages = {1700-1726},
          doi = {10.1093/mnras/stad364},
archivePrefix = {arXiv},
       eprint = {2304.09942},
 primaryClass = {astro-ph.SR},
       adsurl = {https://ui.adsabs.harvard.edu/abs/2023MNRAS.521.1700M}
}

@ARTICLE{Barbary2016,
       author = {{Barbary}, Kyle},
        title = "{SEP: Source Extractor as a library}",
      journal = {The Journal of Open Source Software},
         year = 2016,
        month = oct,
       volume = {1},
       number = {6},
          eid = {58},
        pages = {58},
          doi = {10.21105/joss.00058},
       adsurl = {https://ui.adsabs.harvard.edu/abs/2016JOSS....1...58B}
}

@ARTICLE{Groote1982,
       author = {{Groote}, D. and {Hunger}, K.},
        title = "{Shell and photosphere of OMI ORI E : new observations and improved model.}",
      journal = {\aap},
         year = 1982,
        month = dec,
       volume = {116},
        pages = {64-79},
       adsurl = {https://ui.adsabs.harvard.edu/abs/1982A&A...116...64G}
}

@ARTICLE{Townsend2013,
       author = {{Townsend}, R.~H.~D. and {Rivinius}, Th. and {Rowe}, J.~F. and {Moffat}, A.~F.~J. and {Matthews}, J.~M. and {Bohlender}, D. and {Neiner}, C. and {Telting}, J.~H. and {Guenther}, D.~B. and {Kallinger}, T. and {Kuschnig}, R. and {Rucinski}, S.~M. and {Sasselov}, D. and {Weiss}, W.~W.},
        title = "{MOST Observations of {\ensuremath{\sigma}} Ori E: Challenging the Centrifugal Breakout Narrative}",
      journal = {\apj},
         year = 2013,
        month = may,
       volume = {769},
       number = {1},
          eid = {33},
        pages = {33},
          doi = {10.1088/0004-637X/769/1/33},
archivePrefix = {arXiv},
       eprint = {1304.2392},
 primaryClass = {astro-ph.SR},
       adsurl = {https://ui.adsabs.harvard.edu/abs/2013ApJ...769...33T}
}

@ARTICLE{Johns-Krull2007,
       author = {{Johns-Krull}, Christopher M.},
        title = "{The Magnetic Fields of Classical T Tauri Stars}",
      journal = {\apj},
         year = 2007,
        month = aug,
       volume = {664},
       number = {2},
        pages = {975-985},
          doi = {10.1086/519017},
archivePrefix = {arXiv},
       eprint = {0704.2923},
 primaryClass = {astro-ph},
       adsurl = {https://ui.adsabs.harvard.edu/abs/2007ApJ...664..975J}
}

@ARTICLE{Donati2009,
       author = {{Donati}, J.-F. and {Landstreet}, J.~D.},
        title = "{Magnetic Fields of Nondegenerate Stars}",
      journal = {\araa},
         year = 2009,
        month = sep,
       volume = {47},
       number = {1},
        pages = {333-370},
          doi = {10.1146/annurev-astro-082708-101833},
archivePrefix = {arXiv},
       eprint = {0904.1938},
 primaryClass = {astro-ph.SR},
       adsurl = {https://ui.adsabs.harvard.edu/abs/2009ARA&A..47..333D}
}

@ARTICLE{Townsend2005,
       author = {{Townsend}, R.~H.~D. and {Owocki}, S.~P. and {Groote}, D.},
        title = "{The Rigidly Rotating Magnetosphere of {\ensuremath{\sigma}} Orionis E}",
      journal = {\apjl},
         year = 2005,
        month = sep,
       volume = {630},
       number = {1},
        pages = {L81-L84},
          doi = {10.1086/462413},
archivePrefix = {arXiv},
       eprint = {astro-ph/0503668},
 primaryClass = {astro-ph},
       adsurl = {https://ui.adsabs.harvard.edu/abs/2005ApJ...630L..81T}
}

@ARTICLE{DaleyYates2024,
       author = {{Daley-Yates}, S. and {Jardine}, Moira M.},
        title = "{Simulating stellar coronal rain and slingshot prominences}",
      journal = {\mnras},
         year = 2024,
        month = oct,
       volume = {534},
       number = {1},
        pages = {621-633},
          doi = {10.1093/mnras/stae2131},
archivePrefix = {arXiv},
       eprint = {2409.07297},
 primaryClass = {astro-ph.SR},
       adsurl = {https://ui.adsabs.harvard.edu/abs/2024MNRAS.534..621D}
}

@ARTICLE{vandam2020,
       author = {{van Dam}, Dirk M. and {Kenworthy}, Matthew A. and {David}, Trevor J. and {Mamajek}, Eric E. and {Hillenbrand}, Lynne A. and {Cody}, Ann Marie and {Howard}, Andrew W. and {Isaacson}, Howard and {Ciardi}, David R. and {Rebull}, Luisa M. and {Stauffer}, John R. and {Patel}, Rahul and {Cameron + WASP Collaborators}, Andrew Collier and {Rodriguez}, Joseph E. and {Pojma{\'n}ski}, Grzegorz and {Gonzales}, Erica J. and {Schlieder}, Joshua E. and {Hambsch}, Franz-Josef and {Dufoer}, Sjoerd and {Vanmunster}, Tonny and {Dubois}, Franky and {Vanaverbeke}, Siegfried and {Logie}, Ludwig and {Rau}, Steve},
        title = "{An Asymmetric Eclipse Seen toward the Pre-main-sequence Binary System V928 Tau}",
      journal = {\aj},
         year = 2020,
        month = dec,
       volume = {160},
       number = {6},
          eid = {285},
        pages = {285},
          doi = {10.3847/1538-3881/abc259},
archivePrefix = {arXiv},
       eprint = {2010.11199},
 primaryClass = {astro-ph.SR},
       adsurl = {https://ui.adsabs.harvard.edu/abs/2020AJ....160..285V}
}

@ARTICLE{grinin2008,
       author = {{Grinin}, V. and {Stempels}, H.~C. and {Gahm}, G.~F. and {Sergeev}, S. and {Arkharov}, A. and {Barsunova}, O. and {Tambovtseva}, L.},
        title = "{The unusual pre-main-sequence star V718 Persei (HMW 15). Photometry and spectroscopy across the eclipse}",
      journal = {\aap},
         year = 2008,
        month = oct,
       volume = {489},
       number = {3},
        pages = {1233-1238},
          doi = {10.1051/0004-6361:200810349},
archivePrefix = {arXiv},
       eprint = {0808.1069},
 primaryClass = {astro-ph},
       adsurl = {https://ui.adsabs.harvard.edu/abs/2008A&A...489.1233G}
}

@ARTICLE{rodriguezledesma2013,
       author = {{Rodr{\'\i}guez-Ledesma}, M.~V. and {Mundt}, R. and {Pintado}, O. and {Boudreault}, S. and {Hessman}, F. and {Herbst}, W.},
        title = "{The spectral type of CHS 7797 - an intriguing very low mass periodic variable in the Orion Nebula Cluster}",
      journal = {\aap},
         year = 2013,
        month = mar,
       volume = {551},
          eid = {A44},
        pages = {A44},
          doi = {10.1051/0004-6361/201219748},
archivePrefix = {arXiv},
       eprint = {1301.5913},
 primaryClass = {astro-ph.SR},
       adsurl = {https://ui.adsabs.harvard.edu/abs/2013A&A...551A..44R}
}

@ARTICLE{kearns1998,
       author = {{Kearns}, Kristin E. and {Herbst}, William},
        title = "{Additional Periodic Variables in NGC 2264}",
      journal = {\aj},
         year = 1998,
        month = jul,
       volume = {116},
       number = {1},
        pages = {261-265},
          doi = {10.1086/300426},
       adsurl = {https://ui.adsabs.harvard.edu/abs/1998AJ....116..261K}
}

@ARTICLE{winn2004,
       author = {{Winn}, Joshua N. and {Holman}, Matthew J. and {Johnson}, John A. and {Stanek}, Krzysztof Z. and {Garnavich}, Peter M.},
        title = "{KH 15D: Gradual Occultation of a Pre-Main-Sequence Binary}",
      journal = {\apjl},
         year = 2004,
        month = mar,
       volume = {603},
       number = {1},
        pages = {L45-L48},
          doi = {10.1086/383089},
archivePrefix = {arXiv},
       eprint = {astro-ph/0312458},
 primaryClass = {astro-ph},
       adsurl = {https://ui.adsabs.harvard.edu/abs/2004ApJ...603L..45W}
}

@ARTICLE{chiang2004,
       author = {{Chiang}, Eugene I. and {Murray-Clay}, Ruth A.},
        title = "{The Circumbinary Ring of KH 15D}",
      journal = {\apj},
         year = 2004,
        month = jun,
       volume = {607},
       number = {2},
        pages = {913-920},
          doi = {10.1086/383522},
archivePrefix = {arXiv},
       eprint = {astro-ph/0312515},
 primaryClass = {astro-ph},
       adsurl = {https://ui.adsabs.harvard.edu/abs/2004ApJ...607..913C}
}

@ARTICLE{zhu2022,
       author = {{Zhu}, Wei and {Bernhard}, Klaus and {Dai}, Fei and {Fang}, Min and {Zanazzi}, J.~J. and {Zang}, Weicheng and {Dong}, Subo and {Hambsch}, Franz-Josef and {Gan}, Tianjun and {Wu}, Zexuan and {Poon}, Michael},
        title = "{Two Candidate KH 15D-like Systems from the Zwicky Transient Facility}",
      journal = {\apjl},
         year = 2022,
        month = jul,
       volume = {933},
       number = {1},
          eid = {L21},
        pages = {L21},
          doi = {10.3847/2041-8213/ac7b2d},
archivePrefix = {arXiv},
       eprint = {2206.00813},
 primaryClass = {astro-ph.SR},
       adsurl = {https://ui.adsabs.harvard.edu/abs/2022ApJ...933L..21Z}
}




\appendix

\section{Extra figures}

\begin{figure*}
	\includegraphics[width=0.7\textwidth]{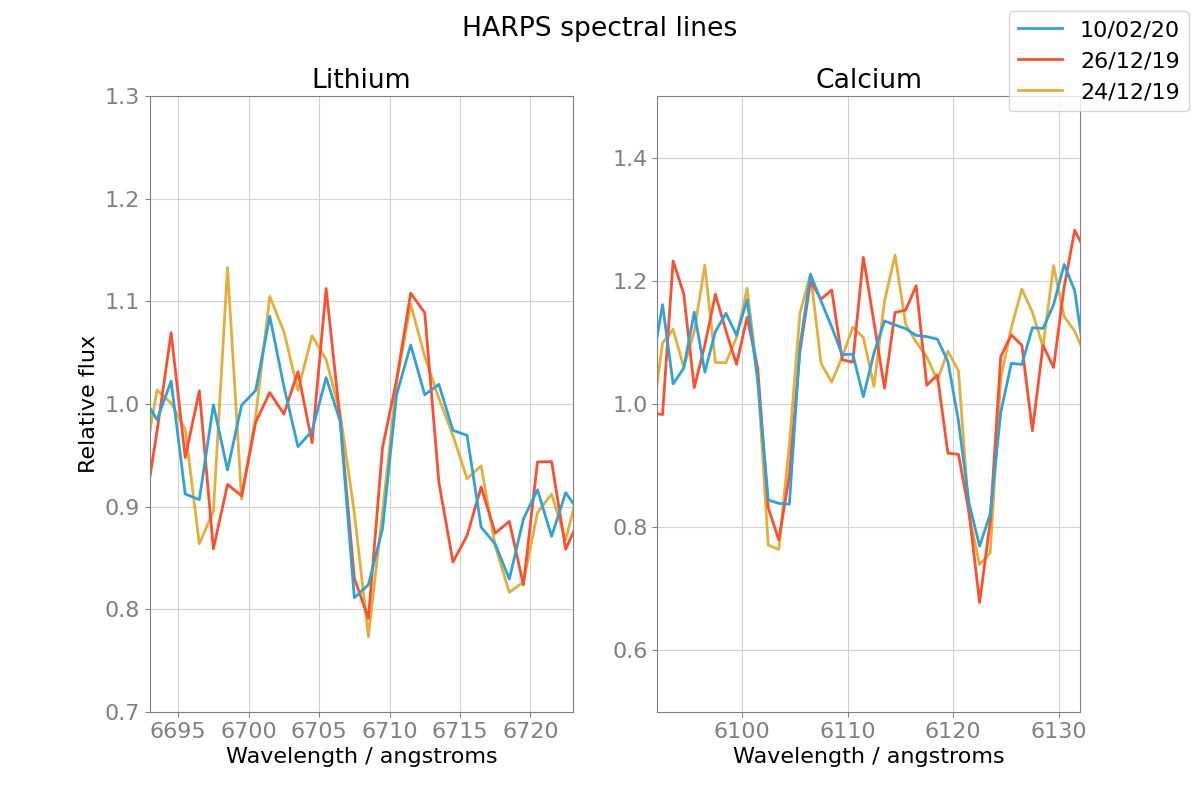}
    \caption{Lithum and Calcium absorption features in the three HARPS spectra. These observations have been sky-subtracted and binned into $\SI{1}{\angstrom}$ chunks.}
    \label{limb lines}
\end{figure*}



\bsp	
\label{lastpage}
\end{document}